\documentclass{aa}  

\usepackage{natbib}
\usepackage{graphicx}
\usepackage{txfonts}
\usepackage{hyperref}
\usepackage{xcolor}

\newcommand{\beforeReferee}[1]{}

\newcommand{\new}[1]{#1}

\newcommand{\ignoreThis}[1]{ }

\newcommand\TODO[1]{} 
\newcommand\TBC[1]{#1}
\providecommand{\dt}[1]{{\tt #1}} 

\newcommand\gaia{\textit{Gaia}}

\newcommand\gdrtwo{\gaia~DR2 }
\newcommand\gedrthree{\gaia~EDR3}

\newcommand\gdr[1]{\gaia~DR#1}         
\newcommand\egdr[1]{\gaia~EDR#1}       
\newcommand\hip{\textsc{Hipparcos}}

\newcommand\tyctwo{{Tycho}-2}

\newcommand\secref[1]{Sect.~\ref{#1}}
\newcommand\figref[1]{Fig.~\ref{#1}}

\newcommand\tabref[1]{Table~\ref{#1}}

\bibpunct{(}{)}{;}{a}{}{,} 

\newcommand\bpminrp{\ensuremath{G_\mathrm{BP}-G_\mathrm{RP}}}

\def\a0{$A_{\rm 0}$}

\def\gmag{$G$}
\def\gbp{$G_{\rm BP }$}
\def\grp{$G_{\rm RP }$}

\def\masyr{\,mas\,yr$^{-1}$}
\def\es{\,e$^-$\,s$^{-1}$}
\def\muas{\,$\mu$as}
\def\logg{$\log g$}

\def\parallax{$\varpi$}
\def\pmra{$\mu_{\alpha* }$}
\def\pmdec{$\mu_{\delta }$}
\def\propm{proper motions }

\begin{document} 

   \title{{\gaia} Early Data Release 3 -- Catalogue validation}



\author{
C.~Fabricius\inst{\ref{inst:ieec}}\fnmsep\thanks{Corresponding author: C. Fabricius\newline
e-mail: \href{mailto:claus@fqa.ub.edu}{\tt claus@fqa.ub.edu}}, 
X.~Luri\inst{\ref{inst:ieec}}, 
F.~Arenou\inst{\ref{inst:gepi}}, 
C.~Babusiaux\inst{\ref{inst:IPAG},\ref{inst:gepi}}, 
A.~Helmi\inst{\ref{inst:kapteyn}}, 
T.~Muraveva\inst{\ref{inst:bologna}}, 
C.~Reyl\'e\inst{\ref{inst:utinam}}, 
F.~Spoto\inst{\ref{inst:cfa}}, 
A.~Vallenari\inst{\ref{inst:padova}}, 
T.~Antoja\inst{\ref{inst:ieec}},
E.~Balbinot\inst{\ref{inst:kapteyn}},
C.~Barache\inst{\ref{inst:SYRTE}}, 
N.~Bauchet\inst{\ref{inst:gepi}}, 
A.~Bragaglia\inst{\ref{inst:bologna}},
D.~Busonero\inst{\ref{inst:torino}},
T.~Cantat-Gaudin\inst{\ref{inst:ieec}},
J.~M.~Carrasco\inst{\ref{inst:ieec}},
S.~Diakit\'e\inst{\ref{inst:utinam}}, 
M.~Fabrizio\inst{\ref{inst:roma},\ref{inst:asdc}},
F.~Figueras\inst{\ref{inst:ieec}}, 
A.~Garcia-Gutierrez\inst{\ref{inst:ieec}},
A.~Garofalo\inst{\ref{inst:bologna}},
C.~Jordi\inst{\ref{inst:ieec}}, 
P.~Kervella\inst{\ref{inst:lesia}},
S.~Khanna\inst{\ref{inst:kapteyn}},
N.~Leclerc\inst{\ref{inst:gepi}}, 
E.~Licata\inst{\ref{inst:torino}},
S.~Lambert\inst{\ref{inst:SYRTE}}, 
P.~M.~Marrese\inst{\ref{inst:roma},\ref{inst:asdc}},
A.~Masip\inst{\ref{inst:ieec}},
P.~Ramos\inst{\ref{inst:ieec}},
N.~Robichon\inst{\ref{inst:gepi}}, 
A.~C.~Robin\inst{\ref{inst:utinam}}, 
M.~Romero-G\'omez\inst{\ref{inst:ieec}},
S.~Rubele\inst{\ref{inst:padova}}, 
M.~Weiler\inst{\ref{inst:ieec}}
}

\institute{
Dept. FQA, Institut de Ci\`encies del Cosmos, Universitat de Barcelona (IEEC-UB), Mart\'i i Franqu\`es 1, E-08028 Barcelona, Spain 
\label{inst:ieec}
\and
GEPI, Observatoire de Paris, Universit{\'e} PSL, CNRS, 5 Place Jules Janssen, 92190 Meudon, France
\label{inst:gepi}
\and
Univ. Grenoble Alpes, CNRS, IPAG, 38000 Grenoble, France
\label{inst:IPAG}
\and
Kapteyn Astronomical Institute, University of Groningen, Landleven 12, 9747 AD Groningen, The Netherlands
\label{inst:kapteyn}
\and 
INAF - Osservatorio di Astrofisica e Scienza dello Spazio di Bologna, via Piero Gobetti 93/3, 40129 Bologna,  Italy                                        \label{inst:bologna}
\and
Institut UTINAM, CNRS, OSU THETA Franche-Comt\'e Bourgogne, Univ. Bourgogne Franche-Comt\'e, 25000 Besan\c{c}on,
France\label{inst:utinam}
\and
Harvard-Smithsonian Center for Astrophysics, 60 Garden St., MS 15, Cambridge, MA 02138, USA\label{inst:cfa}
\and
INAF, Osservatorio Astronomico di Padova, Vicolo Osservatorio, Padova, I-35131, Italy
\label{inst:padova}
\and 
SYRTE, Observatoire de Paris, Universit\'e PSL, CNRS, Sorbonne Universit\'e, LNE, 61 avenue de l'Observatoire, 75014 Paris, France
\label{inst:SYRTE}
\and 
INAF - Osservatorio Astrofisico di Torino, Via Osservatorio 20, 10025 Pino Torinese, Torino,  Italy                           \label{inst:torino}
\and
INAF - Osservatorio Astronomico di Roma, Via di Frascati 33, 00078 Monte Porzio Catone (Roma), Italy
\label{inst:roma}
\and
ASI Science Data Center, Via del Politecnico, Roma
\label{inst:asdc}
\and
LESIA, Observatoire de Paris, Universit{\'e} PSL, CNRS, Sorbonne Universit{\'e}, Universit{\'e} de Paris, 5 place Jules Janssen, 92195 Meudon, France
\label{inst:lesia}
}

\date{ }

\abstract
{
The third {\gaia} data release is published in two stages. The early part, \egdr{3}, gives very precise astrometric and photometric properties for nearly two billion sources together with seven million radial velocities from \gdr{2}. The full release, \gdr{3}, will add radial velocities, spectra,
light curves, and astrophysical parameters for a large subset of the sources, as well as orbits for solar system objects.
}
{
Before the publication of the catalogue, many different data items have undergone dedicated validation processes.  The goal of this paper is to describe the validation results in terms of completeness, accuracy, and precision for the \egdr{3} data and to provide recommendations for the use of the catalogue data.
}
{
The validation processes include a systematic analysis of the catalogue contents to detect anomalies, either individual errors or statistical properties, using statistical analysis and comparisons to the previous release as well as to external data and to models.
}
{
\egdr{3}\ represents a major step forward, compared to \gdr{2}, in terms of
precision, accuracy, and completeness for both astrometry and photometry.
We provide recommendations for dealing with 
issues related to the parallax zero point, negative parallaxes, photometry for faint sources, and the quality indicators.
}
{}


   \titlerunning{\egdr{3} -- Catalogue validation}
   \authorrunning{C. Fabricius et al.}

   \maketitle

%

\section{Introduction}
%
The third data release from the European Space Agency mission {\gaia} \citep{2016A&A...595A...1G, EDR3-DPACP-130} covers observations made between July 2014 and
May 2017. It takes place in two stages, where the first (early) stage,
\egdr{3}, provides the updated astrometry and photometry. For convenience it
also includes (nearly all)  radial velocities from the second data release, \gdr{2}\
\citep{DR2-DPACP-36}. The second stage, the full
\gdr{3}, will include the same sources
as \egdr{3}\ and
add new radial velocities, spectra, light curves, astrophysical parameters,
and orbits for solar system objects, as well as a detailed analysis of
quasars and extended objects, for example.  

This paper describes the validation of \egdr{3} with the aim of facilitating
the optimal use of the catalogue, comprehending its contents, and
especially exposing the known issues.   The approach followed is a
transverse analysis of the properties of the various contents from an internal
as well as external point of view. We also use the previous release, \gdr{2},
as a reference for comparisons in order to quantify the changes and improvements from one
release to the next.

The general properties of the catalogue are described in \secref{sec:comp}.
This includes the completeness in terms of limiting magnitude and
angular resolution and also in terms of high proper motion stars. Likewise, we discuss how sources, and their identifiers, have changed since \gdr{2}.

The new astrometric solution \citep{EDR3-DPACP-128,EDR3-DPACP-133} determines two parameters (position), `2p', five parameters (position, parallax,
proper motion), `5p', or six parameters, `6p', for a source. In the latter
case, the sixth parameter is the source colour (effective wavelength),
listed in the \gaia\ archive as the \dt{pseudocolour}. The use
of three levels of solutions introduces tricky issues in the validation.
Another important topic is the presence of spurious solutions and the means
available to identify them.  The validation of the astrometry is discussed
below in \secref{sec:ast}.

Cycle 3 photometry is described by \citet{EDR3-DPACP-117} both in terms of
the various calibration steps and in terms of data quality. Important
changes have been made to the background modelling, leading to improvements for \gbp\ and
\grp\ photometry, which, however, suffers from other issues in the faint end.
Changes to the overall response modelling have had undesired effects in a few
cases, leading to the elimination of photometry for 5.4~million sources over a
wide range of brightness. The
validation of the photometry is discussed in \secref{sec:phot}.

Finally, \secref{sec:global} presents a statistical approach to comparing
\egdr{3} and \gdr{2}.\ Additionally, we describe the overall results and recommendations 
in \secref{sec:conclusions}.

Updated radial velocities will only appear in \gdr{3} and, as mentioned, \egdr{3}
therefore contains values copied from \gdr{2}. The process for the
identification of sources and the validation of the velocities is described in
detail in \cite{EDR3-DPACP-121} and is therefore not discussed here.
As a result of this process, 14\,800 radial velocities (0.2\%)
were discarded. 

In addition to the papers mentioned above, the \gaia\ archive provides online documentation\footnote{
\url{https://gea.esac.esa.int/archive/documentation/GEDR3/index.html}}
with additional details about the data processing and the description of the
data items in the published catalogue and its various accompanying tables.
\gaia\ jargon is difficult to avoid and we therefore include a short dictionary
of \gaia\ and \egdr{3}\ related terms in Appendix\,\ref{sec:acronyms}.


\section{General tests and completeness }\label{sec:comp}
%

Our general tests cover a wide range of issues from simple, yet indispensable,
checks that the catalogue has  been correctly populated to more sophisticated
statistical tests on completeness.
 
%
\subsection{\gdr{2}\ sources in \egdr{3}}\label{sec:comp_source_id}
\begin{figure}[h]
\includegraphics[width=\hsize]{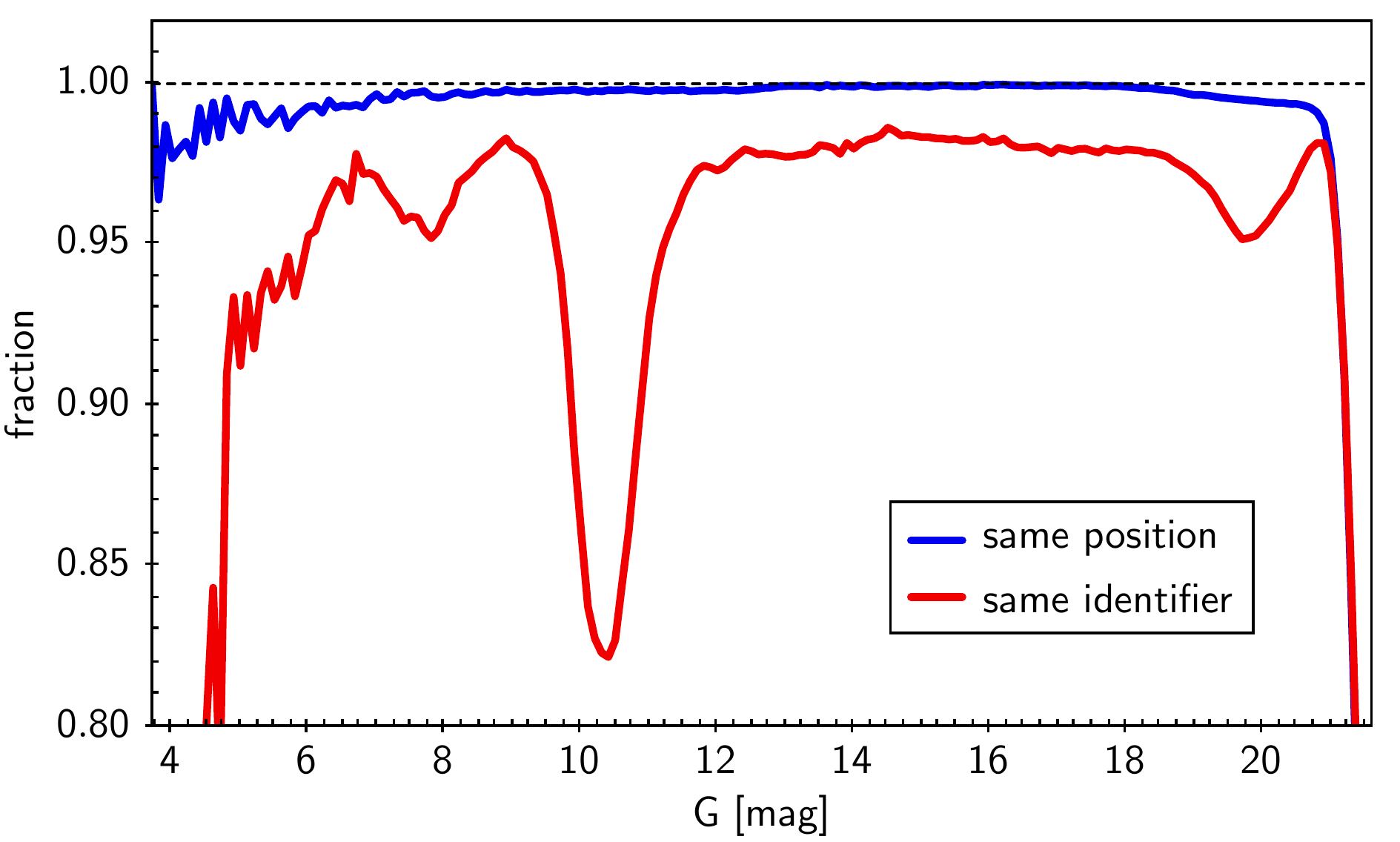}
\caption{Fraction of \gdr{2}\ sources maintaining the same source 
identifier in \egdr{3} (red curve), and the fraction -- irrespective of the identifier -- 
having a counterpart in \egdr{3}\ at the same position within 50~mas (blue curve).  
\label{fig:comp_sust} }
\end{figure}

%
An important question is how to find \gdr{2}\ sources in \egdr{3} and determine whether they are
still present and if they maintain their source identifiers.  In general this
is the case, but there are also many exceptions. In \egdr{3}, we still have 96.2\% of the
\gdr{2} sources at the same position to within 10~mas when taking \egdr{3}\
proper motions into account. If we, in addition, require the same identifier, we
are down to 93.6\%, so close to three percent of the sources now have a
different identifier.
This typically happens when two processes are in conflict. 
\TBC{On-board, the transit of a single star may trigger more than one detection
or a close pair may not be resolved.}
\TODO{The on-board 
detection may trigger the transit of a single star more than once or only occasionally be able to resolve a close
source pair.} Later, on ground, the cross match algorithm has to decide
if it is dealing with one or two sources.
Since the previous data release, more information has become available and algorithms
have been adapted to better handle the difficult cases.
Figure~\ref{fig:comp_sust}, which
is based on a representative subset, shows that sources in the range of
$10<G<11.5$\,mag
are strongly affected by this identifier change. This is a magnitude range where the
on-board detection often detects two sources, rather than just one, especially
in the upper rows of the focal plane, where images tend to be wider, cf.\
\citet[][fig.\ 11]{EDR3-DPACP-73}. \TODO{} 
Measures are being taken in the on-ground processing to essentially eliminate
this issue in future data releases.  Also for the brightest sources, the swarm
of spurious detections they trigger on-board \gaia\ creates problems for the
cross match process, and a large fraction therefore have new identifiers. These
points are discussed in detail in \citet{EDR3-DPACP-124}. 

More remarkable than a change of identifier is perhaps that many \gdr{2}\
sources have no close counterpart in \egdr{3}. If we use a closeness limit of
10\,mas, as many as 3.8\% sources have vanished. This limit may be a little too strict,
for example,\ for faint sources, and if we relax it to 50~mas it is only 0.61\% of the
sources.
Going all the way to 2\arcsec, 
0.18\% are still missing. As shown in \figref{fig:comp_sust}, where a 50~mas
limit is used, 1--2\% of the brightest sources have changed. In the faint end,
the fraction of missing sources is very small at 18\,mag, but it increases 
slowly until 20.7\,mag after which it rises sharply, reaching 20\% at 21.1~mag. 
There can be many reasons for these changes, but binaries, crowding,
and spurious sources are among the likely explanations.

On the other hand, counterparts may be offset for good reasons. For double stars we may, for example, have
only the photocentre in \gdr{2}, but a resolved pair in \egdr{3}. It can also
be that the proper motion is unknown or erroneous, and this can be important
even when propagating positions by the mere 0.5~yr, which is the difference in
epoch between the two catalogues. Counterparts may also be completely missing,
for example,\ if the detections upon which the \gdr{2}\ source was based are now
considered spurious.

A table, \dt{gaiaedr3.dr2\_neighbourhood}, is provided in the \gaia\ archive. \TBC{We recommend using this table}
for looking up \gdr{2}\ sources in \egdr{3}.

%
\subsection{Large-scale completeness of \egdr{3}}\label{sec:comp_large}
%

\begin{figure}[h]
\includegraphics[width=\hsize]{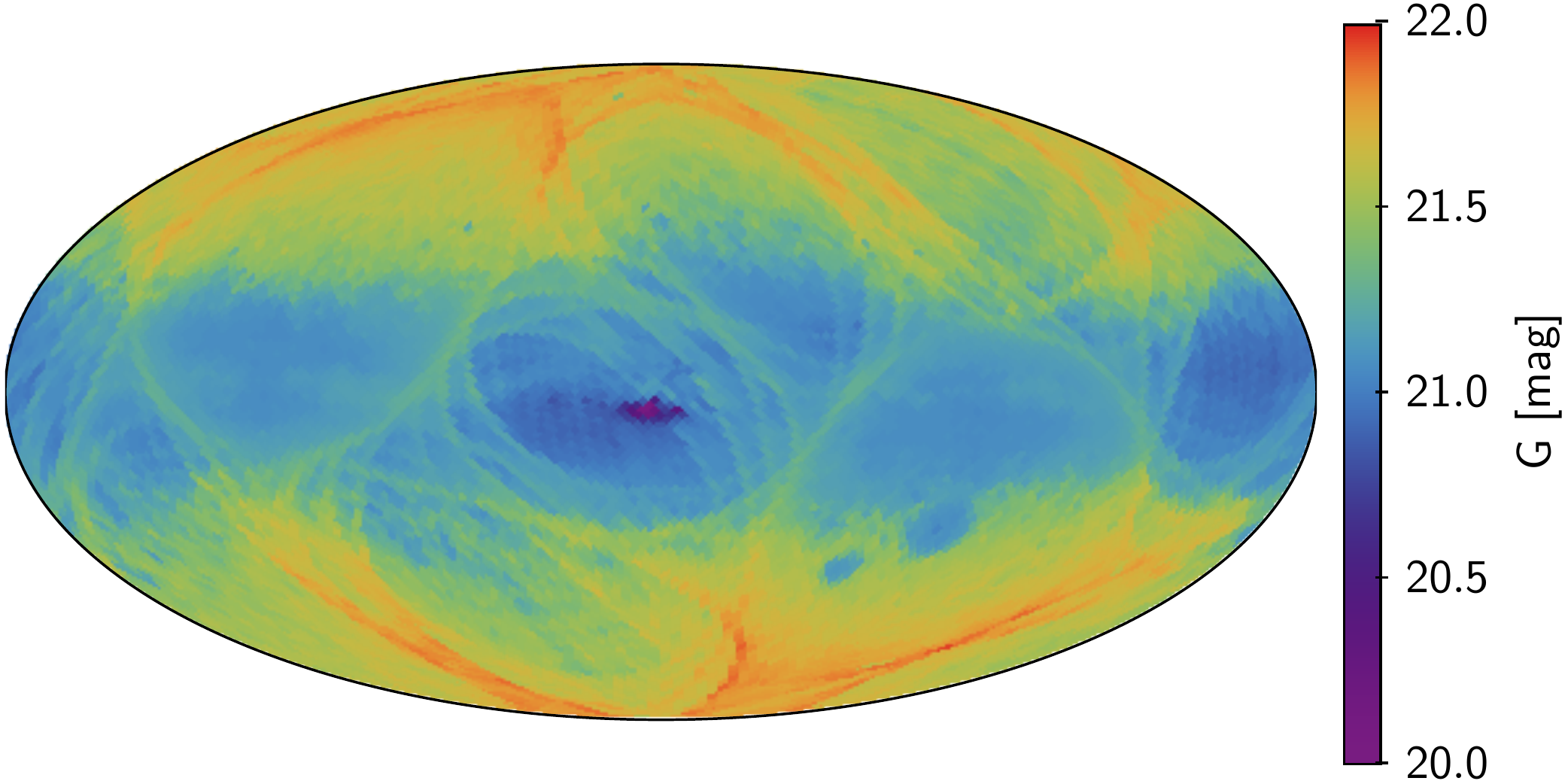}
\caption{
\label{fig:comp_g99}
Map in Galactic coordinates of the 99th percentile of the \gmag\ magnitude
at healpix level 5, i.e.\ in $3.36\,\sq\degr$\ pixels.}
\end{figure}

On-board \gaia, sources are selected for observation based on two criteria:
they must be roughly pointlike and they must be brighter than $G=20.7$\,mag.
The instrument has, however, a limited capacity for the number of simultaneous
observations, cf.\ \citet{2015A&A...576A..74D}, and when scanning close to the
Galactic plane some observations -- in particular the fainter -- are never sent
to ground because of limited mass-memory and telemetry capacity.  As a simple
measure of the actual magnitude limit, \figref{fig:comp_g99} shows the 99th
percentile of the \gmag\ magnitude across the sky using the healpix spatial index \citep{Gorski}. The area with the brightest
limit is Baade's window, unsurprisingly, followed by low Galactic
latitudes in general. Here, the finite on-board resources clearly dominate. The limit is fainter on
higher latitudes, especially along the caustics of the
scanning law, where more transits are available.

\begin{figure}[h]
\includegraphics[width=\hsize]{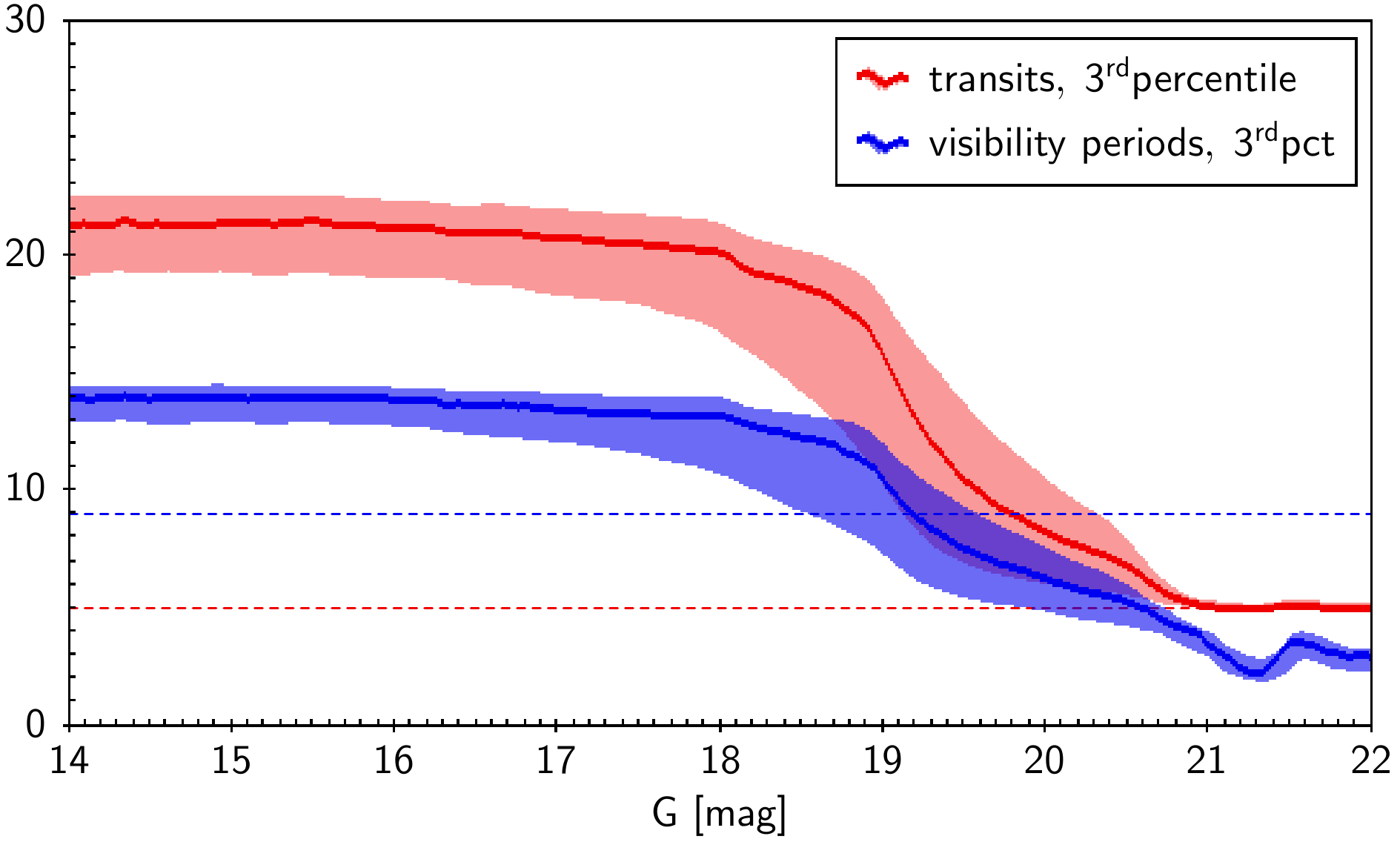}

\includegraphics[width=\hsize]{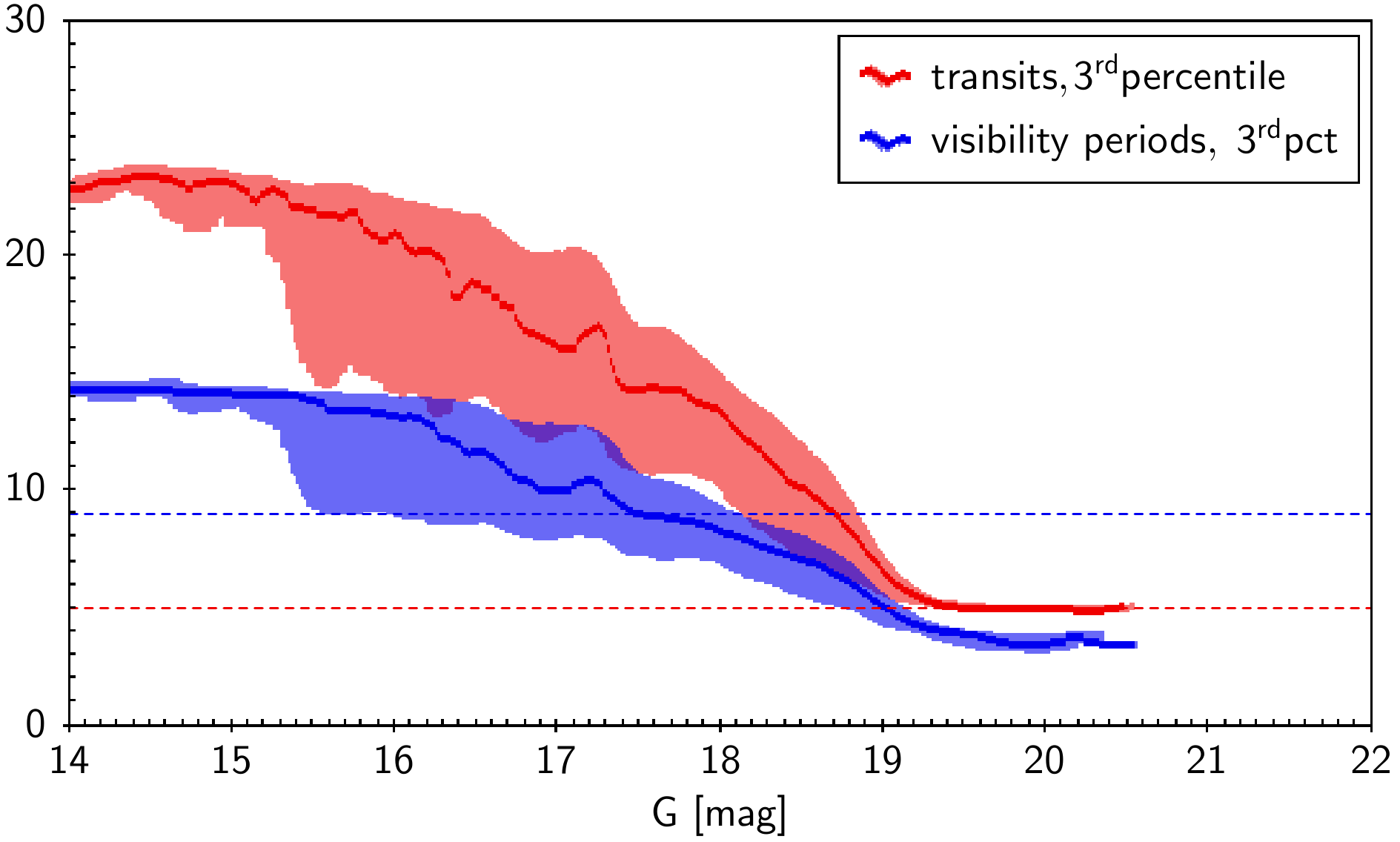}
\caption{\label{fig:visper} Third percentile of the number of transits and
number of visibility periods per source used in the astrometric solution. 
The shaded areas show the range of the first and fifth percentiles.
The limits of five transits for inclusion in the catalogue and nine visibility
periods for a full astrometric solution are also indicated.
{\em Top:} The catalogue in general. {\em Bottom:} A field in Baade's window.
} \end{figure}

Another way to estimate the completeness is to look at how the actual number of
transits obtained for each source depends on the magnitude as illustrated in
\figref{fig:visper}.  It shows the first, third, and fifth percentiles of the
number of transits and the number of visibility periods\footnote{
A visibility period, included in the \gaia\ archive as \dt{visibility\_periods\_used} is the time range when a source is observed without a time
gap of more than four days. From one period to the next, the scan direction has changed and a couple of months may have passed.
} used in the astrometric
solution.  For a magnitude range where the catalogue is nearly complete, we
expect these percentiles for the number of transits to lie well above the
required minimum of five transits for a source.  For the catalogue in general
(top panel), this holds for sources brighter than $G\sim19$\,mag; however, when we reach
$G\sim20$\,mag, the incompleteness is noticeable. For a field in Baade's window (lower
panel), sources are deprived of transits at a much earlier point and the
incompleteness is severe at $G\sim19$\,mag.

The number of visibility periods used for astrometry is also of interest
because a minimum of nine periods is required in order to publish the parallax and
proper motion, cf. \cite{EDR3-DPACP-128}. Here, an insufficient number of
periods is noticeable in the catalogue beyond $G\sim18.5$\,mag and even more severe
after $G\sim19.5$\,mag; whereas for Baade's window, insufficiency sets in about 1.5\,mag
earlier.
Thinking ahead to \gdr{4}, the mission segment covered will be twice
as long as for \egdr{3}, the number of transits and visibility periods will
be double, and a significant improvement in completeness can be expected.

\begin{figure*}[ht]
\begin{center}
        \centerline{
                \includegraphics[width=0.9\textwidth]{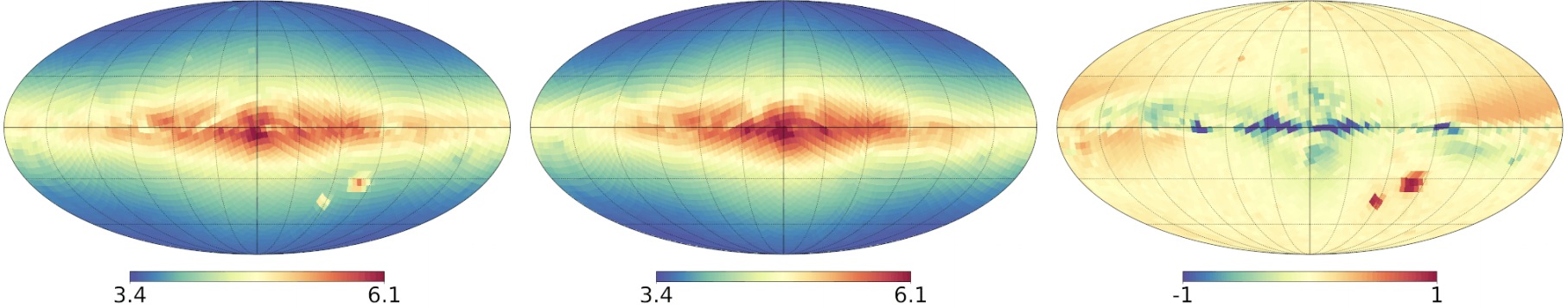}
        }
\caption[Star counts]{Star counts in \egdr{3} (left), GOG20 (middle), and relative difference (right) in the magnitude range of  $17<G<18$. A relative difference, (EDR3-GOG20)/EDR3, of $-1$ (resp. +1) corresponds to an excess  (resp.  a deficit) of 100\% in the GOG20 model with regard to \egdr{3} data. The colour scale is logarithmic in the left and middle panels, and linear in the right panel. The healpix level is 4.}\label{fig:count_map}
\end{center}
\end{figure*}

\begin{figure}[ht]
\begin{center}
\includegraphics[width=0.45\textwidth]{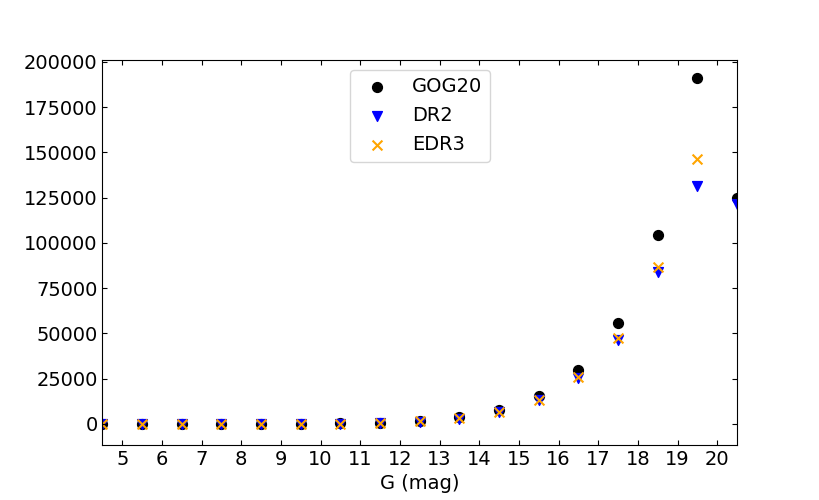}
\includegraphics[width=0.45\textwidth]{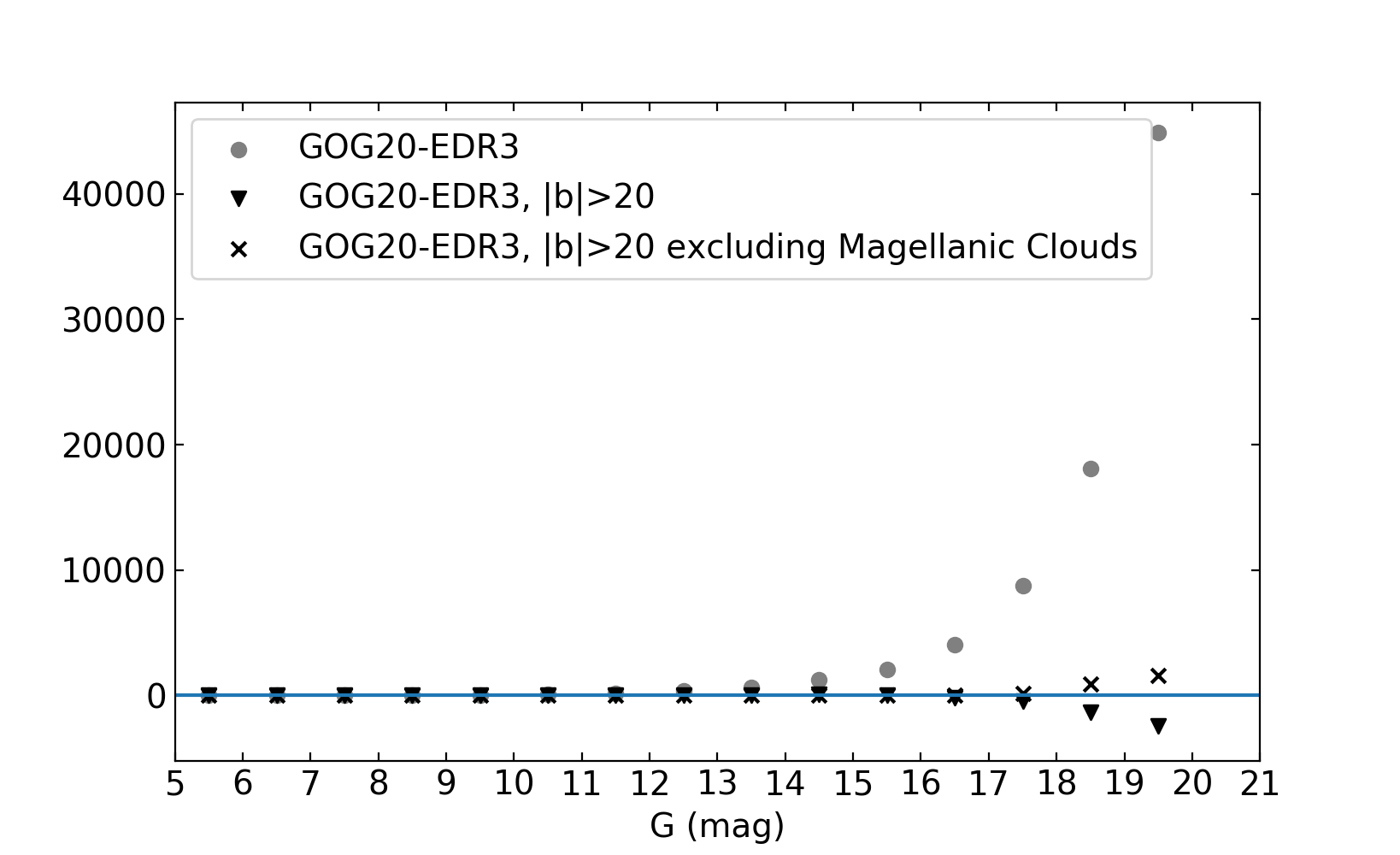}
\caption[{\em Top:} Average star counts]{  \new{Star counts averaged among the healpix bins} over the whole sky as a function of magnitude, in  \egdr{3}  \new{(orange crosses)} and \gdr{2}  \new{(blue triangles)}, compared to GOG20  \new{(black circles)}. {\em Bottom:} Difference in counts between GOG20  and \egdr{3} over the whole sky (circles), excluding the Galactic plane (triangles), and excluding the Galactic plane and the Magellanic Clouds (crosses). The deficit in GOG20 at faint magnitudes is mainly due to the Magellanic Clouds as they are not included in the model.}\label{fig:counts_by_mag}
\end{center}
\end{figure}

Comparisons with models have been performed to check the data for the star density
as a function of the position on the sky and of $G$ magnitude. The reference model
is  GOG20. It is described in detail in the online documentation\footnote{\url{https://gea.esac.esa.int/archive/documentation/GEDR3/Data\_processing/chap\_simulated/}}
and is also released with the \egdr{3} set of 
catalogues.\footnote{GOG20 is published in the \gaia\ archive in the table
\dt{gaiaedr3.gaia\_source\_simulation}} In order to have good statistics in each pixel of the sky maps, the
comparisons are done using  healpix of order 4, corresponding to 3072 pixels
per sky map  (Fig. \ref{fig:count_map}).
The  comparison with the GOG20 simulation shows that the overall picture of the
sky densities are very well comparable in data and model, although
incompleteness may remain towards the inner Galaxy for faint stars. The
\egdr{3} completeness has also improved with respect to \gdr{2} for stars
fainter than 18\,mag (Fig. \ref{fig:counts_by_mag}, upper panel). Still the predicted numbers are higher than the observed, but this is mainly
due to the counts in the Galactic plane, where the extinction is underestimated in GOG20 (Fig. \ref{fig:count_map}, right panel, and Fig. \ref{fig:counts_by_mag}, lower panel).

\begin{figure}[h]
\centering
\includegraphics[width=0.9\hsize]{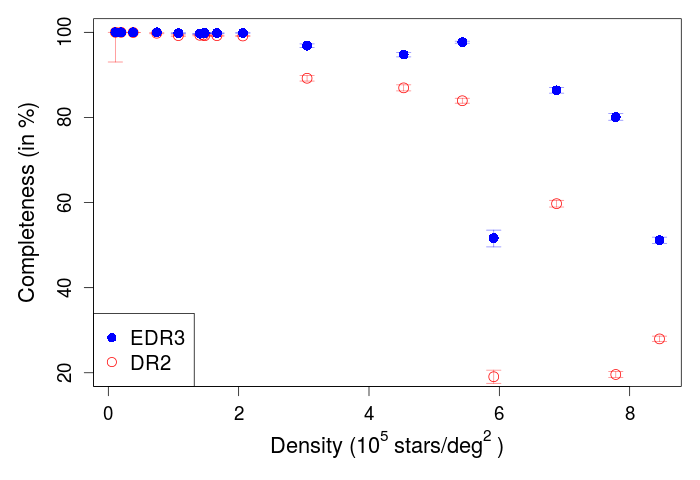}
\caption{
\label{fig:oglecompl}
Improvement of the \gaia\ completeness at \gmag $=20$~mag versus some OGLE fields of different stellar densities\new{ from \gdr{2} (red) to \egdr{3} (blue).}}
\end{figure}

Figure \ref{fig:oglecompl} shows the improvement in the completeness of crowded regions between \gdr{2} and \egdr{3}. Here, we use the OGLE data from \cite{2008AcA....58...69U} which provides only an upper limit to the \gaia\ completeness due to the poorer OGLE spatial resolution. 

For bright sources, detection efficiency starts to drop at $G\sim3$\,mag due
to saturation that is too strong \citep{2016A&A...595A...1G}.  As a consequence,
20\% of the stars brighter than magnitude 3 do not have an entry in \egdr{3}.
A few bright stars which were present in \gdr{2} with rather dubious solutions,
such as\ Polaris, are also missing in \egdr{3}.

%
\subsection{Small-scale completeness of \egdr{3}}\label{sec:comp_small}
%
\begin{figure}[h]
\includegraphics[width=\hsize]{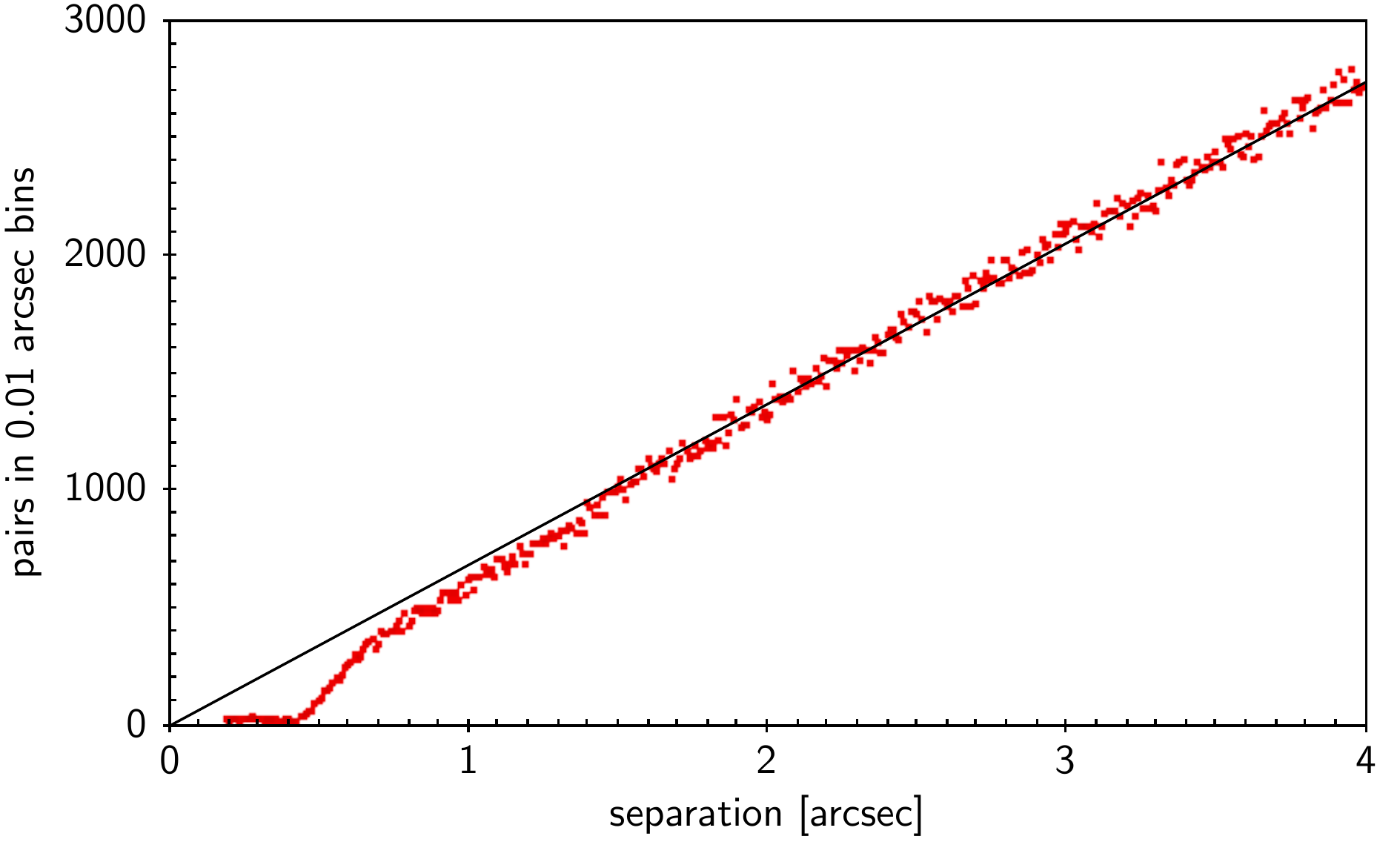}
\includegraphics[width=\hsize]{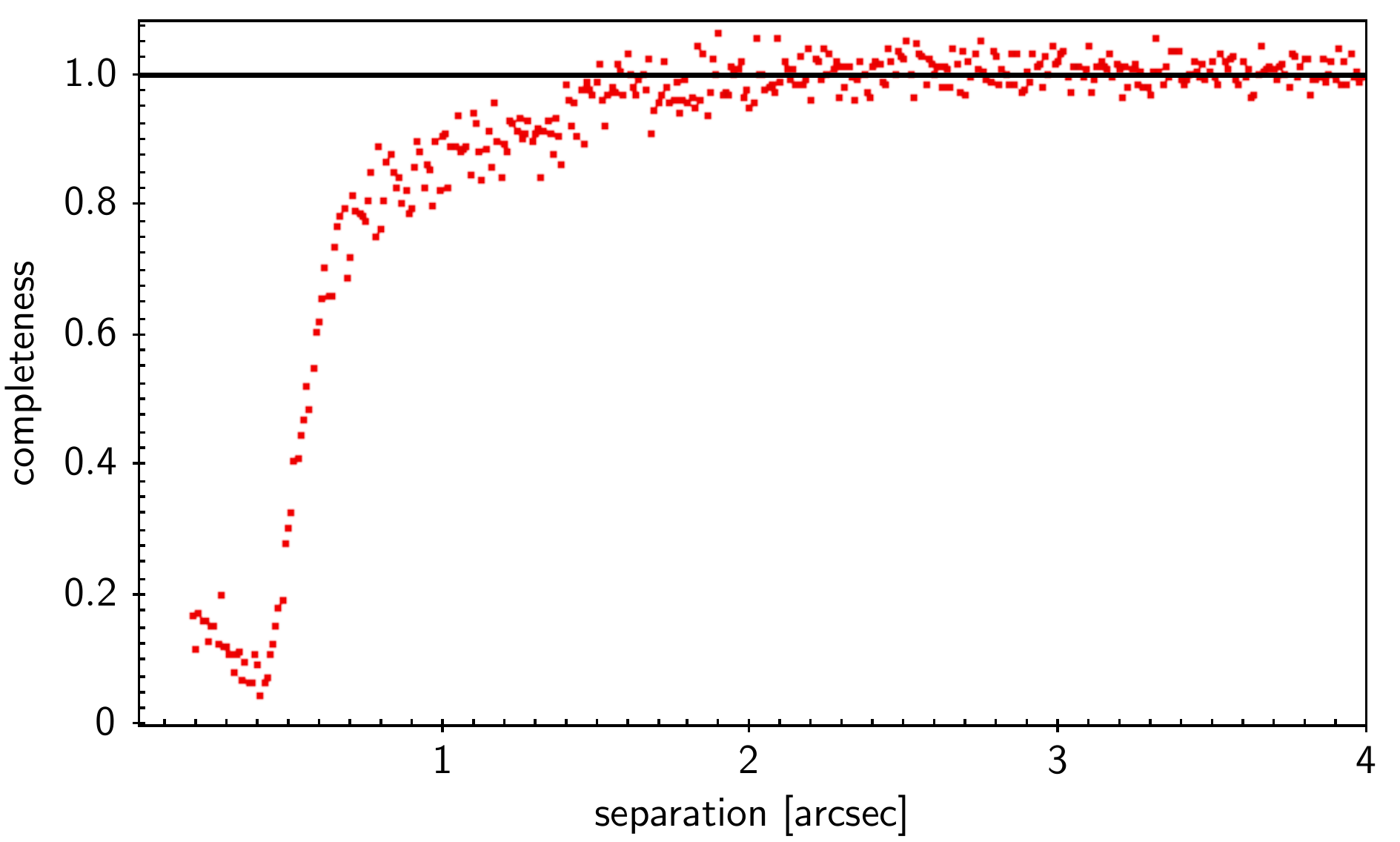}
\caption{\label{fig:densep}
{\em Top:} Histogram of source pair distances in a circular field of radius $0.5\degr$\ 
centred at
$(l,b)=(330\degr,-4\degr)$ with a line showing the expected relation for 
a random distribution of the sources.
{\em Bottom:} Normalised histogram using the expected relation.
}
\end{figure}

The completeness at the smallest angular separations can be tested using a
histogram of source-pair distances in a small dense field near the Galactic
plane. Such a field will be completely dominated by distant field stars and
there will be very few resolved binaries. Figure~\ref{fig:densep} shows (top panel) a
histogram of source separations for such a field, where the black line
indicates the expected relation for a random source distribution. For
separations above 1.5--2\arcsec, the actual distribution closely follows this
line, indicating that we have a high completeness. However, below 1\farcs5, and
especially below 0\farcs7, the completeness falls rapidly. This is expected,
taking the current processing strategy into account, and it is caused by conflicts
between neighbour observations both wanting to use the same pixels. 
Between
0\farcs18 and 0\farcs4, only a few pairs were resolved because
of the particulars of their magnitude difference and their orientation with
respect to the dominating scan directions.
The bottom panel of the figure shows the same distribution but normalised
with respect to the expected relation. This shows an apparently higher
completeness for the lowest separations and the question that is begged is whether spuriously
resolved single sources are at play. We notice that below 0\farcs4, as many
as 74\% of the pairs are composed of 2p solutions, making it difficult to
judge if they are genuine. For separations between 0\farcs4 and 0\farcs5, the
pairs of 2p solutions constitute only 40\%. 

\begin{figure}\begin{center}
\includegraphics[width=\hsize]{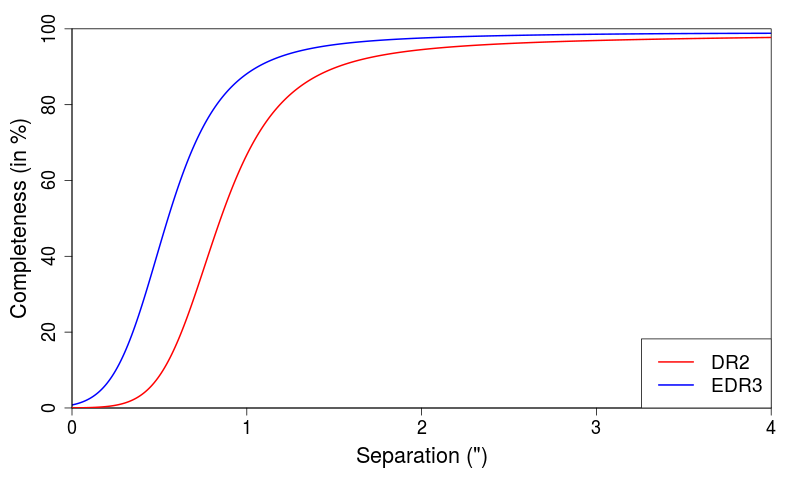}
\caption{Improvement of the completeness (in percent) of visual double stars from the WDS catalogue as a function of the WDS separation between components from {\gdrtwo} (red) to \egdr{3} (blue).}
\label{fig:WDScompleteness}
\end{center}\end{figure}

Figure~\ref{fig:WDScompleteness} shows the improvement in the spatial resolution of \egdr{3} using the Washington \TODO{Visual} Double Star \TBC{Catalog} \citep[WDS;][]{WDS} \TODO{catalogue}. 
It confirms again that incompleteness is severe below 0\farcs7, but it has improved substantially when compared to \gdr{2}.

\subsection{Completeness in crowded regions: Globular clusters}\label{sec:compleGlobular}
As already mentioned, the \gaia\ instrument has a limited capacity for observing
very dense areas, and sources in these fields will get fewer observations and
the limiting magnitude will be brighter. 
We derive the completeness in  a few globular clusters which have various levels of crowding.   The procedure and the sample are
the same as described in \cite{2018A&A...616A..17A}.  \egdr{3} data are compared  with the catalogue of HST photometry by \cite{2007AJ....133.1658S}. 
We recall   that the completeness of HST data is derived using crowding experiments and is higher than 90\% in the whole \gaia\ range.

Table \ref{tab:wp947completeness26gcs} in Appendix \ref{AppendixA} presents the results for the inner and outer regions of the cluster sample and shows the combined completeness of the astrometry and the photometry. Since, by construction, the \gaia\ photometry is only published for sources with an astrometric solution, the photometric completeness cannot be higher than the astrometric completeness. It can, however, be lower, in particular in high density regions for the \gbp\ and \grp\ photometry. Since this photometry is derived from dispersed images, crowding affects these measurements much more than astrometric measurements and \gmag\ band photometry. 
We found that  in globular clusters, a percentage of about 20\%-30\% of  stars  having five or six parameter solutions do not have   \gbp\ and \grp\ magnitudes. One of the  worst cases is NGC~6809 where the percentage of stars without \gbp\ and \grp\ is of 37\%.
Because of the lower level of crowding, open clusters are more favourable cases. In general, the percentage of stars without \gbp\ and \grp\ is of~the order of 1\%-3\%.

\begin{figure}
 \begin{center}
\includegraphics[width=8cm]{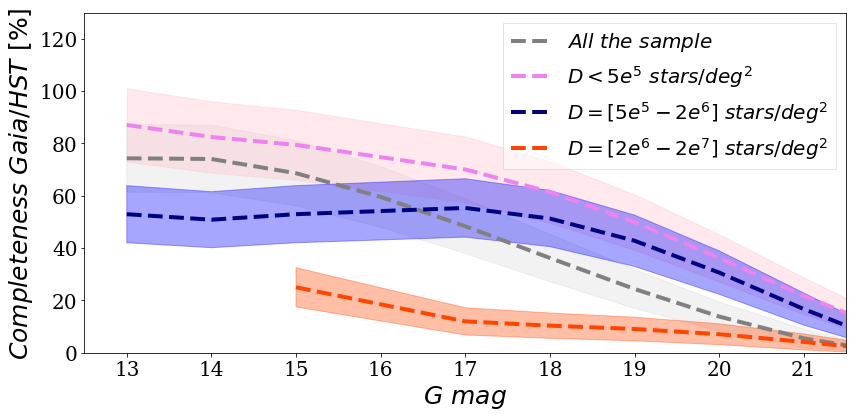}
\end{center}
\caption{Global completeness  as a function of \gmag\ for  the whole sample of globulars (grey line). Pink, blue, and orange lines indicate the completeness in different density ranges D. The shaded areas indicate the uncertainties}\label{fig:comple1Clu}
\end{figure}

In globulars, the completeness is still at the 60\% level for $G\sim 19$\,mag when the density is  of the order of  $ 10^5$ stars/sq deg. The completeness is higher than 20\% at $G\sim 17$\,mag when the density is lower than a few $ 10^7$ stars/sq deg. At the faint end, in favourable cases, the completeness is still at the 80\% level at $G\sim 20$\,mag (see Fig. \ref{fig:comple1Clu}). Inner and outer regions of the globulars have very different levels of completeness. For instance,  in NGC~5053, the completeness in the inner and outer regions are very similar and quite high (60\% at $G\sim 20$\,mag);
in NGC~2298, the inner and outer regions have a very different level of completeness. In the inner region, the completeness is about 10\% at $G\sim 20$\,mag, and 60\% in the outer region. In the very crowded NGC~5286, the completeness in the inner region is 20\% at \gmag$\sim$17\,mag. 
However, the completeness level is still  variable for similar densities and magnitudes, depending on the number of observations among others (see Fig.\ref{fig:comple1Clu}).  

%
\subsection{High proper motion stars}\label{sec:hpm}
%
\begin{figure}[h]
\includegraphics[width=\hsize]{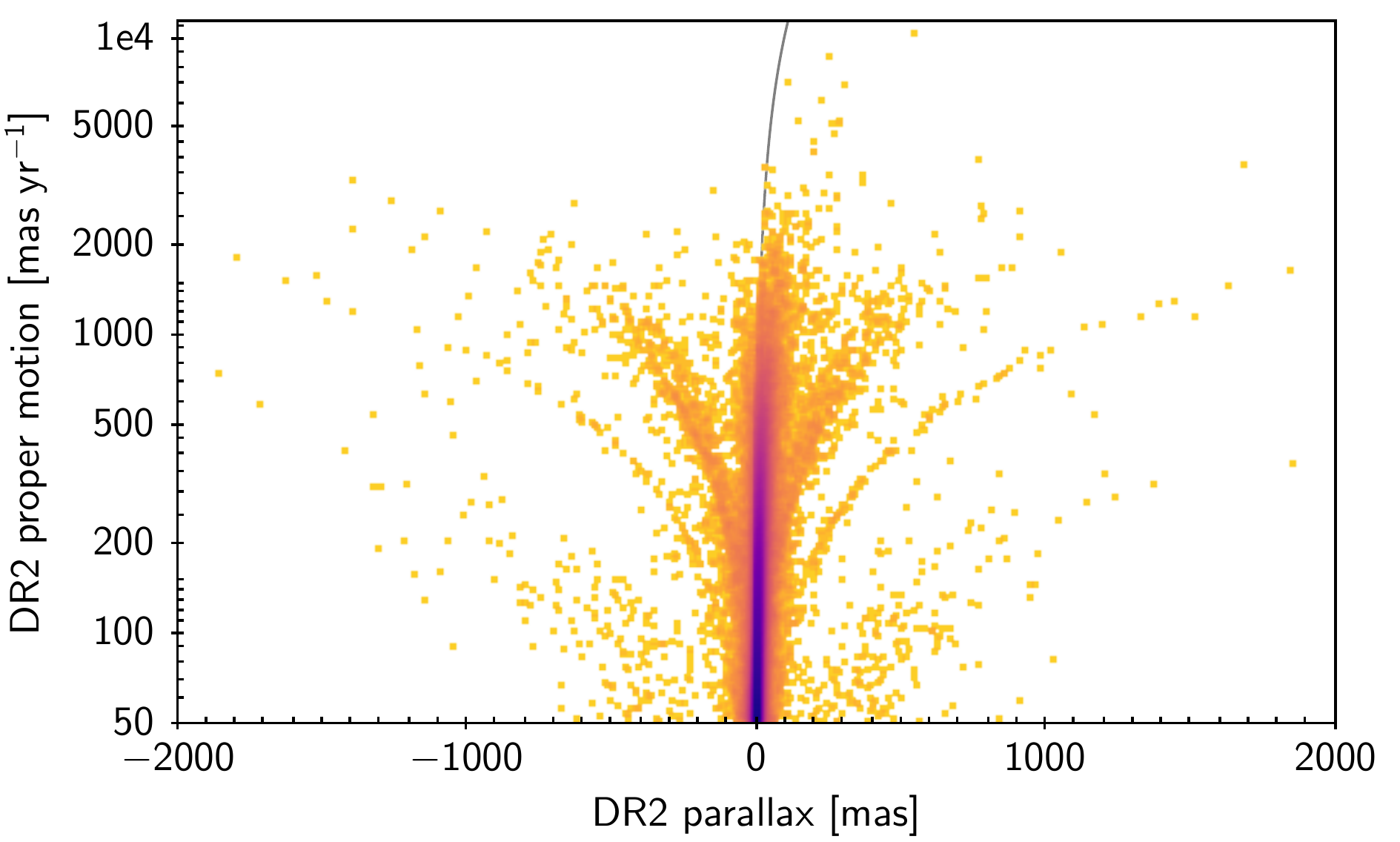}
\includegraphics[width=\hsize]{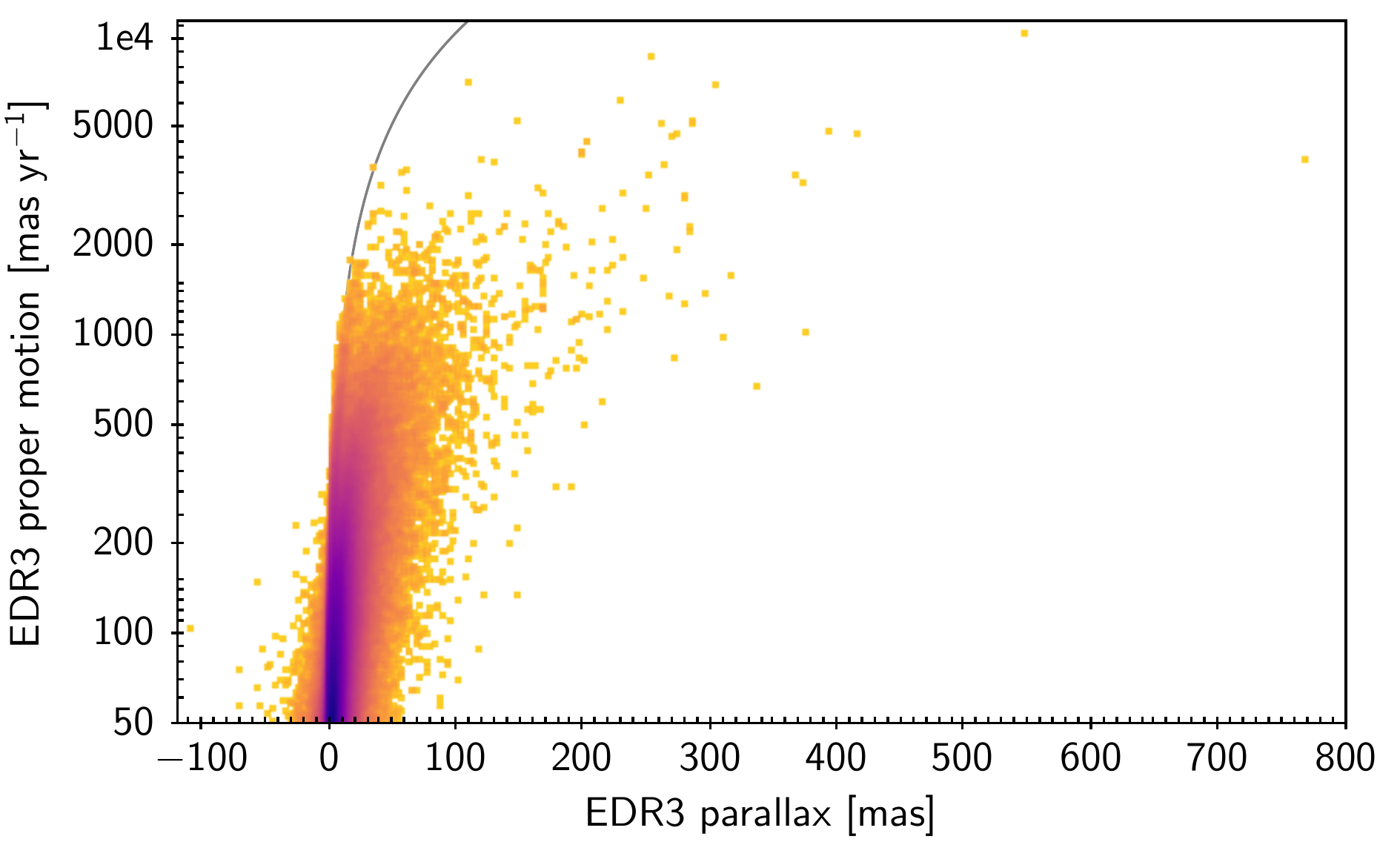}
\caption{\label{fig:hpm_pm_plx}
Proper motion versus parallax for large proper motions. {\em Top:} In \gdr{2}.
{\em Bottom:} In \egdr{3}. The grey line shows the locus of tangential
velocity $500\,{\rm km\,s}^{-1}$.
}
\end{figure}

\gdr{2}\ contained 951 sources with a proper motion above $1000\,{\rm
mas\,yr}^{-1}$ and 3726 with motions above $600\,{\rm mas\,yr}^{-1}$. The
corresponding numbers for \egdr{3}\ are 633 and 2729, respectively, which is\
about 30\% fewer. At first glance, this may look disturbing, but the fall is
largely explained by spurious solutions in \gdr{2}\ having been weeded out.
This point is illustrated in \figref{fig:hpm_pm_plx}. The top panel shows the
3\,280\,360 \gdr{2}\ proper motions above $50\,{\rm mas\,yr}^{-1}$ against
the parallax. There is a large population of negative parallaxes and also
parallaxes approaching two arcseconds in absolute value.  Conditions that trigger a spurious
solution with a negative parallax, such as an\ unresolved duplicity, may equally
well produce a spurious solution with a positive parallax as demonstrated by
the overall symmetrical appearance. We return to this point in Sect.~\ref{sec:ast_spur}.  
The lower panel shows the corresponding
plot for \egdr{3}\ with 3\,273\,397 sources, that is to say\ 6963 fewer. This plot has a 
better 
appearance with much fewer negative parallaxes. In particular, there are no more
negative parallaxes for sources with proper motions above
$300\,{\rm mas\,yr}^{-1}$. 
Sources with a proper motion higher than $1000\,{\rm mas\,yr}^{-1}$ must -- in
the great majority -- be relatively nearby and \TBC{will} only very rarely \TBC{have} \TODO{has} a parallax
below 10\,mas lest their tangential velocity become unrealistically large. We
can therefore safely assume that solutions giving negative parallaxes are
mostly spurious for these large proper motion cases. In \gdr{2,}\ 175 of the 951 sources with large proper motions
had a negative parallax and if we assume that a similar number had a spurious
positive parallax, we are down to about 600 good solutions. This is a number
that compares favourably with the 633 such sources in \egdr{3}. While it is
possible to estimate the number of spurious solutions in \gdr{2}, it is not
easy to identify which ones they are. 
We note that this improvement holds for high proper motion sources because they
benefit more from the longer mission duration. We examine sources of lower
proper motion in Sect.~\ref{sec:ast_spur}.

Comparing high proper motion stars with SIMBAD, we find that 8\% of the SIMBAD stars with a proper motion higher than 600\masyr\ are missing in \egdr{3}.
Those are mostly stars with only a 2p solution in \egdr{3} and stars outside \gaia's magnitude range, and a few show duplicated entries in \egdr{3}.

%
\subsection{Sources without parallax and proper motion}\label{sec:2p}
%

As many as 344 million sources have only a position published from the
astrometric solution and neither parallax nor proper motion. The requirements
for maintaining a full astrometric solution are detailed in
\cite{EDR3-DPACP-128}. A source must, as mentioned, have at least nine
visibility periods, the formal uncertainties \TBC{must be}\TODO{being} sufficiently small, and the
source must be brighter than $G= 21$\,mag according to the photometry
available at the time of the astrometric processing, that is\ in \gdr{2}. The majority
of the 2p sources have simply too few observations and will obtain a full
solution in later releases. 

If we look specifically at sources brighter than
19\,mag, where the completeness is high, we have 8.8 million 2p solutions out
of 575.9 million sources, that is\ 1.5\%. This is a clear improvement over
\gdr{2}, which contained 13.8 million 2p solutions among 568.1 million sources,
that is\ 2.4\%.  For these brighter sources, the problem is only the lack of
observations for less than half of them. The rest have, for a large part, a
problem with a close neighbour. This follows from the various indicators in the
catalogue, such as \dt{ipd\_frac\_odd\_win}, which indicates observation window conflicts for wider pairs,
\dt{ipd\_frac\_multi\_peak}, indicating resolved, closer pairs and
\dt{ipd\_harmonic\_gof\_amplitude}, indicating asymmetric images for the
closest pairs.
The processing approach in \egdr{3}\ did not intend to resolve these pairs.


\section{Astrometric quality of \egdr{3}}\label{sec:ast}
%

\begin{figure*}[h]
\sidecaption
\includegraphics[width=12cm]{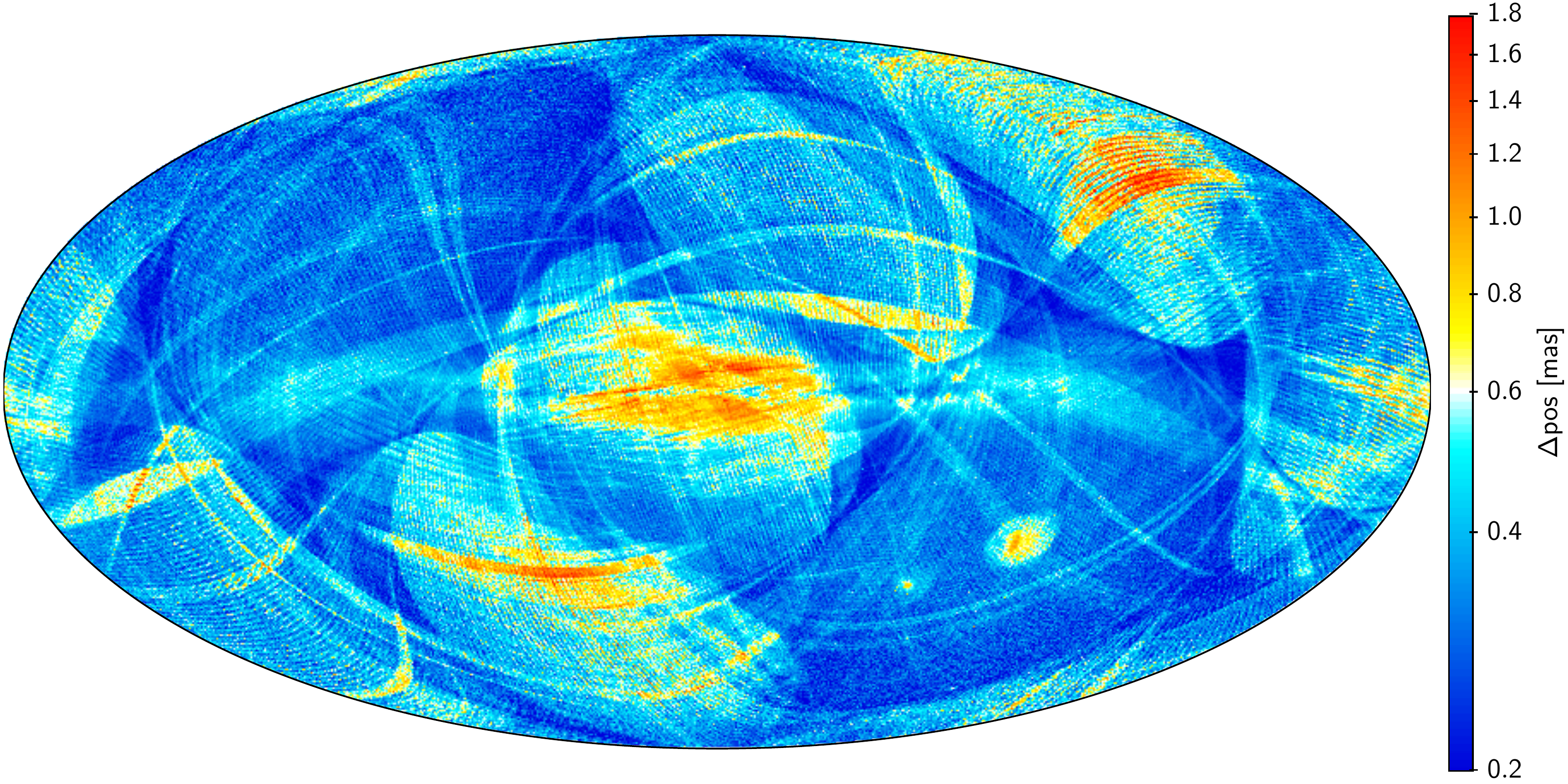}
\caption{Map in Galactic coordinates of the mean positional difference between \egdr{3}\ and
\gdr{2}, having propagated the \egdr{3}\ positions to the epoch of \gdr{2}.
}
\label{fig:ast_dr3dr2}
\end{figure*}

The astrometric solution for \egdr{3}, cf.\ \cite{EDR3-DPACP-128}, is based on
33 months of observations as compared to the 21 months for \gdr{2}. Therefore, there
are reasons to not only expect a much improved precision, but also
a much better ability to disentangle proper motions and parallaxes even for
sources where (some) transits are biased by a close neighbour.

As already mentioned, there are three flavours of astrometric solutions in
\egdr{3}: either 2p, with only a position, 5p, with also parallax and proper
motion, and finally 6p, where also a pseudo-colour is derived. The latter
category encompasses the faintest sources and sources without a known colour
in \gdr{2} or with a clearly biased colour. Therefore, the
more accurate astrometric solutions are those with five parameters (5p).
This is partly the case because of the correlations introduced by deriving a colour
from the astrometric measurements, in part because the photometric colour is
missing precisely for the sources observed in the more difficult conditions,
for example,\ in crowded areas or with a brighter source in the vicinity.
Here, we concentrate on the 5p and 6p solutions, which are the ones of more 
interest. 

We test the astrometry with an emphasis on the parallaxes. Here, large negative
values reveal the presence of spurious solutions; distant objects such as quasars (QSOs) are
suited for testing the parallax zero point; and star clusters and binaries, with
sources located at nearly the same distance, are ideal for finding magnitude
and colour dependent parallax errors.  For the parallax precision, we use the
negative wing of the parallax distribution.
For proper motions, star clusters and binaries are again ideal when looking
for magnitude dependent errors.

 \cite{EDR3-DPACP-132} calculated a parallax zero point correction depending on
the magnitude, colour, and ecliptic latitude, using different tracers (QSOs, red
clump stars, stars from the Large Magellanic Cloud (LMC), binaries).  The paper provides recipes for correcting
the zero point error, at least in a statistical sense.  We therefore also look
into the effect of applying these corrections.

%
\subsection{Imprints of the scanning law}\label{sec:ast_scan}
%
%
The scanning law for \gaia\ \citep[][sect.\ 5.2]{2016A&A...595A...1G} specifies
the direction of the spin axis of the spacecraft -- and thereby also the
great circle being observed -- as a function of time. The axis precesses around
the Sun at an angle of 45\degr\ with a period of 63~days, thereby creating
a characteristic pattern in the number of transits across the sky with high
values at $\pm 45\degr$ ecliptic latitude and in loops in between.

Figure~\ref{fig:ast_dr3dr2} shows the average separation between sources in
\egdr{3}\ and their counterparts in \gdr{2}, taking \egdr{3}\ proper motions
into account. Only separations below 10\,mas and sources with a known proper motion have been used (1458.7 million sources).  The map shows
that the discrepancy is normally at the sub-mas level, but there are also specific areas
of the sky where they approach 2\,mas.  On the one degree scale, there is a
clear scan-law pattern. The plot is dominated by very faint sources and does
not reflect the performance in the bright end.  Although it cannot be deduced
from the map itself,
we believe that it largely shows \new{positional} errors of \gdr{2}.

\begin{figure}[h!]
\begin{center}
\includegraphics[width=0.49\columnwidth]{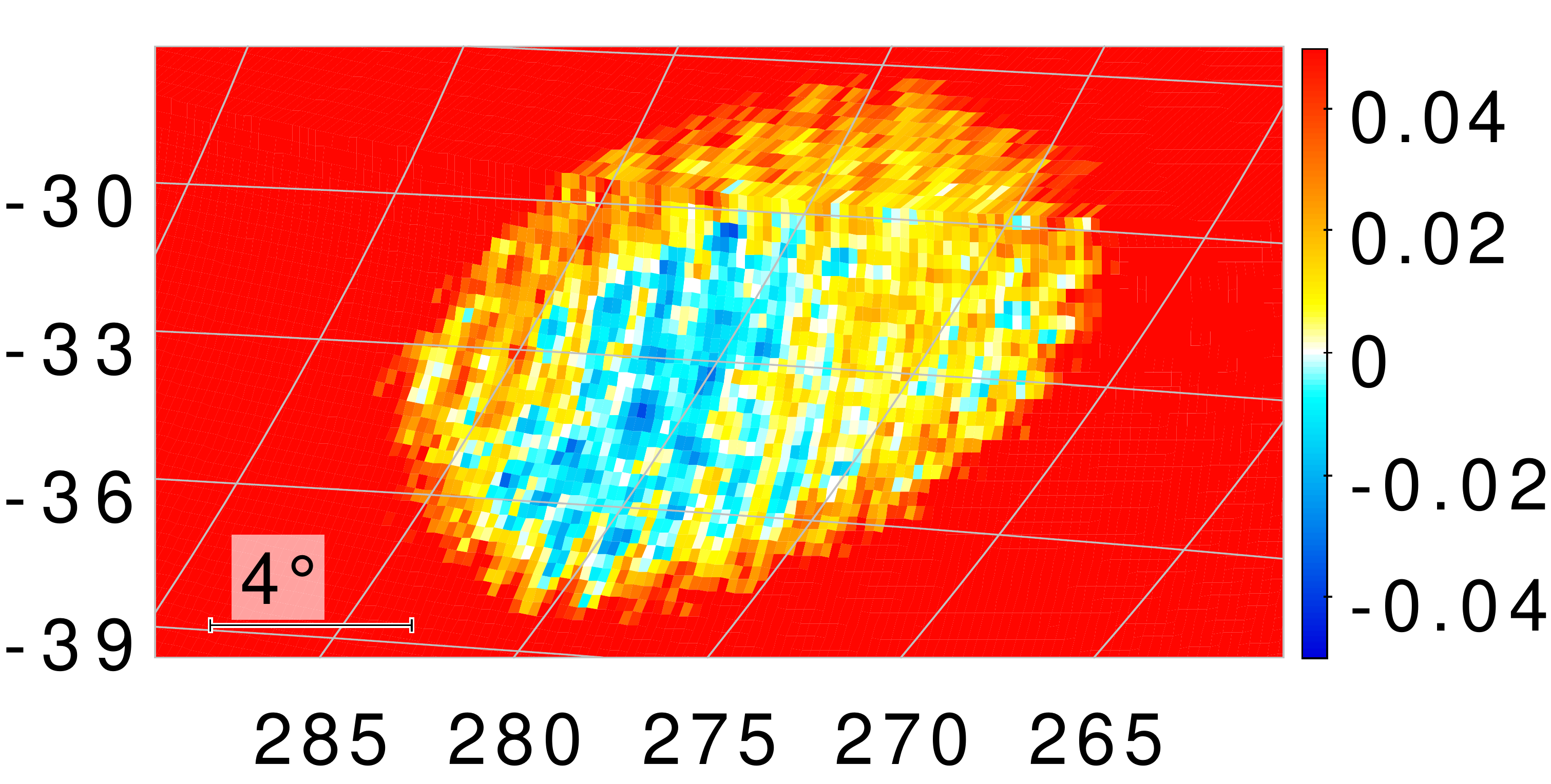}
\includegraphics[width=0.49\columnwidth]{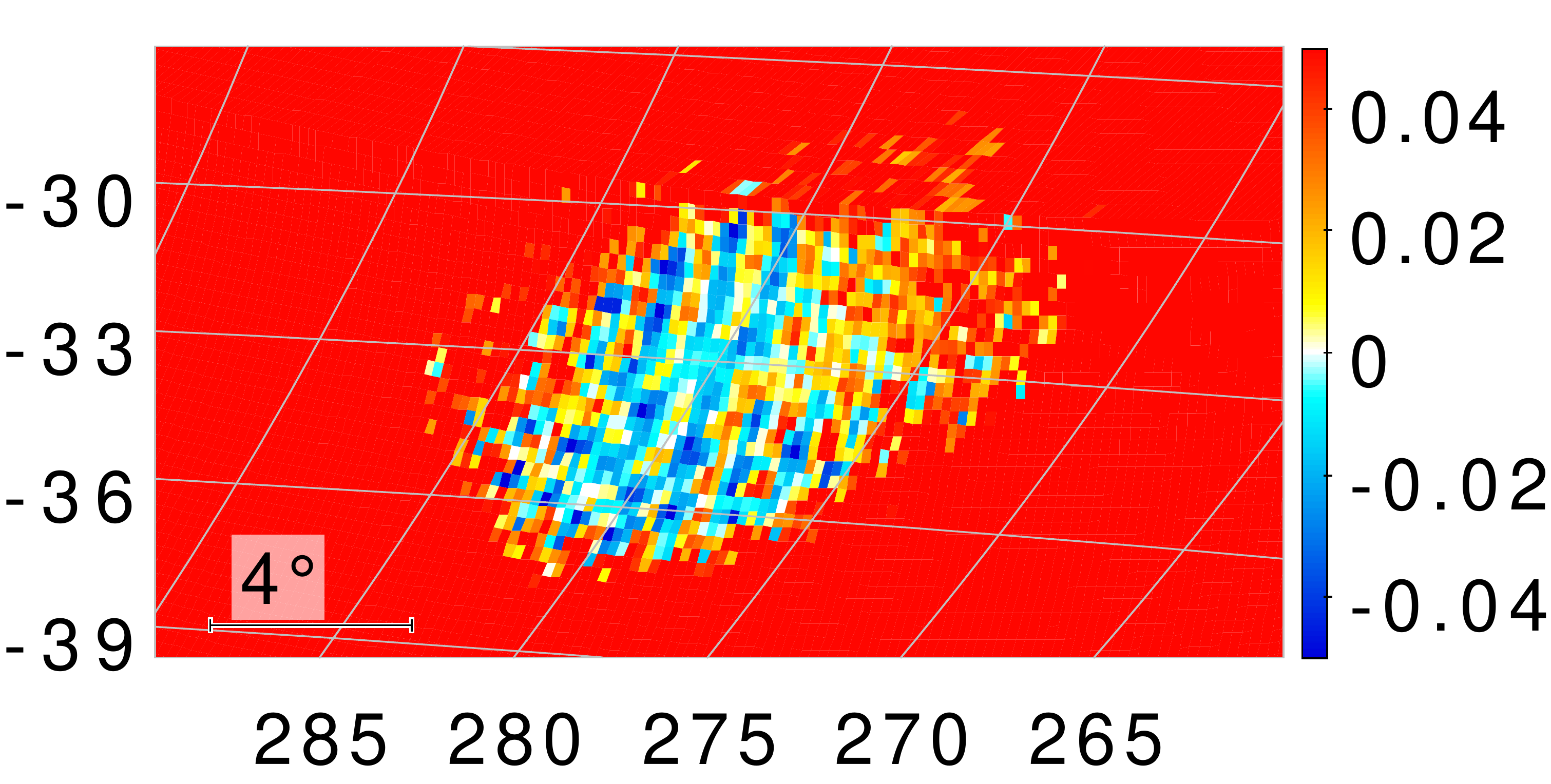}
\includegraphics[width=0.49\columnwidth]{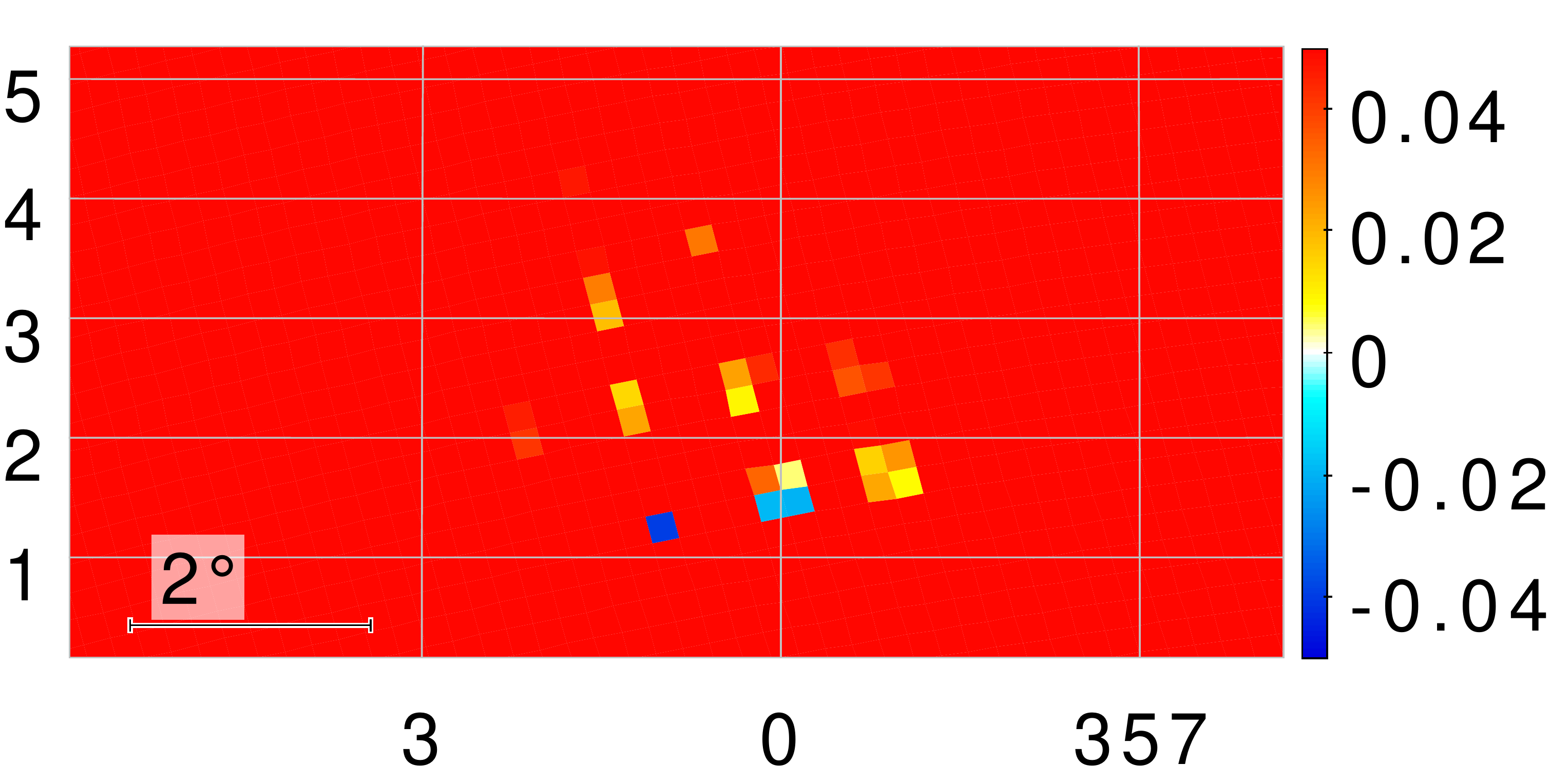}
\includegraphics[width=0.49\columnwidth]{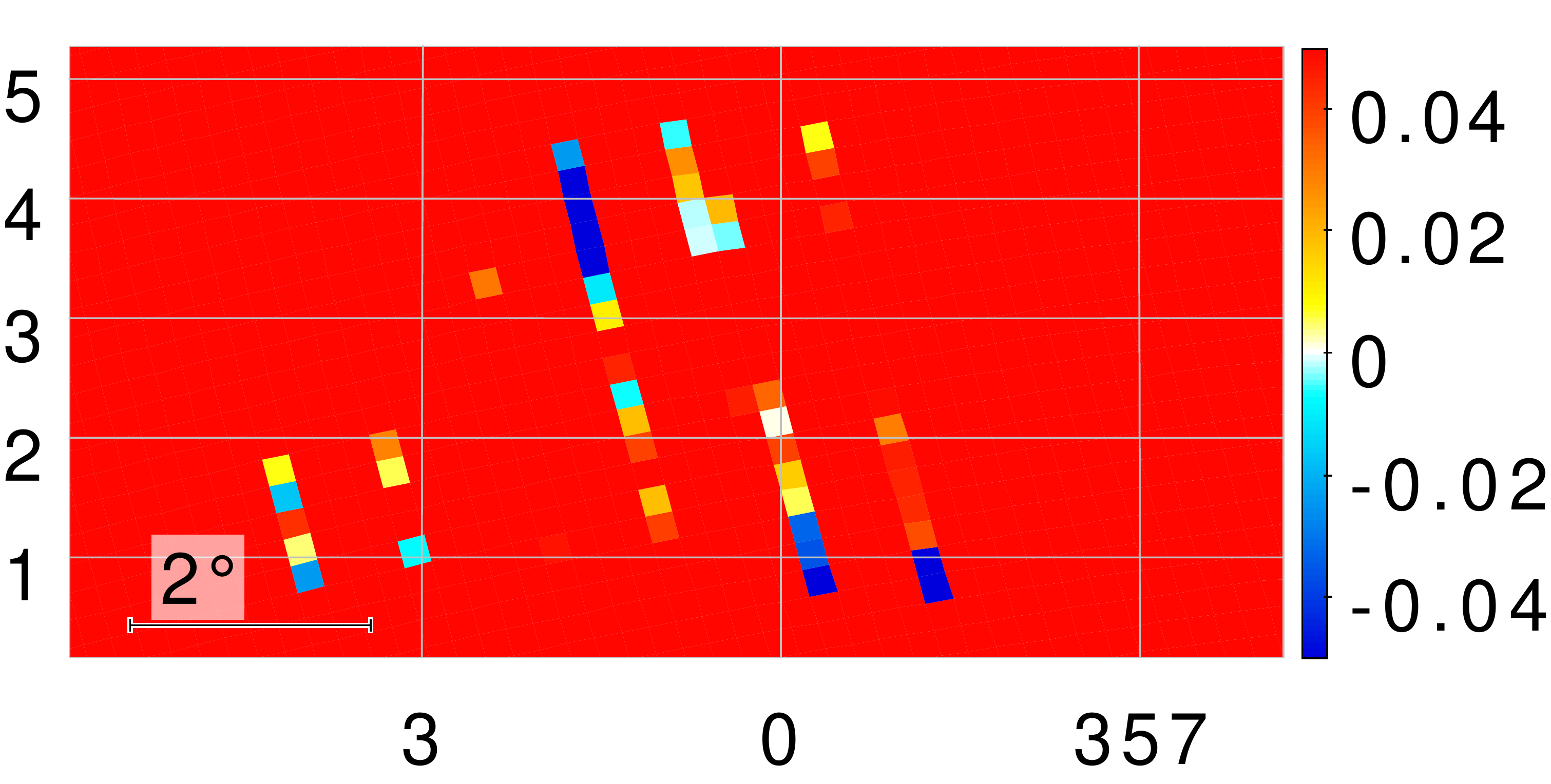}
\end{center}
\caption{\TBC{Maps in Galactic coordinates of} median parallaxes in the direction of the LMC ({\em top}) and 
Galactic centre ({\em bottom}) for
5p ({\em left}) and 6p ({\em right}) solutions. 
To increase the contrast, the represented parallax range is $[-0.05,0.05]$ mas, 
although the median parallax in most of the field is above $0.05$ mas. 
}\label{fig:patternPlx}
\end{figure}

Figure \ref{fig:patternPlx} shows the remaining chequered-pattern
systematics in the parallaxes due to the {\gaia} scanning law in 
the direction of the LMC and of the Galactic centre, where the parallaxes
are small and homogeneous enough so that variations in the median values merely reflect the parallax errors.
The systematics for 6p are larger than for 5p, but they have otherwise decreased
together with the chequered-pattern
since DR2 \citep[cf.][fig.~13]{2018A&A...616A..17A}.

%
%
\subsection{Spurious astrometric solutions}\label{sec:ast_spur}
%

\begin{figure}[h]
\sidecaption
\includegraphics[width=\hsize]{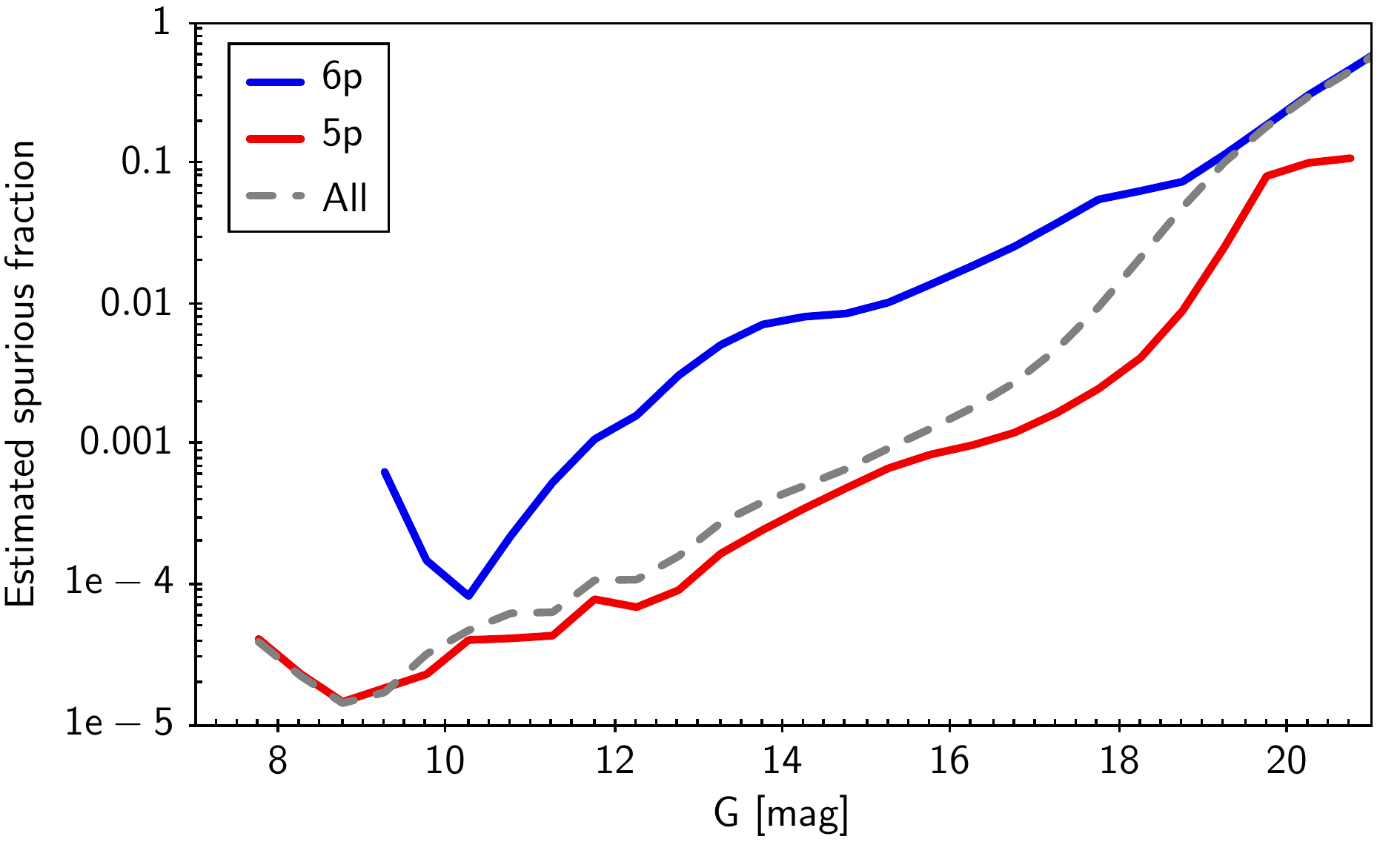}
\includegraphics[width=\hsize]{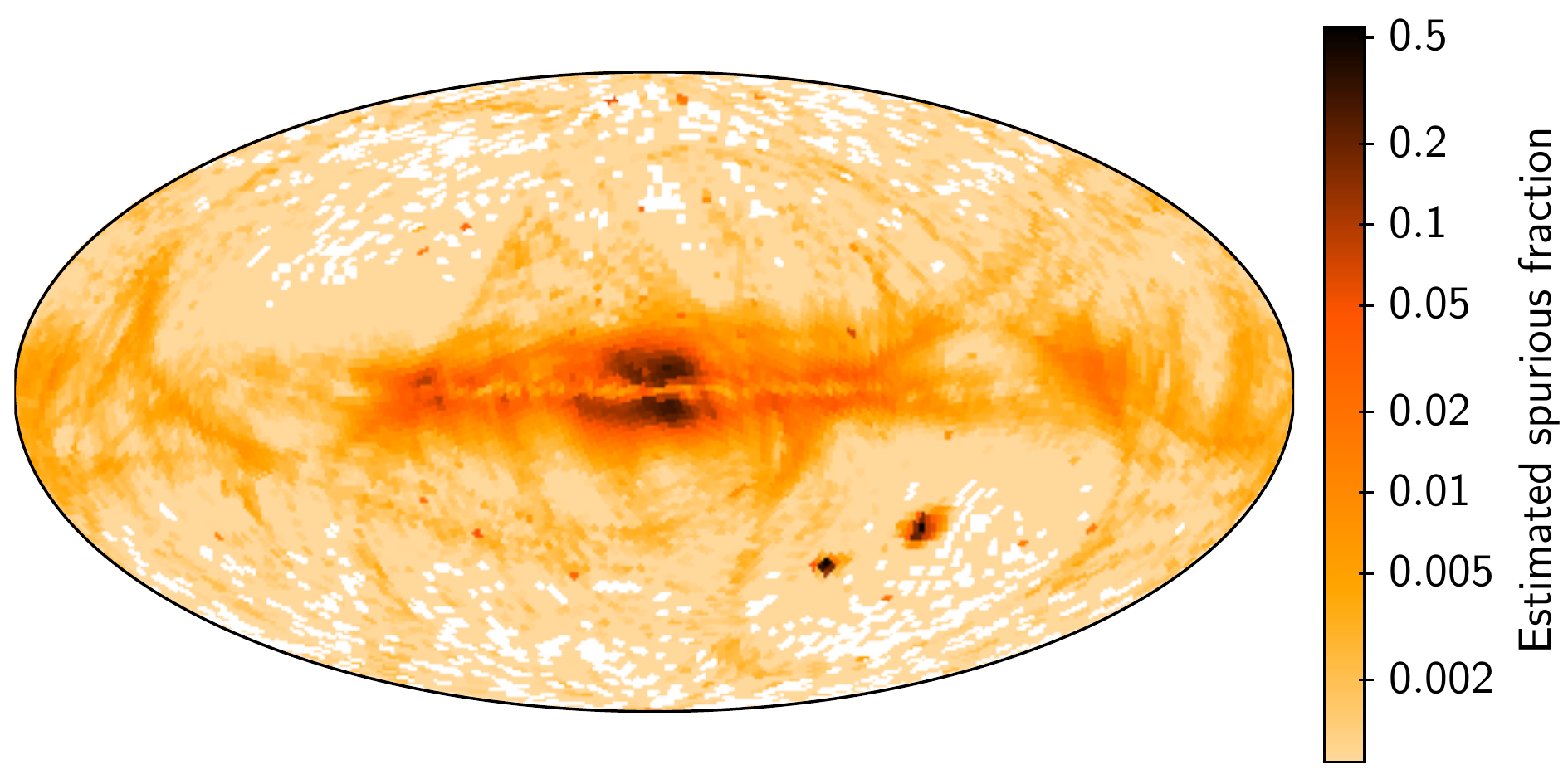}
\caption{Estimated fraction of spurious solutions for sources where the formal
uncertainty is at least five times smaller than the parallax. {\em Top:} 
The fraction by \gmag\ for the whole catalogue (dashed line) as well as separately for 5p (lower, red line) 
and 6p (upper, blue line) solutions. {\em Bottom:} Skymap of the fraction for the whole catalogue 
in Galactic coordinates.
}
\label{fig:ast_sgm5plx}
\end{figure}

An astrometric solution for a source is derived from a cluster of transits
covering nearly three years and associated with the same source identifier. It
is a fundamental assumption that the image parameters for each of these
transits all refer to the same astrophysical source. Ideally, this is an
isolated point source, but it could also be the photocentre of an unresolved
pair. For close source pairs, this assumption breaks down. Depending mainly
on the scan angle, different transits may then give image parameters for one
or the other source, and sometimes for the photocentre. In these cases, there
is a risk that the astrometric solution will produce meaningless proper motions
and parallaxes. Here, we call these spurious astrometric solutions. 
Several different quality indicators for the solution help
to identify such cases, cf.\ \cite{EDR3-DPACP-128}.

A common way of selecting reliable astrometric solutions -- in particular
parallaxes -- is to use only parallaxes larger than some apparently safe
value or only parallaxes much larger than their estimated uncertainties.
At first glance, it does appear safe to use only parallaxes with relatively small
errors, for example\ with \dt{parallax\_over\_error} $ > 5 $. There are 192.21\,million
sources in \egdr{3}\ with such good parallaxes. We use the limit of five as an
illustrative example and not as a recommendation. The point is simply that
the number of negative parallaxes fulfilling the \dt{parallax\_over\_error} $ < -5 $ condition is
expected to be extremely small for a Gaussian error distribution.

Formal uncertainties can, however, be misleading. They are based on the
assumption that the source is undisturbed and can be properly described using a
five-parameter model. This is normally true, but far from always.
One way to find spurious solutions is to count the fraction of very negative
parallaxes, for example for the present example smaller than minus five times the formal uncertainty.
There are
3.04\,million sources with \dt{parallax\_over\_error} $< -5 $\new{. These solutions} are clearly
spurious. 

We can reasonably assume that a disturbance giving rise to a negative (spurious)
parallax, for example\ image parameters affected by duplicity or crowding, could
just as well have produced a spurious solution with a positive parallax and
with roughly the same probability. We therefore get a conservative estimate
of the number of spurious, positive parallaxes by counting the negative ones.
Needless to say, disturbances can also be so small that they merely produce
slightly wrong positive parallaxes, but these cases are harder to find.

We can therefore say that among the 192.21\,million significant, positive
parallaxes, of the order of 3.04\,million are spurious, that is to say\ 1.6\% of this `good'
sample. 
As illustrated in \figref{fig:ast_sgm5plx} (upper panel), the spurious
fraction, determined in this way, strongly depends on magnitude and is much
higher for 6p solutions than for 5p ones. We recall here that 6p solutions are used
for sources where some circumstances prevented good \gbp\ and \grp\ photometry
from being determined in the processing for \gdr{2}. It is reasonable to
assume that it is these very circumstances that have also led to the spurious
astrometry rather than the inclusion of a sixth parameter.  The lower panel of
\figref{fig:ast_sgm5plx} shows that areas such as the LMC and the Galactic centre have a
particularly high fraction of spurious solutions. This is very likely caused by
crowding.
When evaluating parallaxes for a particular sample of sources, where only
positive parallaxes are selected, we therefore 
recommend to also select a similar sample, but with negative parallaxes in
order to evaluate the likely fraction of spurious results.

\begin{figure}[h]
\includegraphics[width=0.9\columnwidth]{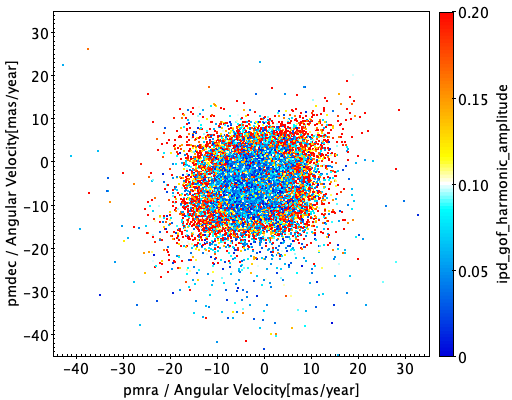}
\includegraphics[width=0.9\columnwidth]{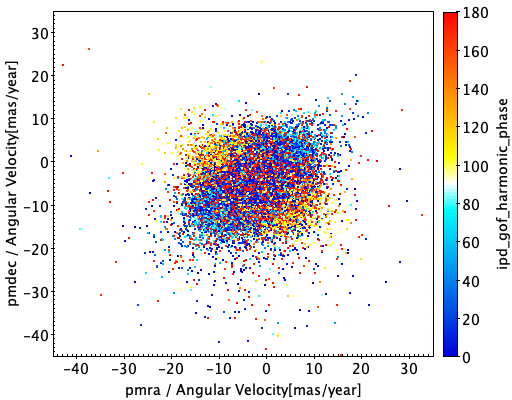}
\caption{Proper motion diagram of sources near the Galactic centre within a $0.5\degr$ radius. {\em Top:} Colour-coded by 
\dt{ipd\_gof\_harmonic\_amplitude}.
{\em Bottom:} Coded by \dt{ipd\_gof\_harmonic\_phase}. 
Reddish points in the top panel reveal potentially spurious solutions.}
\label{fig:spuriousGCpm}
\end{figure}

Thanks to the better angular resolution in \egdr{3}, 
the number of spurious solutions has decreased substantially since \gdr{2}. 
This can be illustrated with a proper motion diagram near the Galactic centre
(Fig.~\ref{fig:spuriousGCpm}). In this region of the ecliptic, with a small 
number of visibility periods, there are mostly two perpendicular scanning 
directions which are now barely visible, but which clearly appeared with spurious proper motions in the
corresponding \gdr{2}\ figure \citep[][fig.~11b]{2018A&A...616A..17A}. 
With many half-resolved doubles in this very dense region, 
distorted image parameters
can explain a large number of spurious solutions, that is to say solutions, which have large proper
motion errors in \gdr{2}.

Compared to \gdr{2}, the dispersion of the proper motions
in Fig.~\ref{fig:spuriousGCpm} is  a factor $>3$ 
smaller, so that one could wonder whether spurious solutions are still present.
Here the
\dt{ipd\_gof\_harmonic\_amplitude}\footnote{\dt{ipd\_gof\_harmonic\_amplitude}
indicates the level of asymmetry in the image, cf.\ Table~\ref{tab:acro}.} 
parameter can be of help: Values above 0.1 for sources
with \dt{ruwe}\footnote{\dt{ruwe} is the renormalised unit weight error (for
astrometry) given in the \gaia\ archive.} larger than 1.4 characterise resolved doubles, which have not been correctly handled yet.  Using this parameter as an explanatory variable 
on \figref{fig:spuriousGCpm} (upper panel), we conclude 
that the corona of relatively large proper motions can be spurious, since
the \dt{ipd\_gof\_harmonic\_phase}\footnote{\dt{ipd\_gof\_harmonic\_phase} indicates the orientation of an asymmetric image.} in \figref{fig:spuriousGCpm} (lower panel)
suggests that these sources were partly resolved along the two principal 
scanning directions.

%
\subsection{Large-scale systematics}\label{sec:ast_large}
%

The quasars are distant enough so that the DR3 measured parallax directly gives the astrometric error, thus QSOs can be used to estimate the large-scale 
variation of the parallax systematics.
The QSO sample used is mostly a subset from outside of the Galactic plane of
the sources listed in the table \dt{agn\_cross\_id} published as part of the {\gaia}
Archive for EDR3. The sample was filtered from potential $5\sigma$ outliers in
parallax or proper motion and from potential non-single objects using: \dt{ruwe} $<
1.4$ and \dt{ipd\_frac\_multi\_peak} $\leq 2$ and
\dt{ipd\_gof\_harmonic\_amplitude} $< 0.1$.

\begin{figure}[h!]
\begin{center}
\includegraphics[width=\hsize,height=0.55\columnwidth]{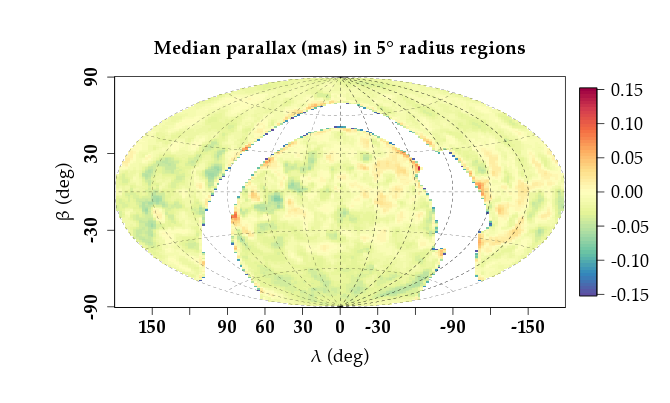}
\includegraphics[width=\hsize,height=0.6\columnwidth]{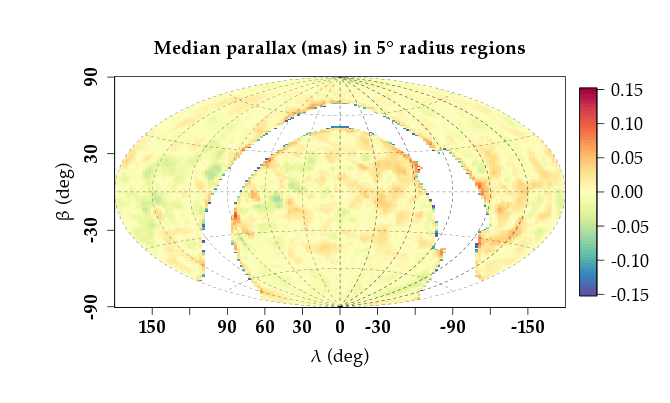}
\end{center}
\caption{Maps in ecliptic coordinates of the variations of QSO parallaxes (mas) in $5\degr$ radius fields.
{\em Top}: 5p solution. {\em Bottom}: 5p with zero point correction.
}\label{fig:eclqsoMedPlx}
\end{figure}

Median parallaxes were computed in overlapping regions of
radius 5\degr\ having 
at least 20 QSOs and are shown \new{in} \figref{fig:eclqsoMedPlx}.  
Compared to the similar plot done for \gdr{2}
\citep[][fig.~15]{2018A&A...616A..17A}, the improvement in the top panel of
\figref{fig:eclqsoMedPlx} is very clear for the 5p solutions. A slight north-south asymmetry appears with, for example, parallaxes
below $\beta < -30\degr$ being about 8{\muas} more negative than those above $\beta > 30\degr$.
Applying the zero point correction from \cite{EDR3-DPACP-132} removes this asymmetry (see bottom
panel of
\figref{fig:eclqsoMedPlx});
some east-west asymmetry along the ecliptic of a few {\muas} may, however, remain.
It is more difficult to conclude about the 6p solutions:
They represent only 20\% of the QSO sample and have larger uncertainties, so
the amplitude of the variations may be more related to random errors
than to systematics.

\subsection{Comparison to external data}\label{sec:ast_ext_src}

We compared the \egdr{3} parallaxes with external catalogues described in
detail in the \egdr{3}\ online documentation.\footnote{
\url{https://gea.esac.esa.int/archive/documentation/GEDR3/Catalogue\_consolidation/chap\_cu9val/sec\_cu9val\_944/}} Those are the same as used in
\cite{2018A&A...616A..17A}, except that we updated the APOGEE catalogue to the DR16 version
\citep{2020ApJS..249....3A}, the ICRF catalogue to its third realisation \citep{ICRF3}, and
added dSph members.  We show the summary of the results in
Table~\ref{tab:cu9val_wp944_summaryplx} without and with the parallax zero point correction of
\cite{EDR3-DPACP-132} applied. The correction significantly
improves the parallax differences, the exceptions being the LMC and SMC \TBC{(Small Magellanic Cloud)} stars
sub-selected by their \gdr{2}\ radial velocities, which are bright, and the two
largest dSph of our sample, Sculptor and Fornax. The parallax difference with
{\hip} is within the expected {\hip} parallax zero point uncertainty
\citep[up to 0.1~mas,][]{1995A&A...304...52A}, but a correlation \TODO{with the magnitude} of the parallax
difference \TBC{with the magnitude} is seen \TBC{for}\TODO{with } \hip\ stars brighter than $G=6$~mag. 
The jump in the parallax zero point at $G\sim13$~mag \citep{EDR3-DPACP-132} is seen in the APOGEE comparison and removed by the application of the parallax zero point correction.
The variation in the QSO parallax with \TBC{the}\TODO{a } magnitude for 5p solutions was also removed by the parallax zero point correction. 
A correlation of the parallax zero point with \TBC{the}\TODO{a } pseudo-colour is seen in the dSph, in particular in Fornax, which was reduced but not fully removed by the parallax zero point correction of \cite{EDR3-DPACP-132}.

\begin{table*}
\caption[Comparison between the \egdr{3}\ parallaxes and the external catalogues]{Summary of the comparison between the {\gaia} parallaxes and the external catalogues.}
\begin{center}
\begin{tabular}{lccccccc} 
\hline\hline
{ Catalogue} & Nb & { Outliers} & $<G>$ & { \parallax~difference} & { \parallax$_{cor}$~difference } & { \parallax$_{cor}$~uwu} & \\ 
 \hline
   {\hip} & 62484 & 0.15\% & 8.2 & \textcolor{violet}{$-$0.089 $\pm$ 0.003} & \textcolor{violet}{$-$0.068 $\pm$ 0.002} & \textcolor{violet}{2.4 $\pm$ 0.4} \vspace{2mm} \\  
  VLBI & 40 & 8\% & 8.3 & \textcolor{black}{ $-$0.01 $\pm$ 0.01} & \textcolor{black}{ 0.02 $\pm$ 0.01} & \textcolor{black}{2.0 $\pm$ 0.4} \\ 
  HST & 49 & \textcolor{violet}{27\%} & 12.2 & \textcolor{black}{0.01  $\pm$ 0.02} & \textcolor{black}{0.03  $\pm$ 0.02} & \textcolor{black}{1.9 $\pm$ 0.3} \\
  RECONS & 427 & 3\% & 12.6 & \textcolor{violet}{$-$0.8 $\pm$ 0.06} & \textcolor{violet}{$-$0.8 $\pm$ 0.06} & \textcolor{violet}{1.54 $\pm$ 0.06} \vspace{2mm}  \\ 
  \gaia\ Cepheids & 1372 & 1\% & 15.7 & \textcolor{violet}{$-$0.028 $\pm$ 0.0007} & \textcolor{black}{0.006 $\pm$ 0.0007}  & \textcolor{violet}{1.22 $\pm$ 0.02}\\ 
  \gaia\ RRLyrae & 318 & 2\%  & 18.1 & \textcolor{violet}{$-$0.030 $\pm$ 0.008}  & \textcolor{black}{$-$0.001 $\pm$ 0.008}  & \textcolor{violet}{1.04 $\pm$ 0.05} \vspace{2mm} \\ 
  APOGEE & 3453 & 2\% & 18.6 & \textcolor{violet}{$-$0.046 $\pm$ 0.003}  & \textcolor{black}{$-$0.007 $\pm$ 0.003}  & \textcolor{violet}{1.19$\pm$0.02} \\ 
  SEGUE Kg & 2491 & 0.04\% & 17.3 & \textcolor{violet}{$-$0.029 $\pm$ 0.002}  & \textcolor{black}{0.004 $\pm$ 0.001}  & \textcolor{black}{1.07$\pm$0.02} \vspace{2mm} \\ 
  LMC & 52795 & 0.7\% & 19.2 & \textcolor{violet}{ $-$0.023 $\pm$ 0.0003} & \textcolor{black}{ 0.003 $\pm$ 0.0003} & \textcolor{violet}{1.3 $\pm$ 0.004} \\ 
  LMC Vr & 318 & 1.6\% & 12.8 & \textcolor{black}{ $-$0.004 $\pm$ 0.001} & \textcolor{violet}{ 0.015 $\pm$ 0.001} & \textcolor{violet}{2.17 $\pm$ 0.09} \\ 
  SMC & 26480 & 1.3\% & 16.4 & \textcolor{violet}{ $-$0.0255 $\pm$ 0.0002} &  \textcolor{black}{0.0055 $\pm$ 0.0002} &  \textcolor{violet}{1.26 $\pm$ 0.006}  \\ 
  SMC Vr & 114 & 9.6\% & 12.5 & \textcolor{black}{$-$0.006 $\pm$ 0.001} & \textcolor{violet}{0.016 $\pm$ 0.001} & \textcolor{violet}{1.5 $\pm$ 0.1} \\ 
  Draco & 427 & 0\% & 19.3 & \textcolor{violet}{ $-$0.024 $\pm$ 0.005} & \textcolor{black}{ 0.0002 $\pm$ 0.006} & \textcolor{black}{ 1.09 $\pm$ 0.04 } \\ 
  Sculptor & 1286 & 0.08\% & 19.1 & \textcolor{violet}{ $-$0.011 $\pm$ 0.005 } & \textcolor{violet}{ 0.015 $\pm$ 0.004 } & \textcolor{black}{ 1.09 $\pm$ 0.03 } \\ 
  Sextans &  528 & 0\% & 19.4 & \textcolor{black}{ $-$0.015 $\pm$ 0.01} & \textcolor{black}{ 0.01 $\pm$ 0.01} & \textcolor{black}{ 1.00 $\pm$ 0.03 } \\ 
  Carina & 865 & 0\% & 19.8 & \textcolor{black}{ $-$0.014 $\pm$ 0.005} & \textcolor{black}{ 0.012 $\pm$ 0.005} & \textcolor{black}{ 1.03 $\pm$ 0.02 } \\ 
  Antlia II & 159 & 0\% & 18.9 & \textcolor{black}{$-$0.025 $\pm$ 0.012} & \textcolor{black}{$-$0.0006 $\pm$ 0.01} & \textcolor{black}{ 0.89 $\pm$ 0.05 } \\ 
  Fornax & 2660 & 0.6\% & 18.8 & \textcolor{violet}{ $-$0.013 $\pm$ 0.003} & \textcolor{violet}{ 0.011 $\pm$ 0.003} & \textcolor{violet}{1.16  $\pm$  0.03 } \\ 
  LeoII &  185 & 0\% & 19.8 & \textcolor{black}{ 0.005 $\pm$ 0.03} & \textcolor{black}{ 0.02 $\pm$ 0.03} & \textcolor{black}{1.0 $\pm$  0.05 } \\ 
  LeoI & 328 & 0.3\% & 19.6 & \textcolor{violet}{ $-$0.063 $\pm$ 0.02} & \textcolor{black}{ $-$0.05 $\pm$ 0.02} & \textcolor{violet}{ 1.11  $\pm$  0.04 } \\ 
  all dSph & 7174 & 0.3\% & 19.0 & \textcolor{violet}{ $-$0.017 $\pm$ 0.002} & \textcolor{black}{ 0.008 $\pm$ 0.002} & \textcolor{black}{ 1.09 $\pm$ 0.01} \vspace{2mm} \\ 
   ICRF3 & 3172 & 0.06\% & 18.9 & \textcolor{violet}{ $-$0.023 $\pm$ 0.002} & \textcolor{black}{ 0.001 $\pm$ 0.002} & \textcolor{black}{1.10 $\pm$ 0.01} \\ 
   LQRF & 8231 & 0.03\% & 19.1 & \textcolor{violet}{ $-$0.024 $\pm$ 0.0006} & \textcolor{black}{ $-$0.0015 $\pm$ 0.0006} & \textcolor{black}{1.063 $\pm$ 0.003} \\ 
   RFC2016cnoU &  3705 & 0.05\% & 18.9 & \textcolor{violet}{ $-$0.022 $\pm$ 0.002}  & \textcolor{black}{ 0.002 $\pm$ 0.002}  & \textcolor{black}{1.090 $\pm$ 0.01} \\ 
 \hline
\end{tabular}
\tablefoot{
We present the total number of stars used in the comparison (Nb) as well as the percentage of outliers excluded (at 5$\sigma$, in violet if higher than 10\%) as well as the median {\gmag} of the sample.  The parallax differences ($\varpi_G-\varpi_E$, in mas), the parallax difference with the correction of \citet{EDR3-DPACP-132} applied (\parallax$_{cor}$) and the unit-weight uncertainty (uwu) that needs to be applied to the uncertainties to adjust the differences are indicated in violet when they are significant (p-value limit: 0.01) and higher than 10\,{\muas} for the parallax difference and 10\% for the uwu. For the {\hip} catalogue, in addition to the uwu, a systematic uncertainty of 0.5~mas has yet to be quadratically added. 
}
\end{center}
\label{tab:cu9val_wp944_summaryplx}
\end{table*}

Concerning the proper motions, we looked in particular at the difference between the \gaia\ proper motion and the proper motion derived from the positions of \gaia\ and \hip. By construction \citep[sect.~4.5 of][]{EDR3-DPACP-128}, the global rotation between those proper motions, seen in {\gdr{2}} \citep{DR2-DPACP-51,2018ApJS..239...31B}, is not present anymore. However, a variation of this rotation with magnitude and colour is still present but smaller than for \gdr{2} (the maximum variation reaching 0.1\,\masyr\ for bright or red sources). We note that between \gaia\ and \hip\ proper motions, a global rotation is still present with $w=$($-0.120$, 0.173, 0.090) $\pm$ 0.005 {\masyr}. \TBC{This is}\TODO{and } a deviation \TODO{is } well within the estimated accuracy of the \hip\ spin.

\subsection{Comparison to a Milky Way model}\label{sec:ast_model}
\begin{figure}
\begin{center}
\includegraphics[width=0.5\textwidth]{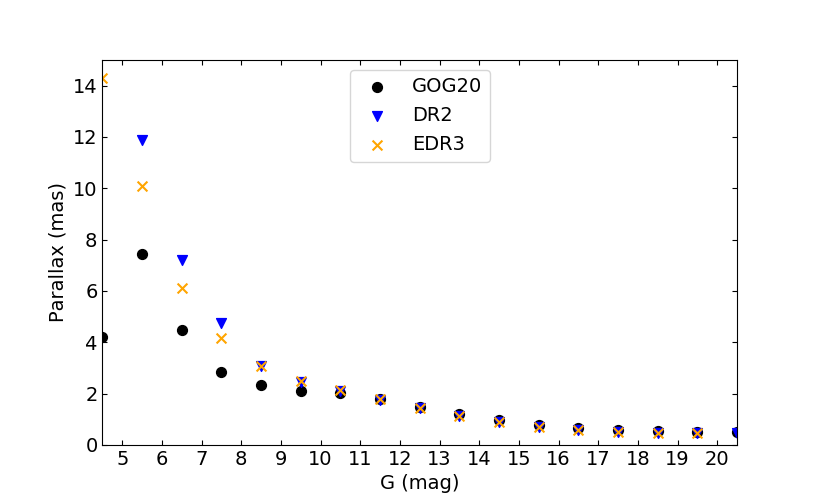} 
\caption[Average parallax per magnitude]{Parallaxes averaged \new{among healpix bins} over the whole sky as a function of magnitude for \egdr{3} \new{(orange crosses)}, GOG20 \new{(black circles)}, and \gdr{2}  \new{(blue triangles)}.
}\label{fig:par-gmag}
\end{center}
\end{figure}

We compared the astrometric data to that of the GOG20 simulation in order to investigate potential systematic errors.  This was done by computing the median of the parallaxes and the median of the proper motions in each healpix bin of the sky map for all of the data and the model.  The comparison for the median parallaxes are shown in Fig.~\ref{fig:par-gmag} as a function of magnitude for \egdr{3}, \gdr{2,}\ and the GOG20 simulation.

The median parallaxes are generally in very good agreement between  \egdr{3} and GOG20, specially at magnitudes larger than ten. However, there is a systematic difference which, in absolute value, depends on magnitude, and it is quite high on the bright side and more than 1 mas. This systematic difference between the data and model simulation is a bit reduced in \egdr{3} compared to \gdr{2}. At $G > 10$\,mag, the difference goes below 0.1 mas. Regarding the proper motions, the model and data present approximately similar patterns in all magnitude ranges. However, there are systematics, as was already noted in the validation of \gdr{2}. Overall, \egdr{3} data are as expected from our knowledge of the Galactic
kinematics up to very faint magnitudes and it is probably the model which suffers from systematics,
or it does not account for asymmetries. Indeed, we note that the change of kinematics prescriptions from GOG18 to
GOG20 generally allows for a better agreement with the data.

\subsection{Astrometric zero point  and precision of the parallaxes  from cluster analysis}

\begin{figure*}
 \begin{center}
\includegraphics[width=14cm]{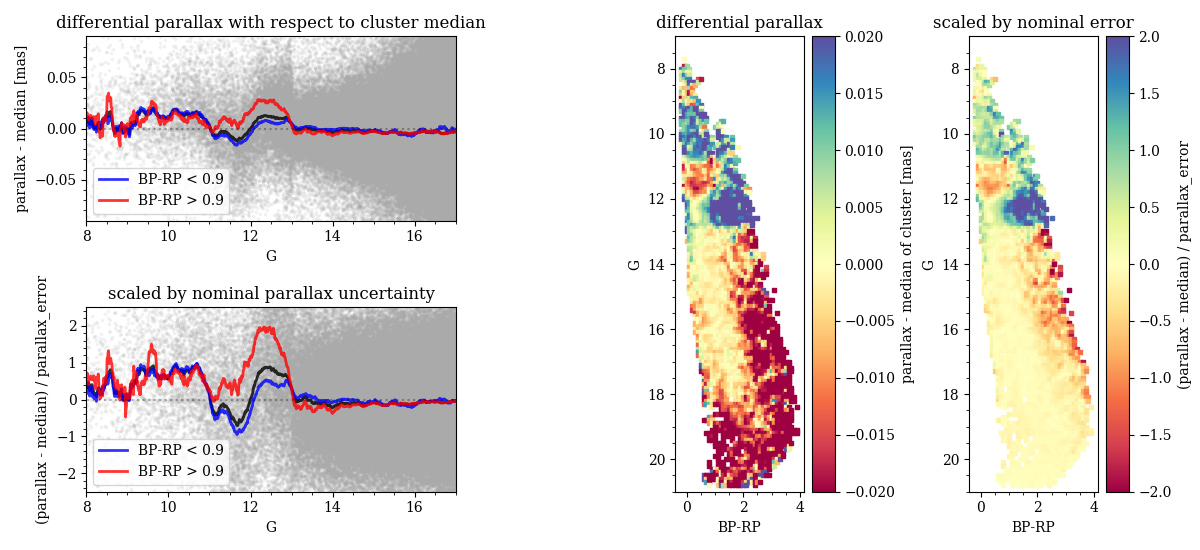}
\end{center}
\caption{{\em Left:} $\Delta_{\varpi}$ (top) and scaled  to the nominal uncertainties $\Delta_{\varpi}$ (bottom) versus {\gmag} for the whole sample of clusters. The solid lines show the 
LOWESS (locally weighted scatterplot smoothing) of the stars bluer (redder) than \bpminrp\ $=0.9$ (blue and red lines), while the black line is for the whole sample. {\em Right:} CMD of the whole sample where the colour shows the differential parallax to the median.}\label{fig:ScaledParColor}
\end{figure*}

The zero point of the parallaxes was verified using three external reference catalogues of open clusters.  We made use of \cite{2014A&A...564A..79D} (hereafter DAML),  \cite{2013A&A...558A..53K} (hereafter MWSC), and finally \cite{2020A&A...640A...1C}  based on \gdrtwo   parallaxes. We selected the most suitable clusters for this aim:  a selection of  about  200 clusters with well known parameters (hereafter best sample) for a total of about 70\,000 stars; and a
wider sample of 2043 clusters including 250\,000 stars (hereafter whole sample).  The best sample is the same sample that was already used to validate \gdrtwo  in \cite{2018A&A...616A..17A}. The whole sample is based on  \cite{2020A&A...640A...1C} cluster identification and parameters. Cluster members for this analysis were obtained  using {\gedrthree} proper motions selected within  one $\sigma$ from the mean value.
Clusters closer than 1000 pc show an intrinsic  internal dispersion in the parallaxes and are not suitable for estimating the zero point. When we used clusters located farther away, we derived an average zero point difference 
($\varpi_{\rm Gaia}-\varpi_{\rm reference})=$ $-0.059$\,mas for MWSC and $-0.091$\,mas for DAML, but with a large $\sigma \sim 0.2$\,mas.  Looking at the trends  of the zero point with colour and magnitude, we find a complex pattern.
In Fig.~\ref{fig:ScaledParColor}, we plotted the differential parallax  to the cluster median  $\Delta_{\varpi}$  for the whole sample of clusters located farther than 1000 pc, which were normalised and not normalised  to the nominal parallax uncertainties. We note that $\Delta_{\varpi}$ gives an indication about zero point changes. When plotted versus \gmag, we detected discontinuities in the zero point at \gmag\ $\sim$ 11, 12, and 13\,mag. Strong variations are evident for  stars bluer and redder than \bpminrp\ $\sim 0.9$\,mag.   Figure~\ref{fig:ScaledParColor} (right panel) presents the variation of $\Delta_{\varpi}$ in the colour-magnitude diagram, showing a number of discontinuities and a complex pattern.  At faint magnitudes, red stars have a higher dispersion; however, the effect can be due  to the less reliable membership, while at red colours, the large variations can reflect poor statistics.      When divided by the nominal uncertainty, these patterns are still present with a reduced amplitude, implying that nominal uncertainties on the parallax do not account for the zero point variation, that is to say\ nominal uncertainties are underestimated.

\begin{figure*}
 \begin{center}
\includegraphics[width=0.75\textwidth]{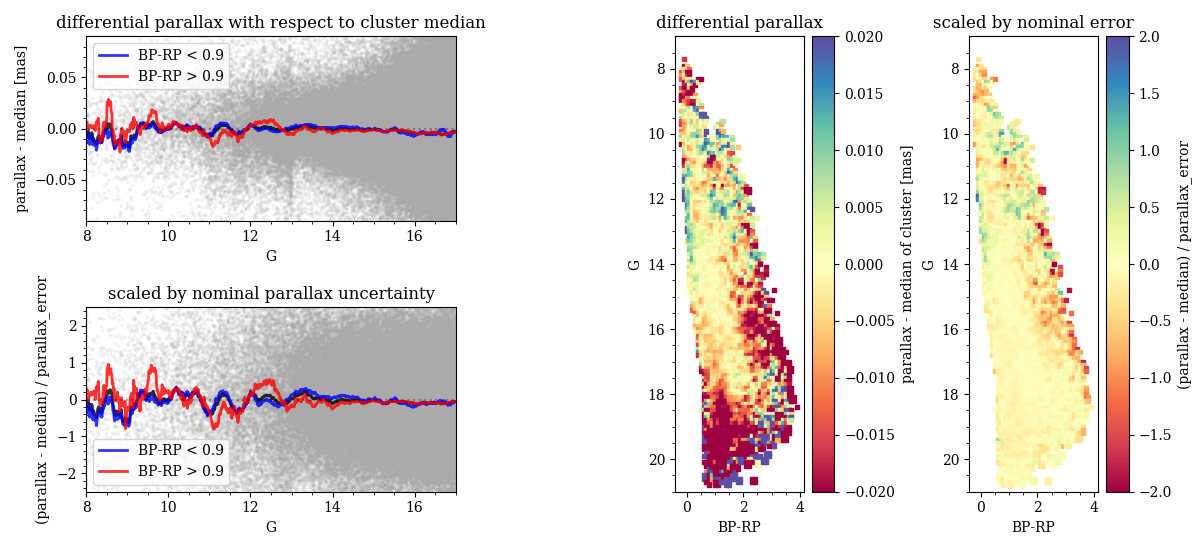}
\end{center}
\caption{{\em Left:}  $\Delta_{\varpi}$ (top) and  the scaled  $\Delta_{\varpi}$ (bottom) versus magnitude for the whole sample after the  correction to the parallax zero point is applied. The solid lines show the LOWESS of the stars bluer (redder) than \bpminrp\ = 0.9\,mag (blue and red lines). The black line is the for the whole sample. {\em Right:} CMD of the whole sample where the colour shows the differential parallax to the median, and the analogous scaled to the nominal parallax uncertainties
after parallax zero point correction (see text for details). }\label{fig:ScaledParColorCorr}  
\end{figure*}

Clusters are very good targets to test the quality of the parallax correction
from \cite{EDR3-DPACP-132} since all the stars are expected to have the same
parallax. In addition, clusters can be found close to the Galactic plane, where
no calibration tracers are located. This allows for an independent
verification.  We applied this correction to the cluster data, using the
Matlab code\footnote{A python code is distributed with  \egdr{3}\ at
\url{https://www.cosmos.esa.int/web/gaia/edr3-code}.} 
provided in the \cite{EDR3-DPACP-132} paper.

The results are shown in Fig.\ref{fig:ScaledParColorCorr}  where we plotted the $\Delta_{\varpi}$   and  the scaled $\Delta_{\varpi}$, that is  to say the analogue was scaled  to the  nominal uncertainties on parallaxes   as a function of the \gmag\ magnitude, and finally  the residuals to the median in the colour magnitude diagram.  This correction reduces the dispersion inside the clusters at bright magnitudes and bluer colours, while at faint magnitudes ($G>18$\,mag) or a redder colour, the dispersion is still high.  The median values scaled to the nominal uncertainties are always $<1$, which indicates that nominal uncertainties account for the residual variations.  Clearly this positive result should be taken with caution. It refers to a specific range of colours and positions in the sky.

Finally, we compare  the parallaxes  of single stars in  \gdrtwo and \egdr{3}\ for the whole cluster sample. The median difference is (\parallax$_{\rm \gdr{3}}-$\parallax$_{\rm \gdrtwo})= 0.017$ mas (with 16th percentile $= -0.047$ mas; 84th percentile $=0.082$ mas)  with a dependence on the magnitude. 

%
\subsection{Uncertainty of the astrometric parameters}
\label{sec:ast_precision}
%
We evaluate the actual precision of the astrometric parameters partly from
parallaxes and proper motions of QSOs and of stars in the LMC and partly
from deconvolution of the negative parallax tail.
As discussed by \cite{LL_VIENNA}, it is useful to describe the true external
parallax uncertainty, $\sigma_{\rm ext}$, as the quadratic sum of the formal catalogue uncertainty
(\dt{parallax\_error}) times a multiplicative factor ($k$) plus a systematic error ($\sigma_{\rm s}$),
\begin{equation}
\sigma_{\rm ext}^2 = k^2\sigma_i^2+\sigma_{\rm s}^2.
\label{eq:sigma}
\end{equation}
In addition to this, the catalogue uncertainties incorporate part of the excess noise
of the solution when present.\ Consequently, large uncertainties typically
correspond to both fainter sources and/or non-single stars.

\subsubsection{Uncertainty of parallax and proper motion from distant objects}

\begin{figure}
\centering
\includegraphics[width=0.8\columnwidth]{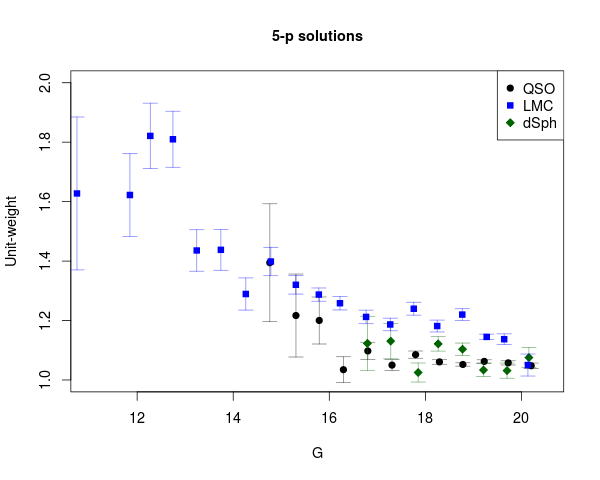}
\includegraphics[width=0.8\columnwidth]{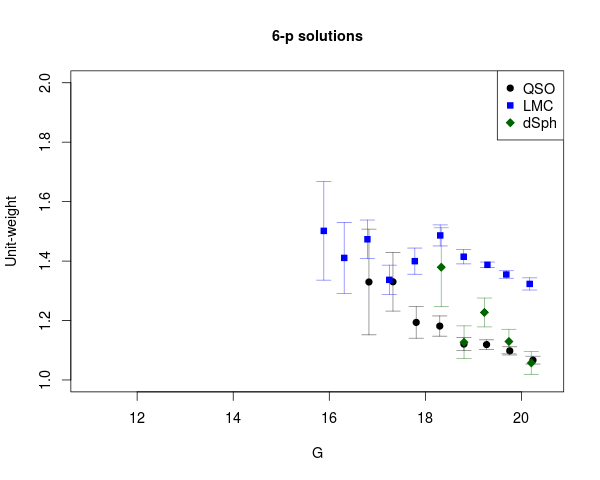}
\caption{Unit-weight uncertainty (uwu) that needs to be applied to the \gaia\ parallax uncertainties to be consistent with the residual distribution (after zero point correction and removing stars with \dt{ruwe}\ $<1.4$) versus LQRF QSOs, LMC, and dSph stars for the 5p \new{(top)} and 6p \new{(bottom)} solutions, as a function of magnitude.}
\label{fig:wp944_varfact}
\end{figure}

\begin{figure} \centering
\includegraphics[width=0.7\columnwidth]{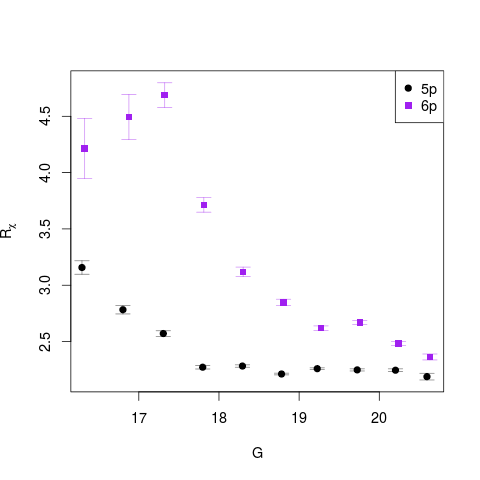}
\caption{$\chi^2$ test of the LQRF QSOs proper motions as a function of $G$
magnitude for the 5p \new{(black circles)} and 6p \new{(purple squares)} solution. The residual $R_\chi$ should follow a
$\chi^2$ with 2 degrees of freedom. The correlation observed here is due to the
underestimation of the uncertainties as a function of magnitude.
\label{fig:wp944_qsopmchi2} } \end{figure}

Similarly to what was found for \gdr{2}\ by \cite{2018A&A...616A..17A}, QSOs show
that the uncertainties are slightly underestimated and that this
under-estimation increases with magnitude. The under-estimation is lower than
for \gdr{2} for the 5p solution, but larger for the 6p
solution. This is seen for a parallax using the unit-weight uncertainty (uwu) in Fig.~\ref{fig:wp944_varfact} and for
proper motion using a $\chi^2$ test in Fig.~\ref{fig:wp944_qsopmchi2}. The
trend with magnitude is also seen with LMC stars, although the under-estimation
of uncertainties is higher, as presented in Fig.~\ref{fig:DeconvPlx}, which is most
probably due to the crowding.  The increase in the under-estimation is also
seen in the uwu presented in Table~\ref{tab:cu9val_wp944_summaryplx}. 
The uwu reported is the $k$ term of Eq.~\ref{eq:sigma}, assuming a negligible systematic error term ($\sigma_{\rm s}$) except for the \hip\ comparison which is the only catalogue for which both terms could be clearly separated. 
The uwu and the residual $R_\chi$ were
computed after applying the parallax zero point correction of
\cite{EDR3-DPACP-132} and removing stars with \dt{ruwe}\ $>1.4$.

%
%
\subsubsection{Parallax uncertainties by
deconvolution}\label{sec:plx_deconvolution}
%

\begin{figure}[h!]
\begin{center}
\includegraphics[width=\columnwidth, height=0.5\columnwidth]{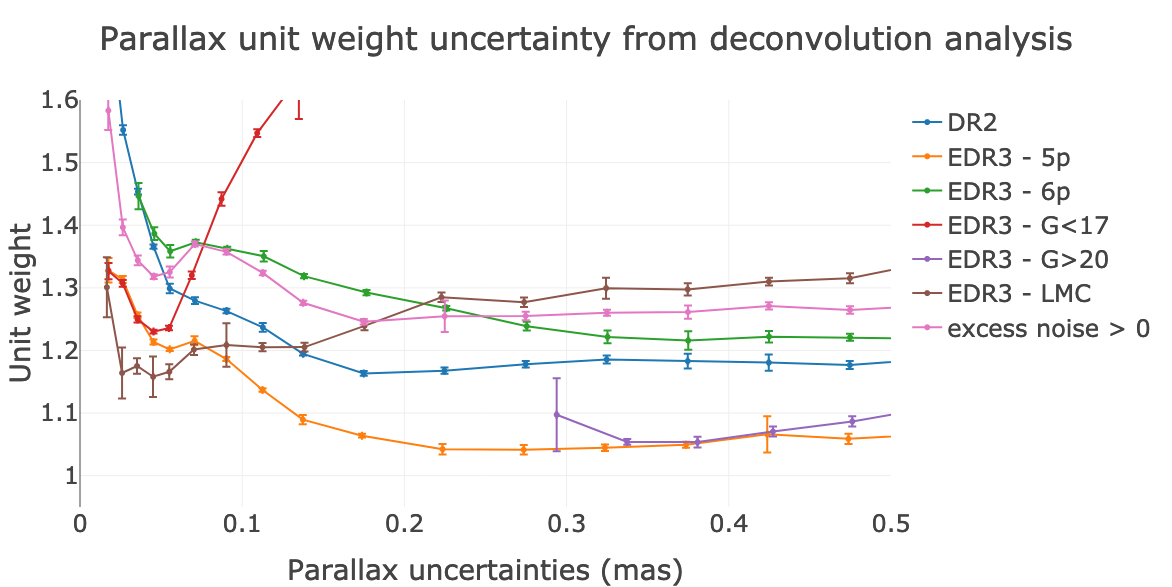}
\end{center}
\caption{Uwu of the parallaxes estimated by deconvolution
versus \dt{parallax\_error} (mas) for several subsets: DR2 and EDR3 for 5p, 6p, G < 17\,mag, G > 20\,mag, LMC, 
and non-zero excess noise.}\label{fig:DeconvPlx}
\end{figure}

The `true' dispersion of the parallaxes was estimated using a
deconvolution method, which was applied on the negative tail of the parallaxes \citep[see
][sect.~6.2.1 for details]{2017A&A...599A..50A}, and the uwu ratio of the external over the internal uncertainty was computed.  Figure~\ref{fig:DeconvPlx} shows the uwu as a function of the
catalogue uncertainties for several illustrative subsets and it can give insights
into the underestimation factor, the systematics, and the contamination
by non-single sources.  

On the right side of the figure, the asymptotic uwu is mostly flat and it gives
the multiplicative factor: It is about 1.05 for 5p solutions (improved from
DR2), slightly more for very faint stars, 1.22 for 6p solutions, and larger for
sources with non-zero excess noise or those in the LMC. While the uwu is in general
slowly increasing with uncertainty due to the contamination by non-single sources, it increases
sharply for sources brighter than 17\,mag, most probably due to half-resolved
doubles (as indicated by the average \dt{ipd\_frac\_multi\_peak} or
\dt{ipd\_gof\_harmonic\_amplitude}), which suggests that
the uncertainties of these non-single bright stars are underestimated quite a bit.
Then, the left part of the curves shows the influence of the systematics. As
confirmed by other tests, the systematics have decreased compared to DR2,
except for 6p solutions, and they are largest for sources with non-zero excess
noise, which is due to either calibration errors or to perturbation of non-single stars.

\subsection{Magnitude dependence from binary stars}\label{sec:ast_mag}
%
One way to check the magnitude variations of the parallax zero point is to use
resolved binary stars. When the period of the binary system is long enough, the
proper motion of the two components is similar, or at least the differences are
smoothed out when a large sample is used.  Potential common proper motion pairs
have been selected over the whole sky; this has been restricted to primaries up to $G < 15$\,mag
and secondaries up to $G < 18$\,mag only: Selecting fainter secondaries would increase the fraction of optical doubles in dense fields too
much, thus biasing the
parallax differences. 
We computed the differences between the two components
of their parallax and proper motion, then the norm of this difference accounting
for its covariance was determined, and a pair was considered as a true binary
if this $\chi^2$ had a p-value above 0.01 and the linear separation
was below $10^4$ au.

\begin{figure}[h!]
 \begin{center}
\includegraphics[width=0.7\columnwidth]{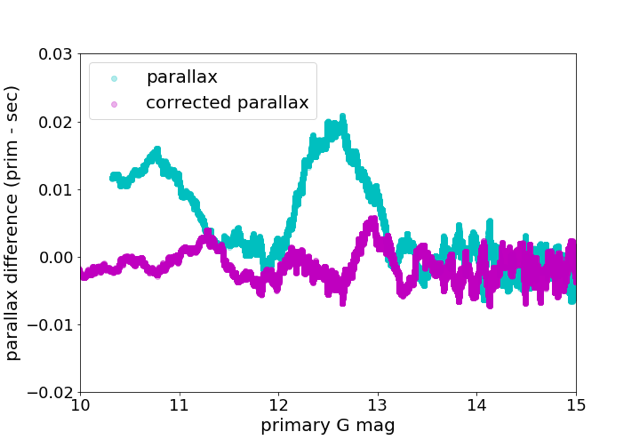}

\includegraphics[width=0.7\columnwidth]{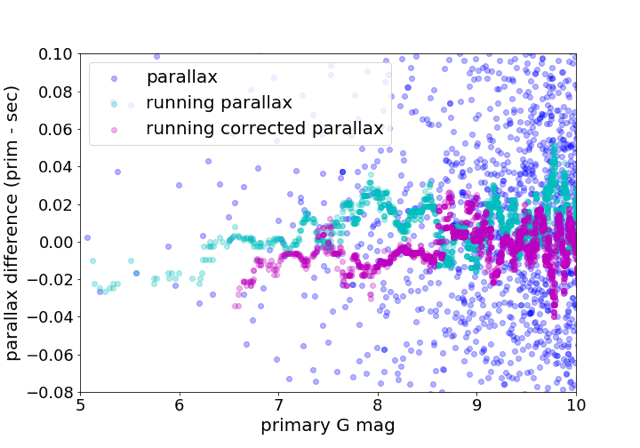}

\includegraphics[width=0.7\columnwidth]{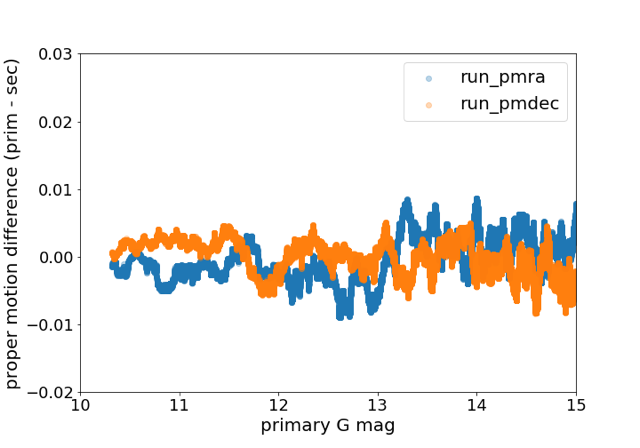}
\end{center}
\caption{Differential astrometry for common proper motion pairs shown in the sense primary minus secondary 
component. 
{\em Top:} Differences between parallaxes (mas, running medians over 3000 sources) before (blue)
and after (magenta) zero point correction, 
as a function of the primary magnitude, $10<G<15$\,mag.
{\em Middle:} Primary minus secondary parallaxes before and after zero point correction
for stars brighter than $G=10$\,mag (running medians over 100 sources).
{\em Bottom:} Differential proper motion (\masyr)
in right ascension (blue) and declination (orange) versus
magnitude, running medians over 3000 sources. }\label{fig:DiffBinaryPlx}
\end{figure}

The differences of the parallax and proper motion between primary and 
secondary components are represented in the top panel of
\figref{fig:DiffBinaryPlx} versus the $G$ magnitude of the primary.
For $12 < G < 13$\,mag, a 20\,{\muas} increase for the parallax zero point differences
appears clearly, which was not present with DR2 parallaxes. 
A significant improvement is seen after the zero point correction 
from \cite{EDR3-DPACP-132}.
Although within the uncertainty (below 0.007 mas, approximately constant),
a small residual effect near $G\approx 11.5$\,mag or $G\approx 13$\,mag may perhaps still
be present. 
It should not come as a surprise that these variations occur in the magnitude interval where there is 
a change in the gating scheme or in the window size \citep[][sect.~2.1]{EDR3-DPACP-132}. 
For primaries brighter than 7\,mag (middle panel of \figref{fig:DiffBinaryPlx}), the zero point
may be more negative, as can also be seen using known WDS binaries.


We tested the consistency between the parallax of two components of wide
physical binaries, which were selected from the WDS (separation larger than 0.5\arcsec\ and flag
'V').  
We again found how the magnitude dependence
of the parallax zero point present in \egdr{3}\ is nicely removed by applying
the parallax zero point correction of \cite{EDR3-DPACP-132}, thus confirming the
results shown in \figref{fig:DiffBinaryPlx} (middle panel).
Similarly,
following \cite{2019A&A...623A.117K}, we used their catalogue of Cepheids and
RR Lyrae resolved common proper motion pairs to check the compatibility of
their parallaxes. With the parallax zero point correction of
\cite{EDR3-DPACP-132}, and opposite to the \gdr{2}\ results for Cepheids, no systematic offset nor strong outlier (outside 5$\sigma$) were found.

While no variation appears for $\mu_{\delta}$ in the bottom panel of \figref{fig:DiffBinaryPlx}, there is an
increase of about 0.01\masyr\ at $G=13$\,mag for \pmra. Although
this may not look statistically significant (uncertainty 
$\approx 0.007$\masyr), this effect is probably real, as
an empirical 0.1 mas rotation correction was applied to the proper motion
system for $G<13$\,mag \citep[][sect.~4.5]{EDR3-DPACP-128}.

%
\subsection{Pseudo-colour dependence}\label{sec:ast_6p}
%
\begin{figure}[h!]
\begin{center}
\includegraphics[width=0.8\columnwidth]{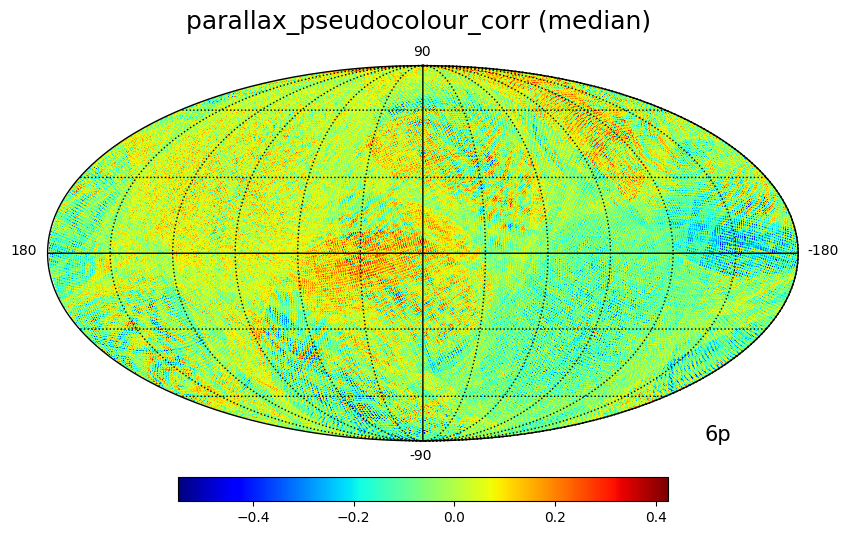}

\includegraphics[width=0.8\columnwidth]{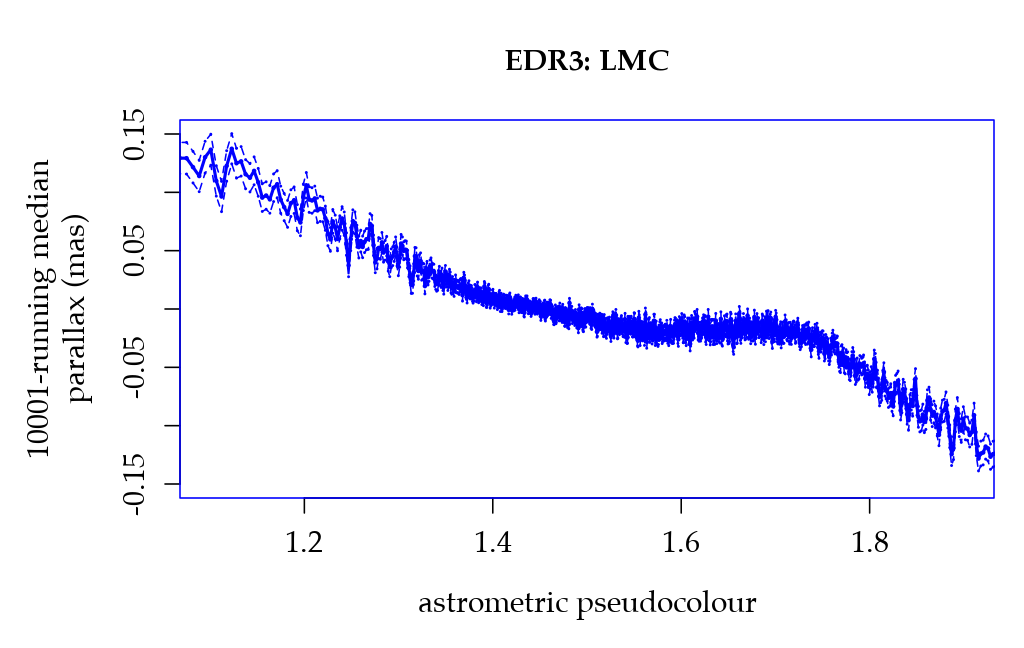}

\includegraphics[width=0.8\columnwidth]{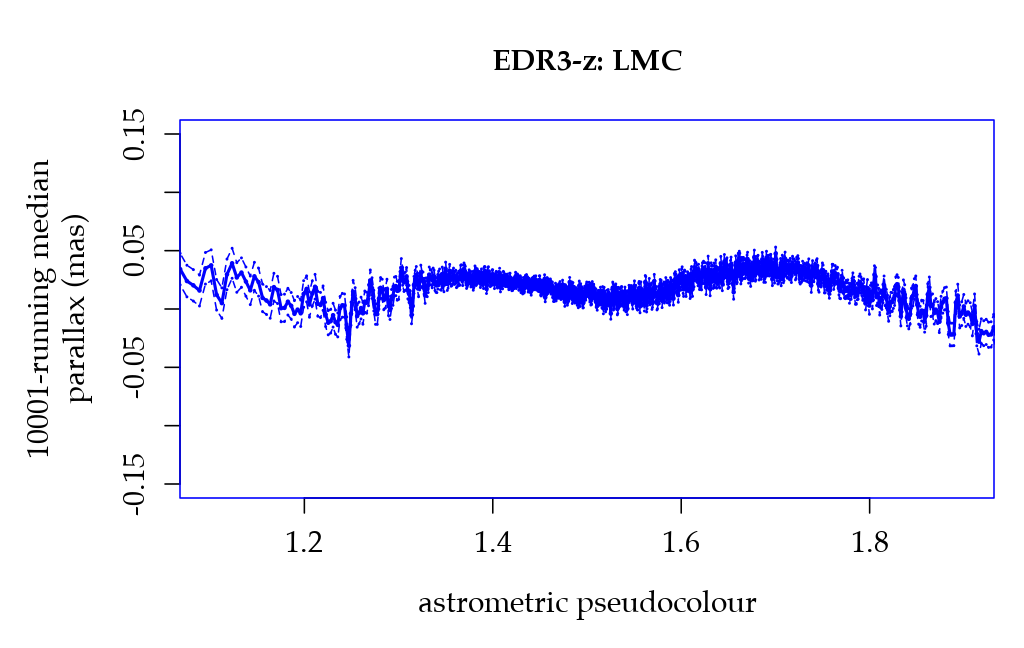}
\end{center}
\caption{{\em Top:} Correlation between parallax and pseudo-colour. {\em Middle:} Running median with uncertainty for LMC parallaxes where the negative correlation translates into systematics for the bluest and reddest sources. {\em Bottom:} LMC parallaxes after zero point correction.}\label{fig:pseudoplx}
\end{figure}

The 6p solutions derive both the astrometric parameters and the pseudo-colour.
Although the correlation between the parallax and pseudo-colour is, in general, not
large (top panel of \figref{fig:pseudoplx}), this nevertheless implies that (random or systematic) 
errors in the pseudo-colour translate into systematics on the parallax, which 
should not, in general, be interpreted as a colour effect.
On average, the correlation is, for example, positive near the Galactic centre and 
negative in the LMC, giving a very large peak-to-peak amplitude of 0.2 mas
(\figref{fig:pseudoplx}, middle panel);
although it is twice as small as for DR2 \citep[][Fig.~18]{2018A&A...616A..17A}.
After the \cite{EDR3-DPACP-132} correction of the zero point, the variations
were considerably reduced (\figref{fig:pseudoplx}, bottom panel).

\subsection{Proper motions from star cluster analysis}

\begin{figure*}
 \begin{center}
\includegraphics[width=14cm]{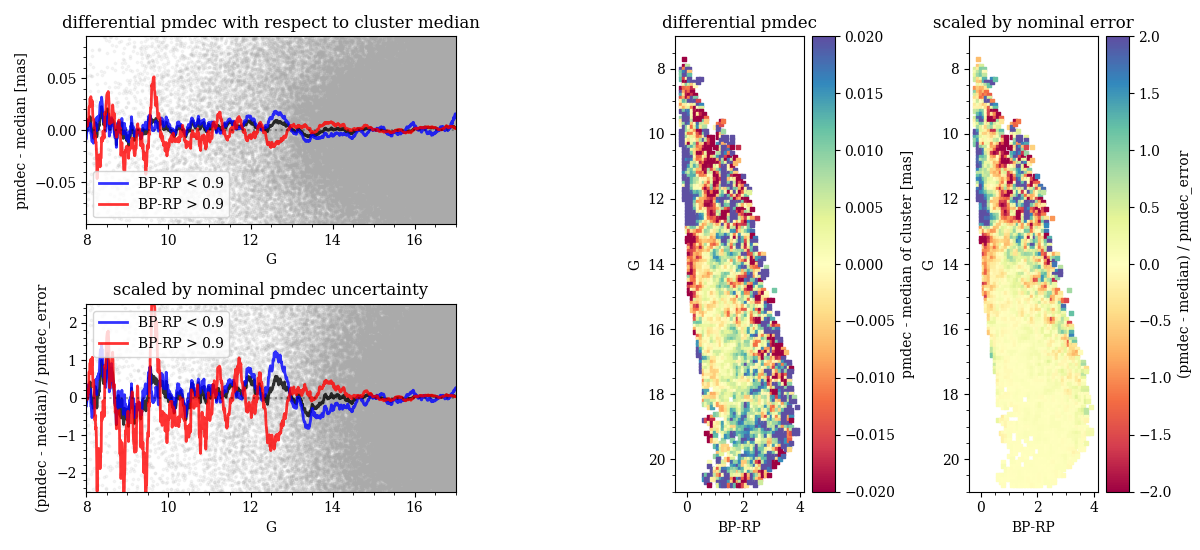}
\end{center}
\caption{{\em Left}: Difference between proper motions in declination and median value (top) and  scaled (bottom) to the nominal uncertainties as a function of {\gmag} for the whole cluster sample. The solid lines indicate the LOWESS of the sample; blue corresponds to stars bluer than \bpminrp\ $=0.9$\,mag and red is for objects redder than this limit.  {\em Right}: Colour-magnitude diagram of the whole cluster sample  where the colour indicates the difference between the star \pmdec\ and the cluster median. The rightmost panel is analogous to the left-hand panel, where the difference is scaled to the nominal uncertainties. }\label{fig:Wp947pmdec}
\end{figure*}

\begin{figure*}
 \begin{center}
\includegraphics[width=14cm]{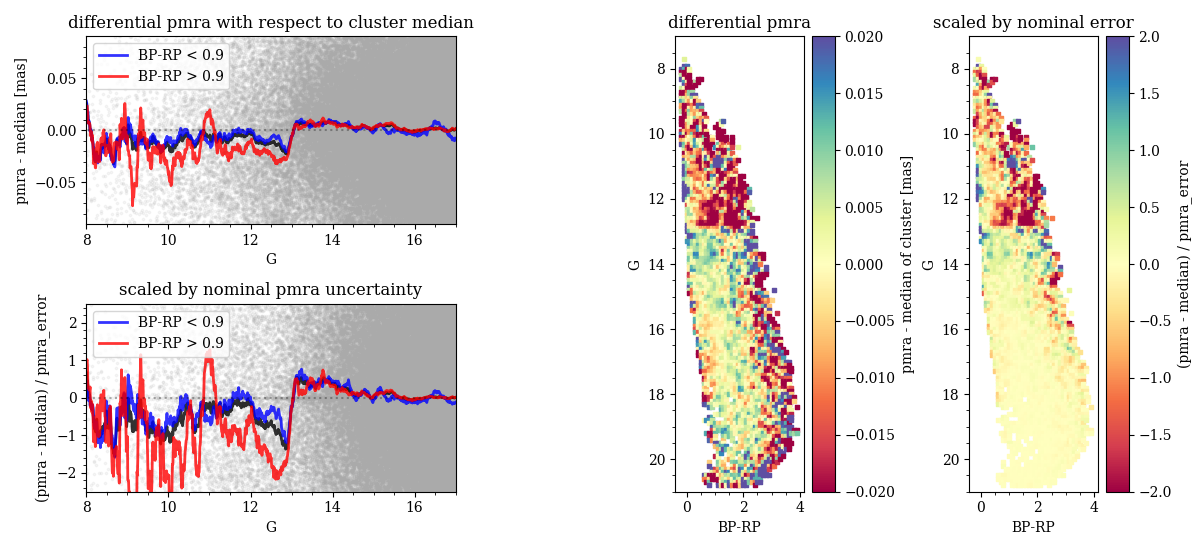}
\end{center}
\caption{Same as Fig.\ref{fig:Wp947pmdec}, but for \pmra.   }\label{fig:wp947pmra}
\end{figure*}

For each star,  we derived the differences of  the proper motions \pmra, \pmdec\ to the cluster median. This provides information about the zero point variations and about the uncertainties.
Figures \ref{fig:wp947pmra} and \ref{fig:Wp947pmdec} present  the results,  which were scaled and not scaled to the nominal errors as function of  \gmag, \bpminrp, and in the colour-magnitude diagram. As already found for the parallaxes, a complex dependence of the difference with magnitude and colour is clearly visible. This reflects a variation in the zero point. In particular, a shift is present at $G \sim 13$\,mag in \pmra. When this difference is normalised to the nominal uncertainties, we find that these patterns are still present for brighter stars. This  means that  nominal uncertainties are underestimated in this magnitude and colour range.  This effect is more evident in \pmra.  Residuals are also present for very red stars  (\bpminrp\ $\sim 1.5-2$\,mag), where the results are less significant, since  the statistics is relatively poor. This region can be contaminated by field stars.

We calculated  the proper motion differences (median, \pmra$_{\rm EDR3}-$\pmra$_{\rm DR2})$  for the whole cluster sample  in \gdrtwo and \egdr{3}. The zero point median difference is $-0.006$\masyr\  with a third quartile of 0.06\masyr\  in right ascension and in declination. 

\subsection{Proper motion precision in crowded areas}

\begin{figure*}
 \begin{center}
\includegraphics[width=16cm]{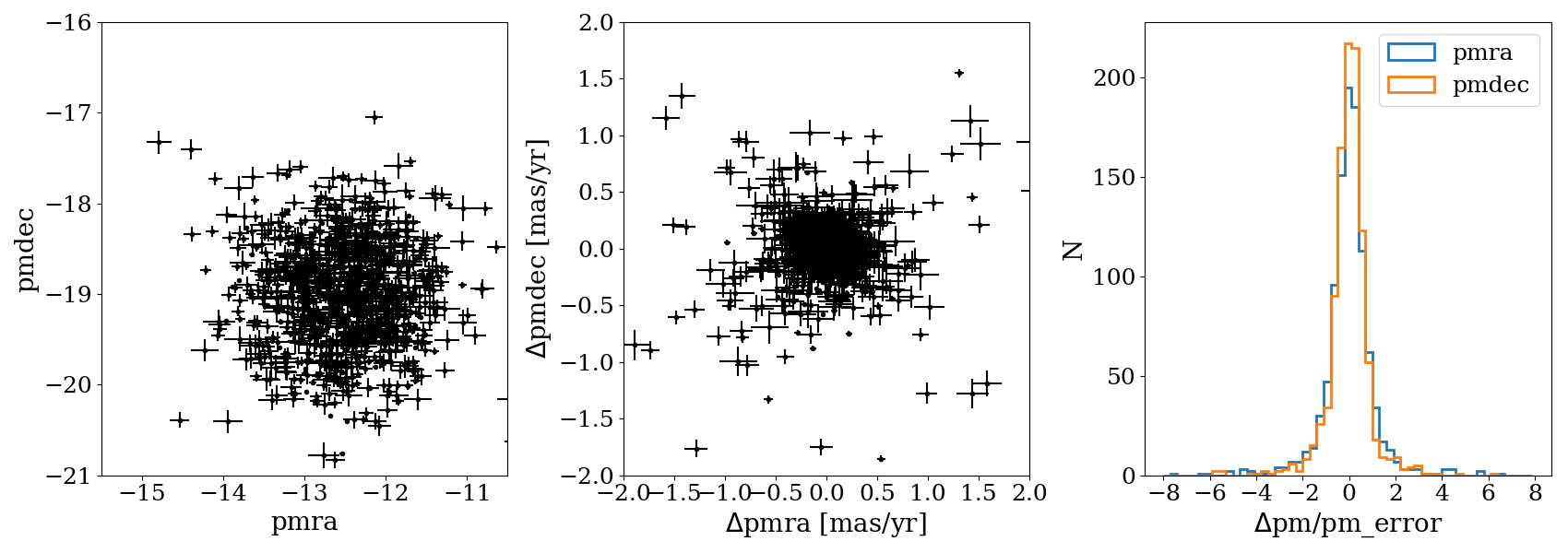}
\end{center}
\caption{ {\em Left:}  Gaia proper motion  distribution in M4. Only stars with a proper motion error $< 0.2$ \masyr and flux excess parameter C $<2 $ are plotted.
{\em Centre:}   Differences between \propm in Gaia and in HST data. {\em Right:}  Scaled dispersion for the stars having Gaia \propm errors $< 0.2$ \masyr and flux excess parameter C $< 2$.}\label{fig:WP947m42}
\end{figure*}

We compared the  \propm in the centre of M4 with external HST data \citep{2014MNRAS.442.2381N}, where high-quality relative \propm  are available.
The precision of HST proper motions is  of the order of 0.33 \masyr. The observed field  is affected by crowding. We used the flux excess parameter C=(\dt{phot\_bp\_mean\_flux}+\dt{phot\_rp\_mean\_flux})/ \dt{phot\_g\_mean\_flux},\footnote{
In the \egdr{3}\ archive, this is \dt{phot\_bp\_rp\_excess\_factor}. } 
as a tracer of  crowding contamination. 
In Fig.~\ref{fig:WP947m42}, we present the \gaia\ \propm distribution,  the differences between \gaia\ and HST proper motions, and the scaled dispersion for the stars having \gaia\ nominal proper motion uncertainties $< 0.2$ \masyr  and a  moderate flux excess, C $<2$. The scaled dispersion is very close to one for both \pmra\ and \pmdec, indicating that the nominal uncertainties were estimated correctly. Stars in this sample are not affected by crowding. Stars with a high level of contamination (C $> 5$) are, on average, at  1.6-1.8 error bars from the expected value, that is to say \gaia\ nominal uncertainties on \propm are  underestimated for faint and/or contaminated stars. In any case, this represents a substantial improvement over \gdr{2}, where  the nominal uncertainties were underestimated by a factor from two to three.

%
\subsection{Goodness of fit for very bright stars}\label{sec:vbs}
%
\begin{figure}[h!]
\begin{center}
\includegraphics[width=0.49\columnwidth]{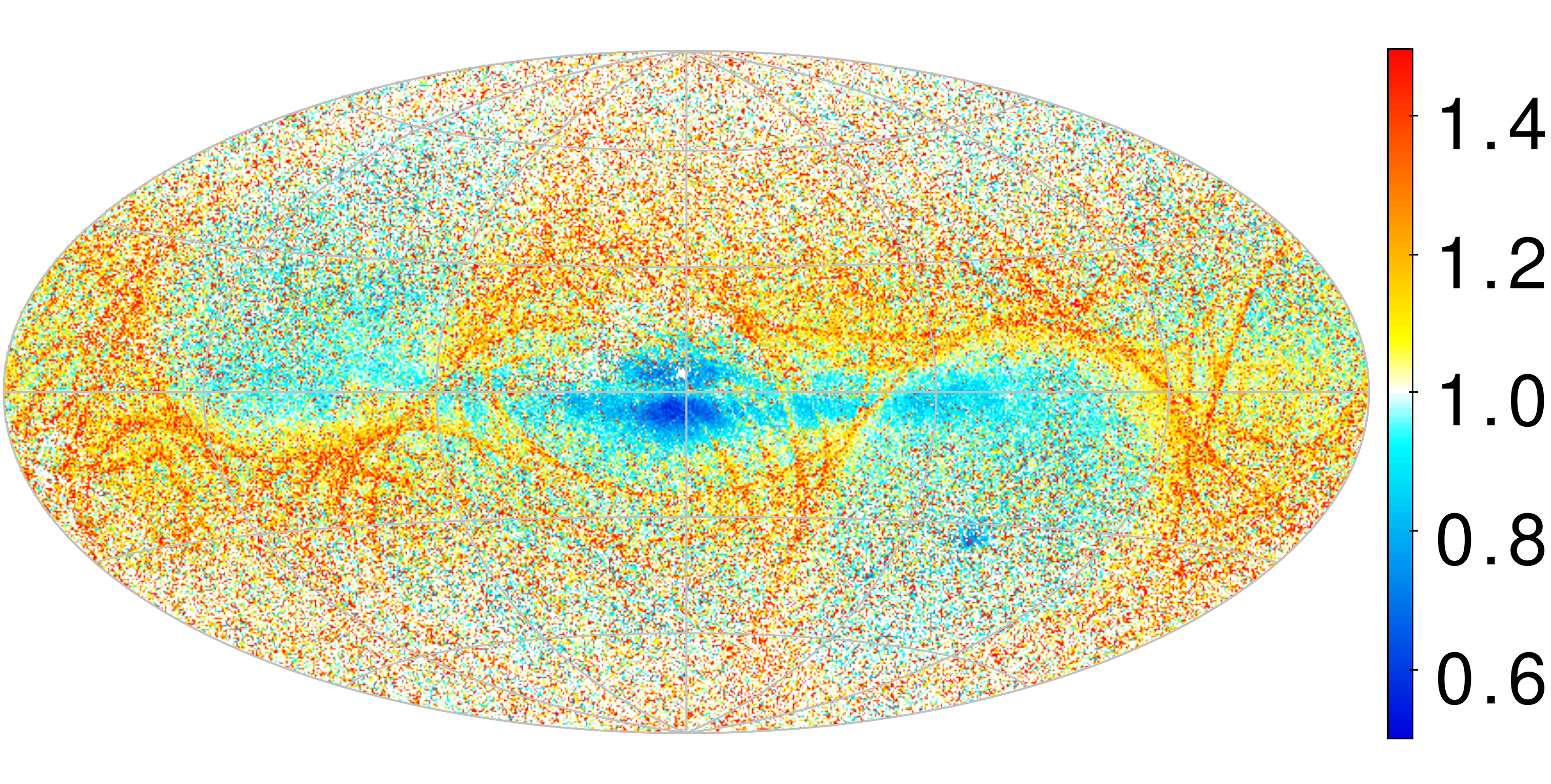}
\includegraphics[width=0.49\columnwidth]{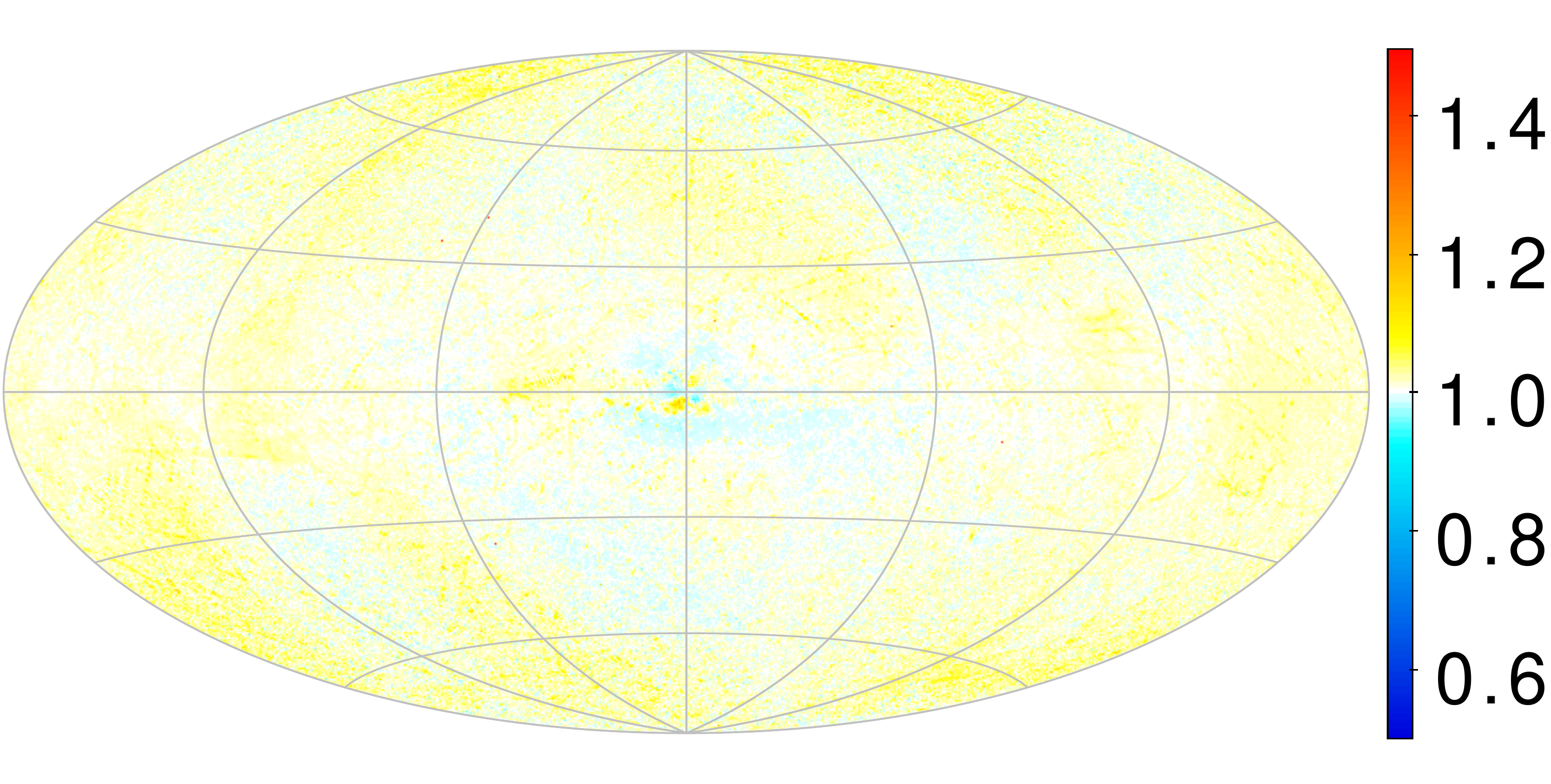}
\end{center}
\caption{Re-normalised unit-weight error for the astrometric solution, \dt{ruwe}, for sources brighter than $G=11$ ({\em left}) and brighter than $G=17$ ({\em right}).}\label{fig:ruwebt11}
\end{figure}

Although it is known that calibration errors or stellar duplicity may
frequently make the quality of the astrometric solutions decrease, it may appear surprising to complain about a solution quality that is too good. Figure~\ref{fig:ruwebt11} (right panel) shows
that the \dt{ruwe} of sources brighter than 17\,mag is near the expected value of one on average. However, it is smaller than one
near the Galactic centre, and twice as small for stars
brighter than $G=11$\,mag (\figref{fig:ruwebt11}, left panel). The interpretation is the following:
In this region, \new{because} most of the sources suffer from crowding,  most
sources should have some excess noise \new{and not just}  a small fraction of them, as in other
regions. Consequently, the attitude excess noise may have absorbed this source
excess noise, leading to source solutions \new{that appear} much better than \new{they truly are} in reality.
Consequently, the caveat is that the \dt{ruwe} of bright sources in large crowded 
areas, and thus their \dt{astrometric\_gof\_al}\footnote{\dt{astrometric\_gof\_al} is the goodness of fit of the astrometric solution.} too, may be much smaller than they should be.


\section{Photometric quality of \egdr{3}}\label{sec:phot}
%

The photometry in {\egdr{3}}  consists of three broad bands: a {\gmag}
magnitude for (almost) all sources (1.8 billion) and a {\gbp} and {\grp} for
the large majority (1.5 billion). The photometry and its main validation is
described in \cite{EDR3-DPACP-117}, and here we present some additional
validation. These tests include internal comparisons, comparisons with external
catalogues, and some simulations.

Among the issues described by \cite{EDR3-DPACP-117}, there are two issues for
the user to be aware of. The first concerns sources where a reliable colour was
not known at the beginning of the \egdr{3}\ processing and which therefore
have not benefited from an optimal processing.
These are all the 6p sources and many 2p sources, but none of the 5p sources. 
\cite{EDR3-DPACP-117} propose a correction to the
catalogue \gmag\ magnitude of the 6p sources that is typically of the order of a
hundredth of a magnitude. For 2p sources, it is unfortunately not
possible to know with certainty if a correction is needed.
The second issue concerns the faintest
\gbp\ sources, where a source with practically no signal, for example\ the
\gbp\ flux for a faint, red source, will still have a significant flux
assigned. This issue, which also applies to a small population of \grp\ fluxes, is discussed below in Sect.~\ref{phot_thres}.

%
\subsection{Spurious photometric solutions}\label{phot_spur}
%

\begin{figure}
        \begin{center}
        \includegraphics[width=0.4\textwidth]{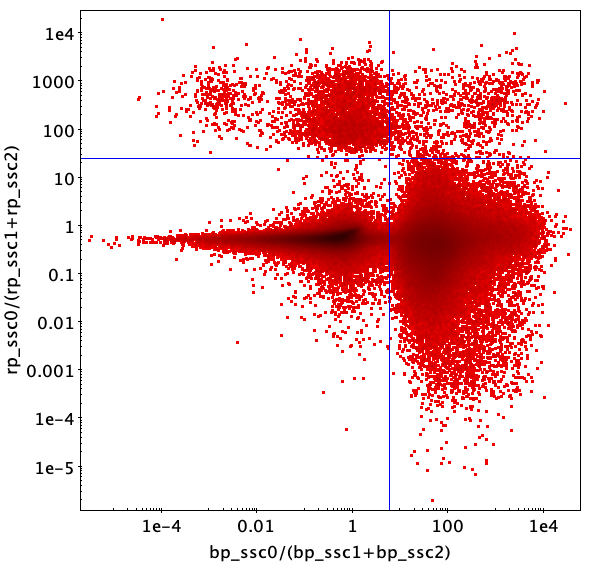}
        \caption{Quotients of blue and red spectral shape coefficients for the set of sources
         fainter than \gmag\ $\sim 21.7$~mag. The photometry for sources with quotients larger than 
         the thresholds (blue lines) were filtered for the publication 
          (see text).}
        \label{fig:wp942_phot_ssc_ratii}
        \end{center}
\end{figure}

There were many unrealistically faint sources, down to $G\sim 25.6$ mag, in a
pre-release version of the catalogue.
Such faint magnitudes can only arise from processing problems. After 
investigations in the photometric pipeline, \dt{PhotPipe}, the root cause was traced to the way poor input 
spectral shape coefficients (SSCs) disturb the application of the photometric calibration 
\citep[see][sect.\,4.4, eqs.\,1--3]{EDR3-DPACP-117}.  From the BP spectrum, four coefficients were derived:
\dt{bp\_ssc0},. . . , \dt{bp\_ssc3}, and four similar from RP.  Quotients, such as
\dt{bp\_ssc0/(bp\_ssc1+bp\_ssc2),} are used in the calibration model without
taking into account that the denominator in certain cases may become extremely
small and the quotient therefore extremely large. This is illustrated in
Fig.~\ref{fig:wp942_phot_ssc_ratii}, where we can see that in the majority of
cases (in fact 82\% for this sample of very faint sources) the principal blue
and red quotients have reasonable values (lower left), and that these values
are well separated from cases of very large values. This has allowed us to
establish well defined thresholds, and the photometry of the sources with
quotients larger than those thresholds is not published in the main catalogue
because it is considered unreliable. The 5.4 million sources 
sources affected were reprocessed by \dt{PhotPipe} using 
default SSCs,  and the $G$ flux derived this way 
is published in a separate file accessible in the 
\egdr{3}\
\href{https://www.cosmos.esa.int/web/gaia/edr3-known-issues}{known issues}\footnote{\url{https://www.cosmos.esa.int/web/gaia/edr3-known-issues}} page.

Colours in {\gdr{2}} were shown to be too blue for faint red sources and this 
also happens in {\egdr{3}}.
The \dt{phot\_bp\_rp\_excess\_factor} parameter gives a measure of the
coherence among $G$, $G_{\rm BP}$, and $G_{\rm RP}$ and can be used to identify
the problematic cases. For those cases, the use of the colour $G-G_{\rm RP}$
instead of $G_{\rm BP}-G_{\rm RP}$ may be more useful.
\cite{EDR3-DPACP-117} identify the root cause of the problem to be a flux
threshold \TBC{applied to the}\TODO{including } individual transits \TBC{when computing}\TODO{in } the mean photometry. We explore
this question below.

%
%
\subsection{Transit-level flux threshold}\label{phot_thres}
%
\begin{figure}[ht]
\begin{center}
\includegraphics[width=7.5cm]{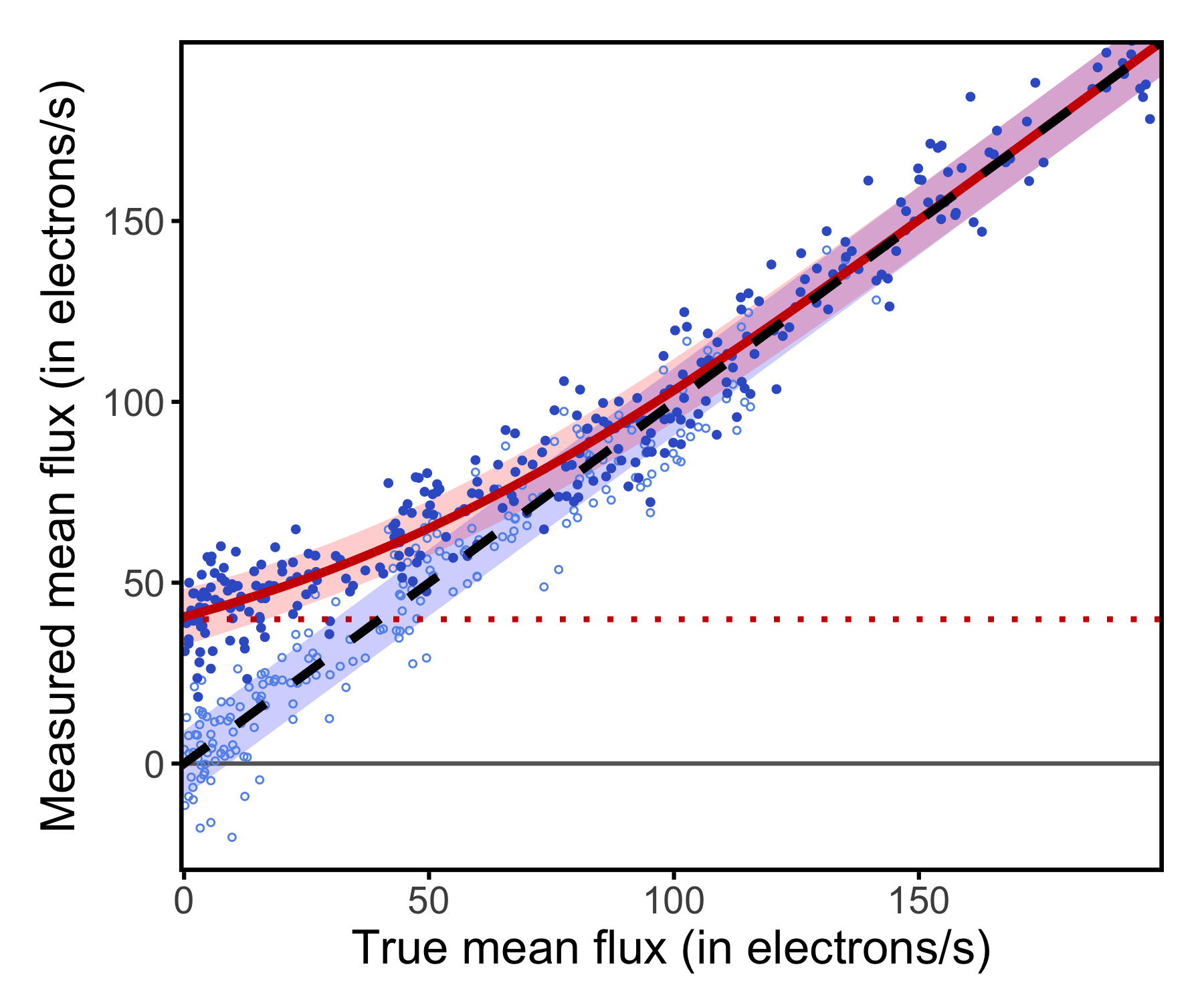} 
\end{center}
\caption{
Simulated mean fluxes as a function of the true mean flux in the presence of the 1\es\ threshold (dark filled symbols) and without (light open symbols). The mean and the 1$\sigma$ confidence intervals are shown as lines and shaded regions. The dotted line indicates the lower bound of the mean in the presence of the threshold.
}
\label{fig:lowFlux1}
\end{figure}

\begin{figure}[ht]
\begin{center}
\includegraphics[width=7.5cm]{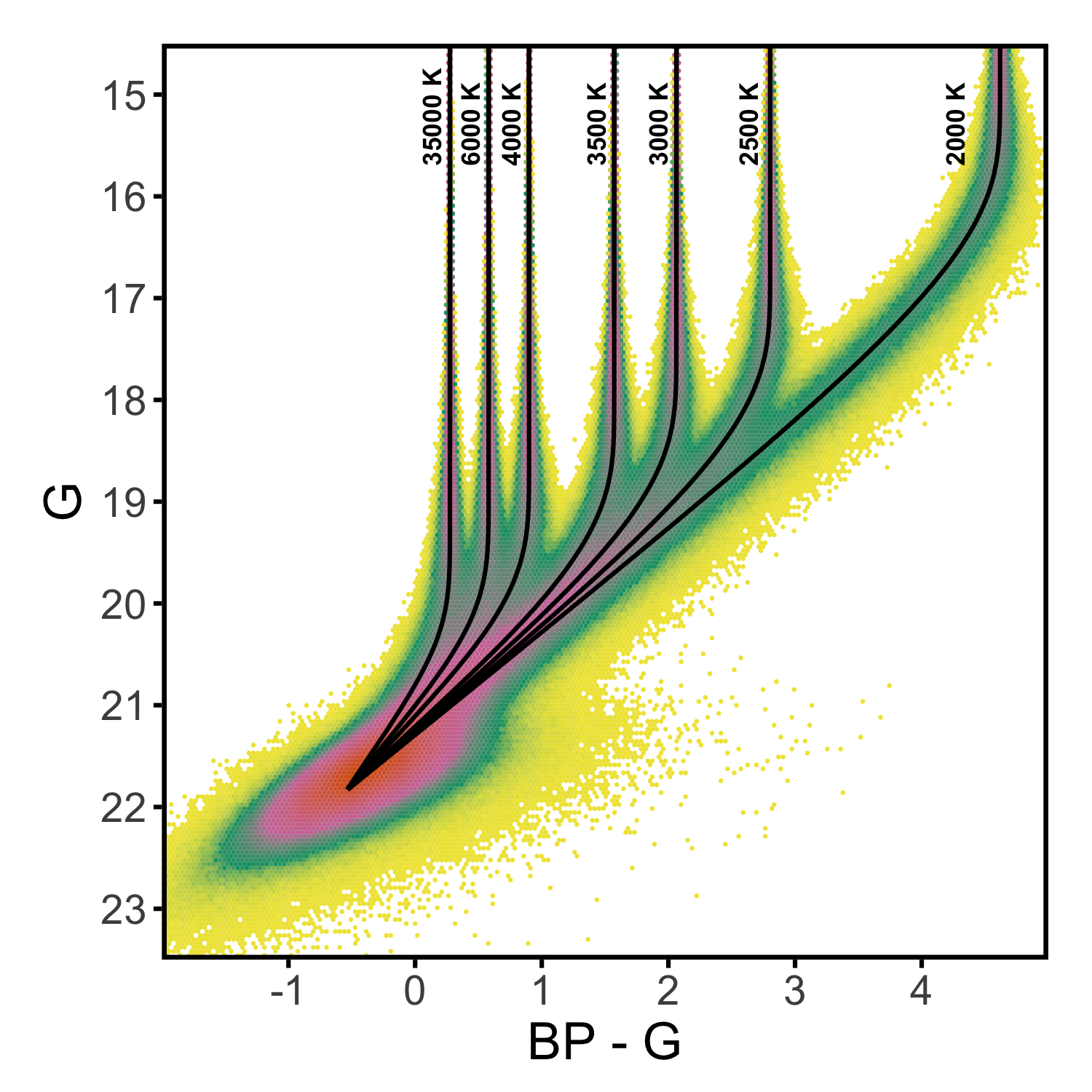} 
\includegraphics[width=7.5cm]{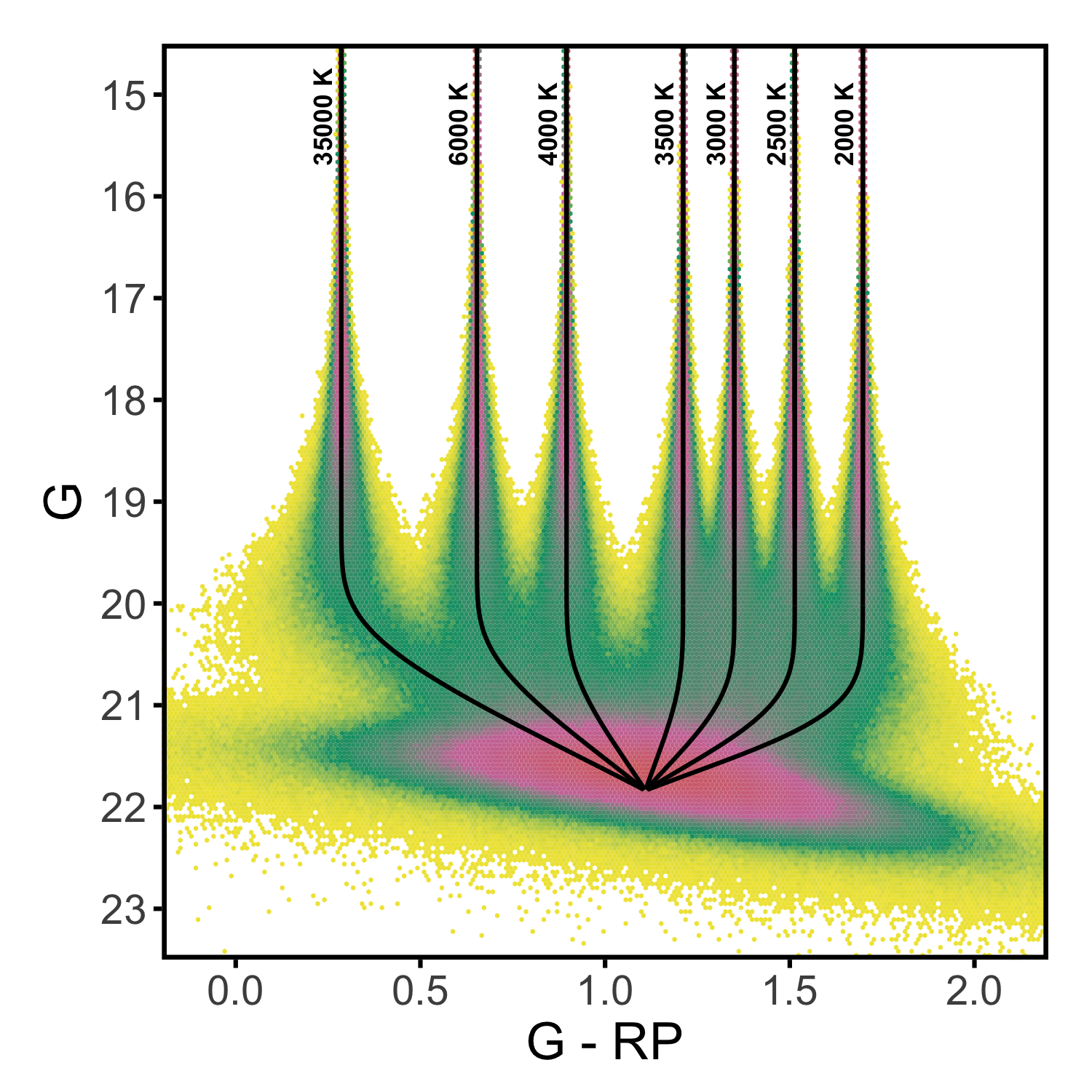} 
\end{center}
\caption{
Distribution of 5 million simulated mean magnitudes for BaSeL spectra with effective temperatures from 2000\,K to 35000\,K, in the presence of a 1\es\ threshold, in the {\gbp} $-$ {\gmag} versus {\gmag} (upper panel) and the {\gmag} $-$ {\grp} versus {\gmag} (lower panel) colour-magnitude diagrams. The solid lines indicate the mean.
}
\label{fig:lowFlux2}
\end{figure}

The \gbp\ and \grp\ mean fluxes were derived from the epoch fluxes at the
several transits of a source across the focal plane during the mission.
As discussed by \cite{EDR3-DPACP-117} (sect.~8.1), in the computation of the
mean fluxes, a threshold of 1\es\ was introduced for the
individual transits. Transits with fluxes below that threshold were
excluded from the computation of the mean fluxes. This introduces a bias in the
mean fluxes for the faintest sources, as the distribution of noise-affected epoch
fluxes becomes truncated when the flux of a source becomes so low that the
distribution of epoch fluxes reaches the threshold. To illustrate the effect,
we simulated mean fluxes by applying and not applying the threshold, under the
simplifying assumption of the same normally distributed background noise of 
50\es\  and 30 transits. As background noise, we subsume here all
noise contributions other than the source's photon noise, that is\ to say contributions
from detector and electronics effects, the sky background, and stray light. In
reality, this background noise can vary significantly between different
transits. Simulations of measured mean fluxes for sources by applying and not 
applying the threshold as a function of true source flux is shown in
Fig.~\ref{fig:lowFlux1}, together with the means, and the 1$\sigma$ confidence
intervals. As the source fluxes become low, the measured mean fluxes
systematically deviate from the true fluxes, and the mean of the distributions
of mean fluxes meet a lower bound. For a threshold much smaller than the
background noise level, this lower bound on the mean is approximately 0.8 times
the standard deviation of the background noise.\par Sources are detected based
on an estimate of their {\gmag} band magnitude, with a detection limit well
above the 1\es\ threshold. Also since astronomical sources do not
become arbitrarily blue, the flux of detected sources measured in the {\grp}
passband cannot become arbitrarily smaller than the flux in the {\gmag} band.
But as astronomical sources can become arbitrarily red, the flux in the {\gbp}
passband can become smaller than the {\gmag} band flux without bound.
Sufficiently red sources can therefore have {\gbp} fluxes that even fall below
the detection limit while having significant fluxes in the {\gmag} and {\grp}
passbands. As a consequence, the bias on the mean flux for faint sources
resulting from the threshold mostly affects the {\gbp} fluxes, and it only has a
much smaller effect for the {\grp} fluxes.\par As a consequence of the {\gbp}
fluxes being far more biased towards larger values than the {\gmag} and {\grp}
fluxes for faint sources, a 'turn' in the colour-magnitude diagram involving
{\gbp} magnitudes results for faint sources, as shown by \cite{EDR3-DPACP-117}.
To illustrate the effect further, we simulated BaSeL spectra
\citep{1997A&AS..125..229L} with different effective temperatures (for solar
metallicity and a surface gravity, \logg, of four) in the {\gmag} versus {\gbp}
$-$ {\gmag} and {\gmag} $-$ {\grp} colour-magnitude diagrams. The distribution
of the simulated mean magnitudes, together with the mean of the distributions
as a function of colour, is shown in Fig.~\ref{fig:lowFlux2}. For the {\gmag}
versus {\gbp} $-$ {\gmag} colour-magnitude diagram, the strong turn in the
distributions is visible, which starts at brighter {\gmag} magnitudes as the
source becomes cooler, and thus redder. If the fluxes approach the noise level,
the distributions become independent of the spectral energy distributions of
the sources, and the means of the distributions for all sources converge at the
same location in the colour-magnitude diagram. For thresholds much smaller than the noise level, the position of this convergence
point only depends on the
background noise and, when converting fluxes into magnitudes, on the zero
points of the passbands.\par For the {\gmag} versus {\gmag} $-$ {\grp}
colour-magnitude diagram, the effect of convergence on the same mean position
in the diagram for faint sources is also present.\ However, the turning of the
distributions occurs at fainter magnitudes, and the effect for blue sources is
much smaller than for the red sources in the {\gmag} versus {\gbp} $-$ {\gmag}
case.\par To minimise the effects introduced by the threshold on the
interpretation of photometric data, it is thus advisable to avoid the use of
{\gbp} magnitudes fainter than about 20.5, corresponding to {\gbp} fluxes below
approximately 86\es, if possible.  For {\grp} magnitudes, significant
bias effects occur at values fainter than about 20.0, corresponding to {\grp}
fluxes below approximately 79\es. Since the bias effects are strongest
in {\gbp}, the {\gmag} $-$ {\grp} colour is more reliable than colours
involving {\gbp}.

\subsection{Photometry of the 6p solutions}\label{phot_6p}

\begin{figure}[ht]
 \begin{center}
\includegraphics[width=\hsize]{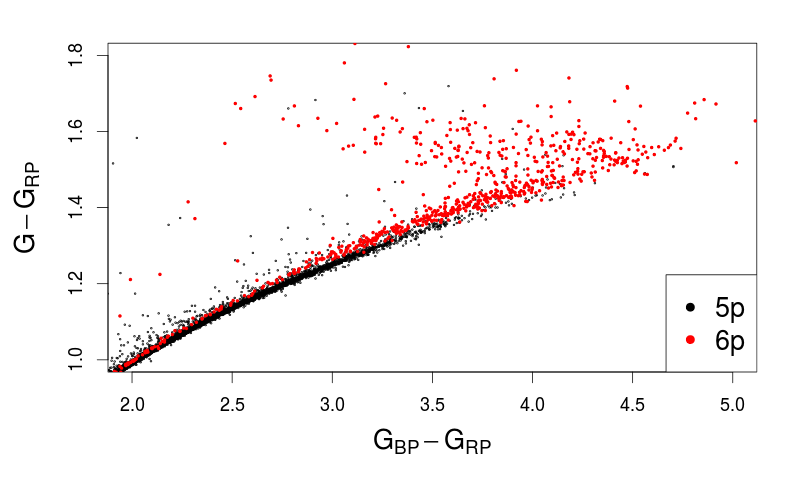} 
\includegraphics[width=\hsize]{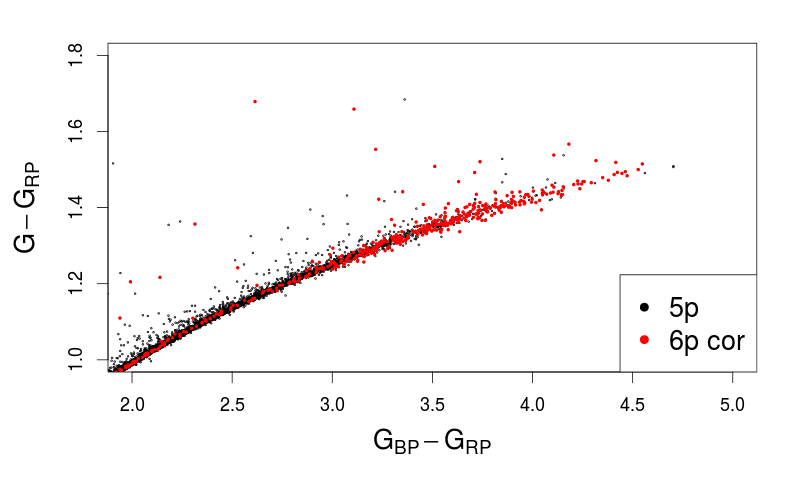} 
\end{center}
\caption{Colour-colour relation of the APOGEE DR16 red clump sample. 
\new{Stars with 5p astrometric solutions are shown with black dots, and the ones with 6p solutions are shown with red dots.} {\em Top:} The full sample. {\em Bottom:}
 After removing stars with {\gbp}~$>20.5$\,mag and correcting the {\gmag} photometry of the 6p solutions using the recipe of \cite{EDR3-DPACP-117}. }
\label{fig:apogee_rc}
\end{figure}

We tested the correction to the {\gmag} magnitude proposed by \cite{EDR3-DPACP-117} for stars with 6p solutions, that is with a default colour in the image parameter determination (IPD), on the red clump sample of APOGEE DR16 \citep{2014ApJ...790..127B}. The variation in the colour-colour relation for those red clump stars is due to the variation in the extinction and it is curved due to the non-linearity of the extinction coefficient with extinction.
In the top panel of Fig.~\ref{fig:apogee_rc}, there are two effects that can be seen. The first one is that of the transit-level flux threshold, discussed in the previous section, seen as a plume of stars becoming bluer than the global relation.  The second effect is the difference between the colour-colour curves of the 5p and 6p solutions (black and red, respectively).
If we were to add the {\gmag} magnitude correction as proposed by \citet[][sect.~8.3]{EDR3-DPACP-117} and remove those sources with {\gbp}~$>20.5$\,mag as proposed in the previous section, we would recover the expected colour-colour relation.

%
%
\subsection{Photometric accuracy}\label{phot_acc}
%

The systematic errors of the photometry can be studied in both internal and
external comparisons as well as from galaxy models of the Milky Way.
In addition, we look at the changes with respect to \gdr{2}.

\subsubsection{Internal comparisons}\label{phot_acc_int}

\begin{figure}[ht]
 \begin{center}
\includegraphics[width=0.8\hsize]{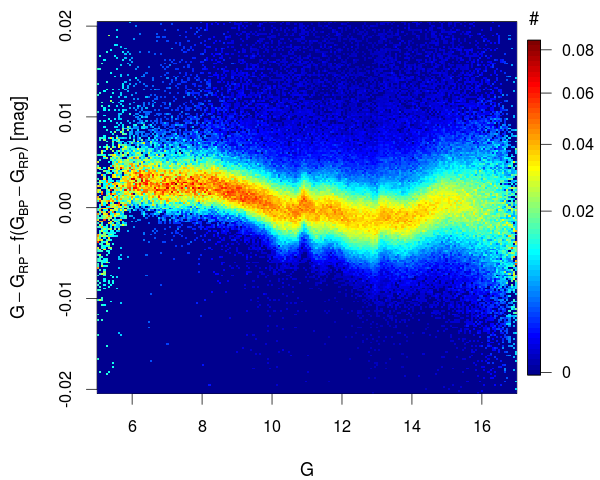} 
\includegraphics[width=0.8\hsize]{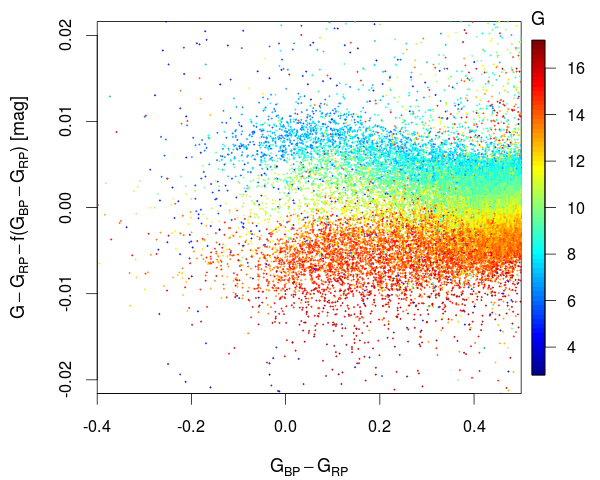} 
\end{center}
\caption{Residuals from a global $G-G_{\rm RP}=f(G_{\rm BP}-G_{\rm RP})$ \new{relation}
for a sample of \new{luminous, low extinction stars} ($A_0<0.05$\,mag, $M_G<4$\,mag).
{\em Top}: As a function of \gmag, colour-coded by the number of stars normalised by the total number of stars per magnitude bin. 
{\em Bottom}: As a function of  \bpminrp\ \new{for a sample of bluer stars}, colour-coded by the magnitude.
}
\label{fig:phot_internal_residuals}
\end{figure}

We studied the internal trends as a function of magnitude.
Figure~\ref{fig:phot_internal_residuals} (top panel) shows the residuals from the median $G-G_{\rm RP}=f(G_{\rm BP}-G_{\rm RP})$ relation as a function of {\gmag} on a sample of stars in the upper part of the H-R diagram with low extinction, that is
$A_0<0.05$\,mag according to the 3D extinction map of \cite{2019A&A...625A.135L} and $M_G<4$\,mag,\footnote{$\dt{parallax}+2 \times \dt{parallax\_error} < \exp((4-\dt{phot\_g\_mean\_mag}+10) \times 0.46)$} taking into account the parallax error at 2 sigma. We further selected only stars with relative photometric errors better than 2\% in {\gmag} and \new{5}\% in {\gbp} and {\grp}. \new{These strict criteria lead to a well behaved colour-colour relation but to less than 800\,000 stars, which are mostly close by and therefore relatively bright.}
There is a small trend with magnitude which is much smaller than in \gdr{2} \citep{2018A&A...616A..17A}.  Discontinuities at $G\sim 10.8$ and 13~mag of only about 0.5 and 1 mmag are also much smaller than in \gdr{2}.
Figure~\ref{fig:phot_internal_residuals} (bottom panel) shows that for blue stars, the trend with magnitude is still stronger (see also sect.~8.4 of \cite{EDR3-DPACP-117}), but the amplitude is about three times smaller than in \gdr{2} \citep[][fig.~35]{2018A&A...616A..17A}.

\begin{figure}[ht]
 \begin{center}
\includegraphics[width=7.5cm]{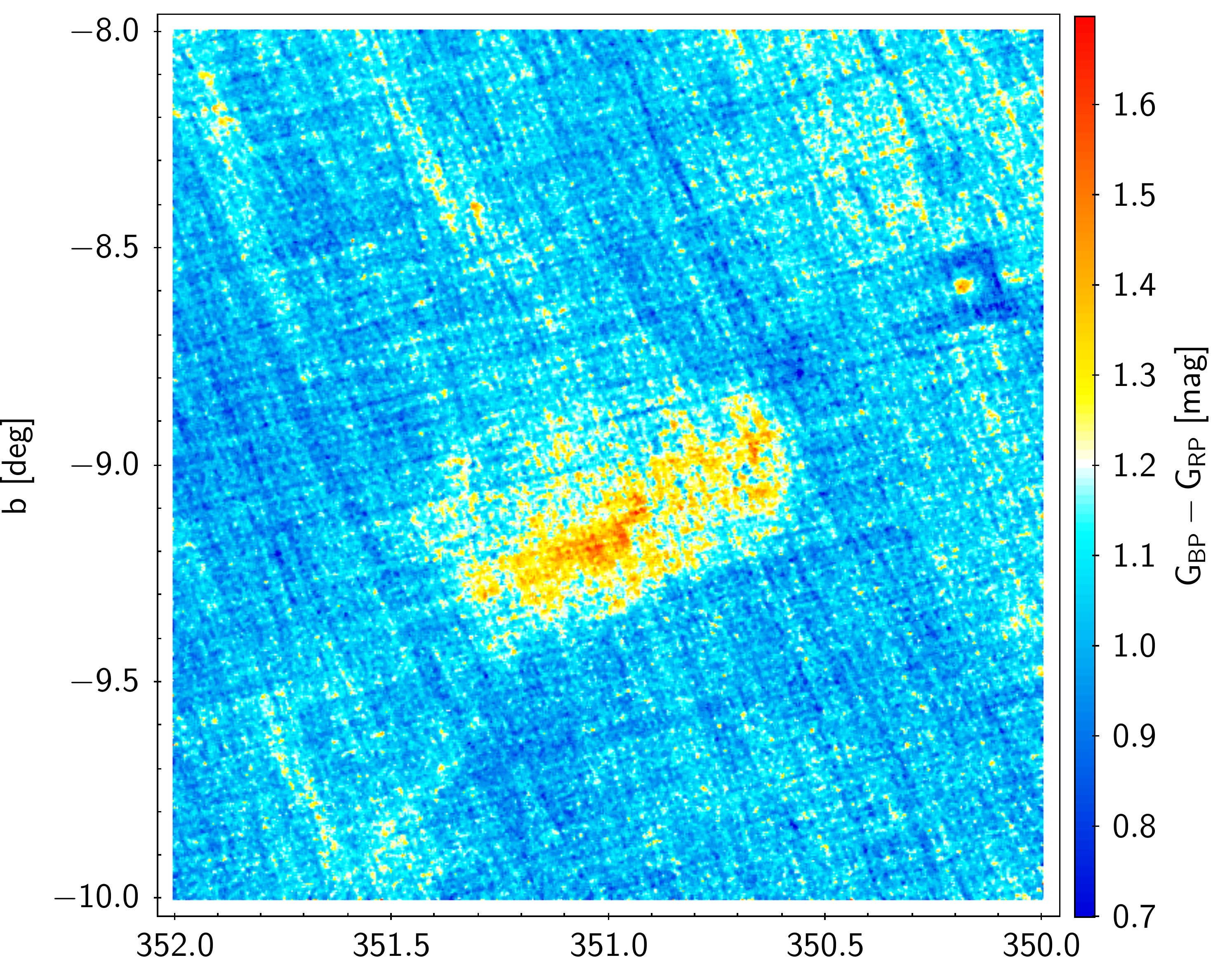}
\includegraphics[width=7.5cm]{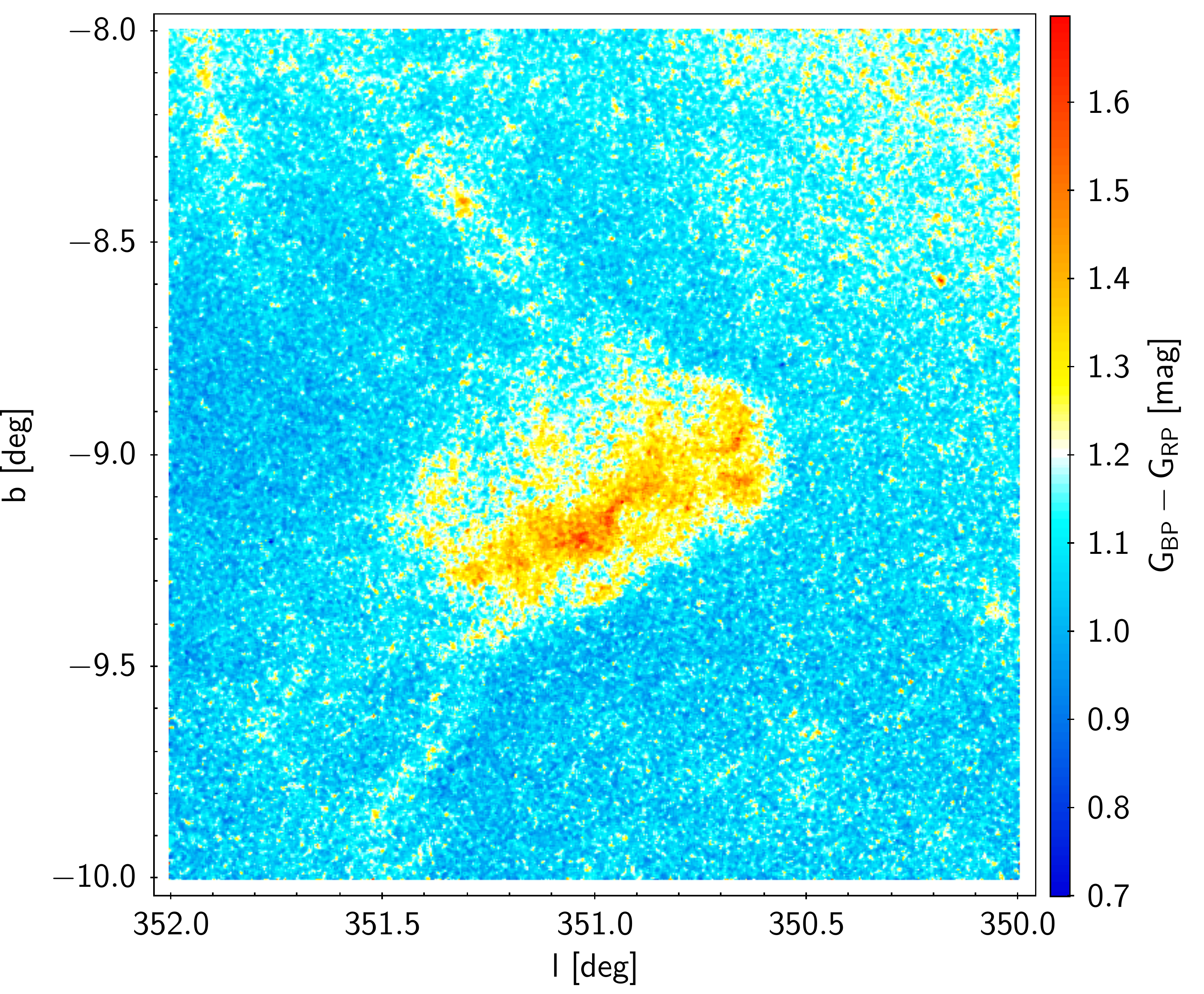}
\end{center}
\caption{Median colours \bpminrp in a dense field (Galactic coordinates)
showing artefacts from the scan pattern in \gdr{2} ({\em top}), which have almost
disappeared in {\egdr{3}}\ ({\em bottom}). 
}
\label{fig:espatial_pattern}
\end{figure}

Figure~\ref{fig:espatial_pattern} shows the median {\bpminrp} colours in a
dense field in the case of \gdr{2} and {\egdr{3}}. One can appreciate that the
artefacts from the scan patterns have decreased quite a bit.

\subsubsection{Comparisons with external catalogues}\label{phot_acc_ext}

\begin{figure}[h]
\centering
\includegraphics[width=0.9\hsize]{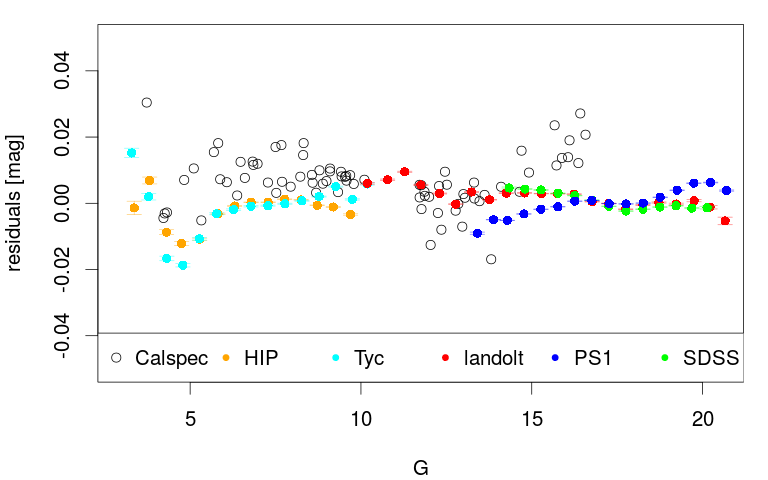}
\includegraphics[width=0.9\hsize]{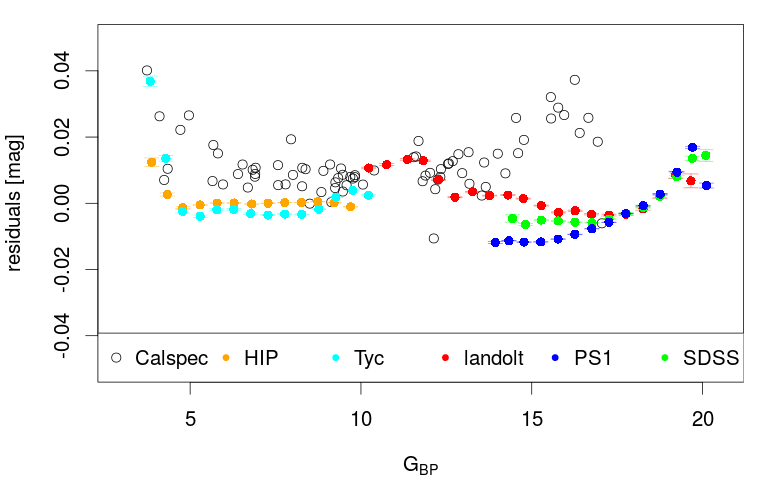}
\includegraphics[width=0.9\hsize]{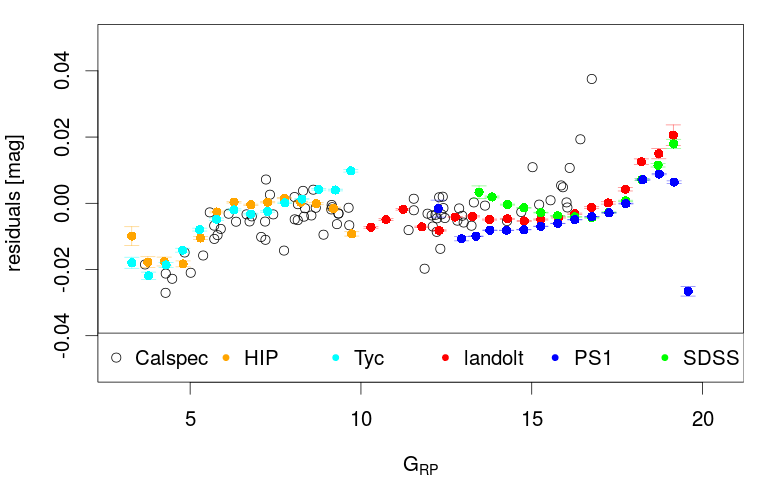}
\caption{From top to bottom, {\gmag}, {\gbp}, and {\grp} photometry (hereafter referred to as $X$) versus external photometry: 
 CALSPEC (black dots) corresponding to $X-X({\rm calspec})$; 
 {\hip} (orange) residuals of the global $X-Hp=f(V-I)$ spline;
 {\tyctwo} (cyan) residuals of the global $X-V_T=f(B_T-V_T)$ spline;
 Landolt (red) residuals of the global $X-V=f(V-I)$ spline;
 and PS1 (blue) and SDSS (green) residuals of the global $X-r=f(g-i)$ spline for SDSS and PS1.
The zero point of those different residuals is arbitrary with the exception of the CALSPEC results. 
\label{fig:wp944_photsummary}
}
\end{figure}

We compared \egdr{3} photometry to the {\hip}, {\tyctwo}, Landolt standards \citep{1983AJ.....88..439L,1992AJ....104..340L,2007AJ....133.2502L,2009AJ....137.4186L,2013AJ....146..131L,2013AJ....146...88C,2016AJ....152...91C}, the SDSS tertiary standard stars of \cite{Betoule13}, and Pan-STARRS1 \citep[PS1,][]{2016arXiv161205560C} photometry. For {\hip}, {\tyctwo}, we selected low extinction stars ($A_0<0.05$\,mag) using the 3D extinction map of \cite{2019A&A...625A.135L} and taking into account the parallax uncertainty. For SDSS and PS1, we selected stars with a high galactic latitude.  For the Landolt catalogue, we selected both on $\vert b \vert>30^{\circ}$ and $A_0<0.05$\,mag for low latitude stars.  
We selected only stars with \dt{flux\_over\_error }$>20$, corresponding to photometric errors of $<0.05$~mag for the external catalogues, to ensure we were working with roughly Gaussian errors in magnitude space. 
An empirical robust spline regression was derived which models the global colour-colour relation. The residuals from those models are plotted as a function of magnitude in Fig.~\ref{fig:wp944_photsummary}. 
We also added in Fig.~\ref{fig:wp944_photsummary} the comparison with the magnitude computed on the CALSPEC \citep[2020 April Update]{2014PASP..126..711B} spectra combined with the \egdr{3} instrument response. The global zero point offset observed is smaller than 1\%, which is in agreement with the CALSPEC expected accuracy \citep{2014PASP..126..711B} and our observation of the variations between different CALSPEC releases. 

Figure~\ref{fig:wp944_photsummary} shows that the strong saturation effect present in \gdr{2} as well as the variation of the $G$ zero point with magnitude have been removed in \egdr{3}. Variations are now smaller than 0.04~mag. The trend as a function of magnitude for bright stars are consistent between the {\hip}, {\tyctwo,} and CALSPEC results. 
\new{On the faint end, a global increase in the residuals is observed in $G_{\rm RP}$ consistently between Landolt, Panstarss, and SDSS, while for $G_{\rm BP}$ those three surveys do not give a consistent amplitude of the variation.}
We recall that applying our procedure to colour-colour relations within the external catalogues' photometric bands leads to global variations of the order of 2 mmag/mag for PS1 and larger for SDSS \citep{2018A&A...616A..17A}.

\subsubsection{Global comparison of \gmag\ to \gdr{2}\label{sec_phot_dr2}}

\begin{figure*}[h]
\sidecaption
\includegraphics[width=12cm]{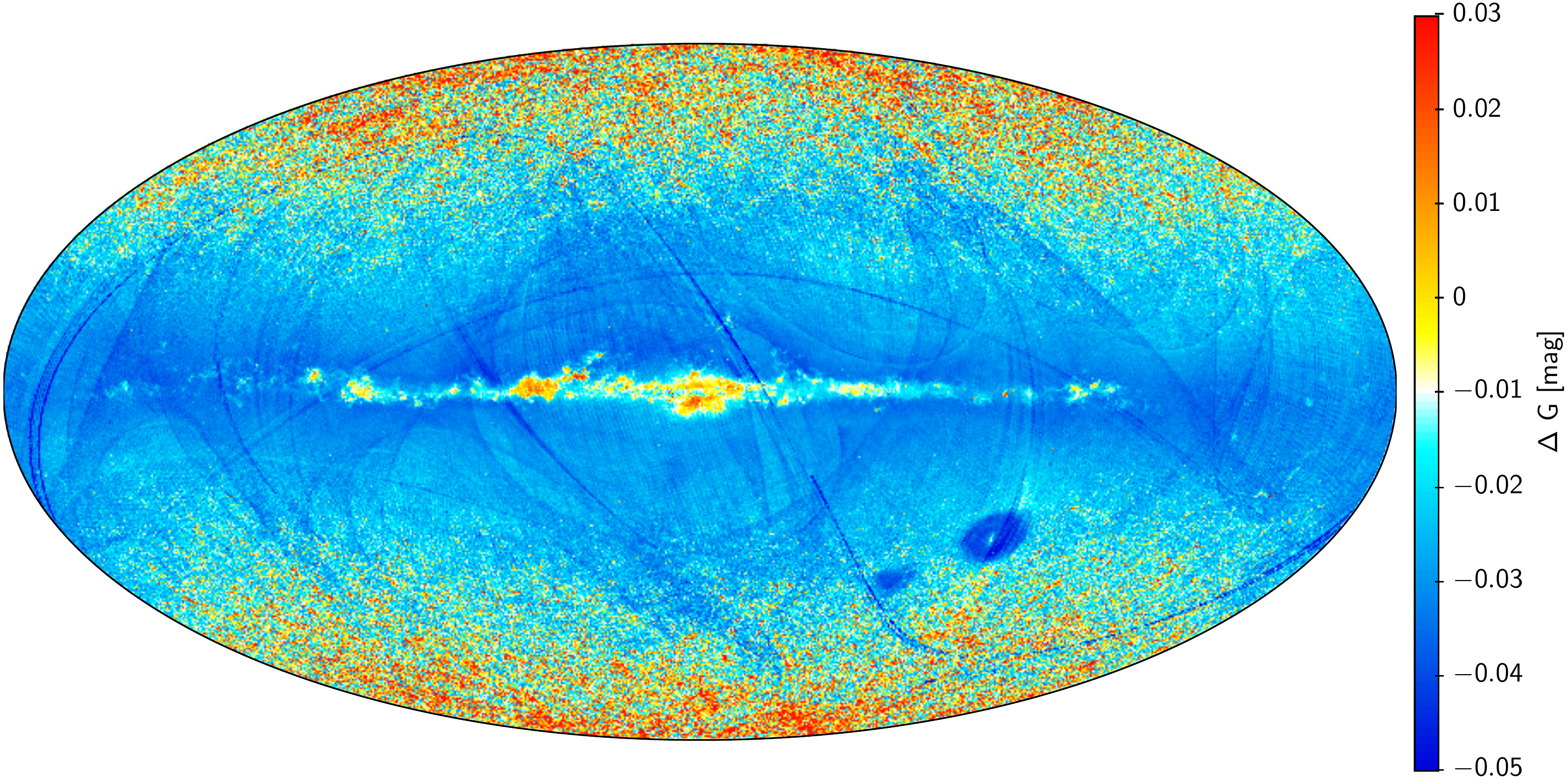}
\caption{
\label{fig:fot_dr3dr2}
Map in Galactic coordinates of the mean difference of the \gmag\ magnitudes 
between \egdr{3}\ and {\gdr{2}}.
}
\end{figure*}

Figure~\ref{fig:fot_dr3dr2} shows the full sky mean difference of the \gmag\
magnitudes between \egdr{3}\ and {\gdr{2}}. Although the comparison between
both releases is not straightforward because their photometric systems are
slightly different, as discussed in \cite{EDR3-DPACP-117}, the map does
illustrate specific areas where the differences are large, though it is 
generally at the level of a couple of hundredths of a magnitude. 
As in the case of
astrometry, we expect that the map mainly shows the errors
of {\gdr{2}} and hence the improvements reached in \egdr{3}.

\subsubsection{Comparisons of \bpminrp\ colours to a Milky Way model}\label{phot_model}
 
The median of the \bpminrp\ colour is computed in each healpix bin of the sky map, for  \egdr{3} data and GOG20 simulations. At intermediate and faint magnitudes, the differences can be large in the Galactic plane and are clearly linked to the extinction. At higher latitudes, the model is in agreement with the data at the level of 0.1 mag. 
However at bright magnitudes ($G<9$\,mag at intermediate latitudes and upwards, $G<12$\,mag in the plane), the data deviate from the values predicted by the model, from $-0.2$ mag to more than $-1$ mag at $G=5$. This discrepancy is slightly reduced compared to \gdr{2}. This could be a possible problem in the colour determination for those bright stars. However, since we are comparing the median value in healpix bins, an underestimate of blue bright stars in the model could rather explain this remaining difference, in particular in the brightest bins affected by large Poisson noise.

%
\subsection{Photometric precision}\label{phot_prec}
%
\subsubsection{Internal comparisons}\label{phot_prec_int}

\begin{figure}
\centering
\includegraphics[width=1.0\columnwidth]{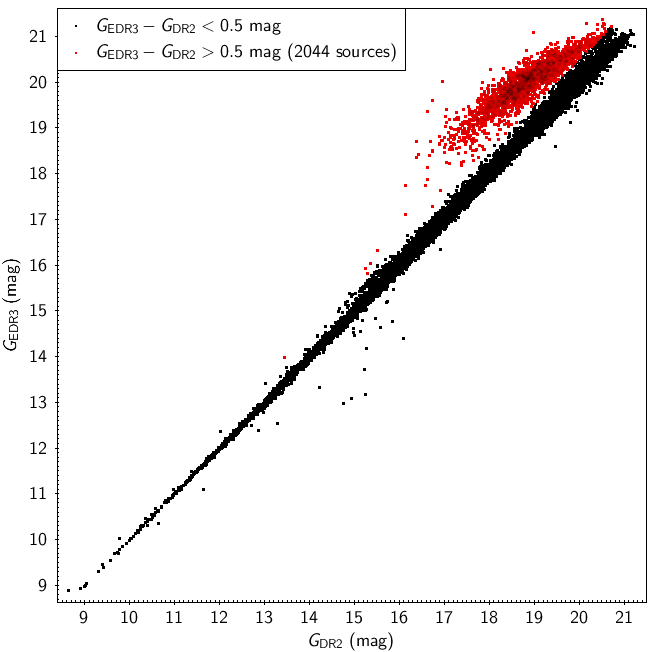}
\caption{Comparison between mean \gmag\ magnitudes provided in the \gdr{2} and \egdr{3} catalogues for 140\,635 RR Lyrae variables. Black and red points show sources with a difference in magnitudes of less and more than 0.5~mag, respectively.\label{fig:wp946_dr2_edr3}}
\end{figure}

A comparison between mean \gmag\ magnitudes provided in the \gdr{2} and \egdr{3} catalogues for a sample of 140\,635 sources classified as RR Lyrae variables in \gdr{2}, for which both estimates of the mean \gmag\ magnitude are available, shows that there is a difference in magnitudes of more than 0.5~mag for 2044 sources, with the \egdr{3} magnitudes being fainter. The \egdr{3} \gmag\  magnitudes plotted versus the \gdr{2} \gmag\ magnitudes for 140\,635  RR Lyrae stars are shown in Fig.~\ref{fig:wp946_dr2_edr3}, where sources which have a difference in magnitude  more and less than 0.5~mag are marked with red and black points, respectively. The figure shows that there is good agreement for a majority of stars, while for 2044 sources shown with red points there is a systematic shift in magnitudes, with the \egdr{3} magnitudes being fainter. Among these 2044 sources, 908  are listed by \cite{2019A&A...622A..60C} in their Table~C1 as known galaxies misclassified as RR Lyrae variables in \gdr{2}. Furthermore, a more detailed analysis of the 2044 stars performed with a dedicated pipeline shows that the vast majority of these sources are extended objects.  
The fainter magnitudes for these 2044 sources are, therefore, most likely due to
an improved background determination in the \egdr{3}\ processing.
              
%
\subsection{Photometric quality indicators}\label{phot_qual}

\begin{figure}[h]
\sidecaption
\includegraphics[width=7.0cm]{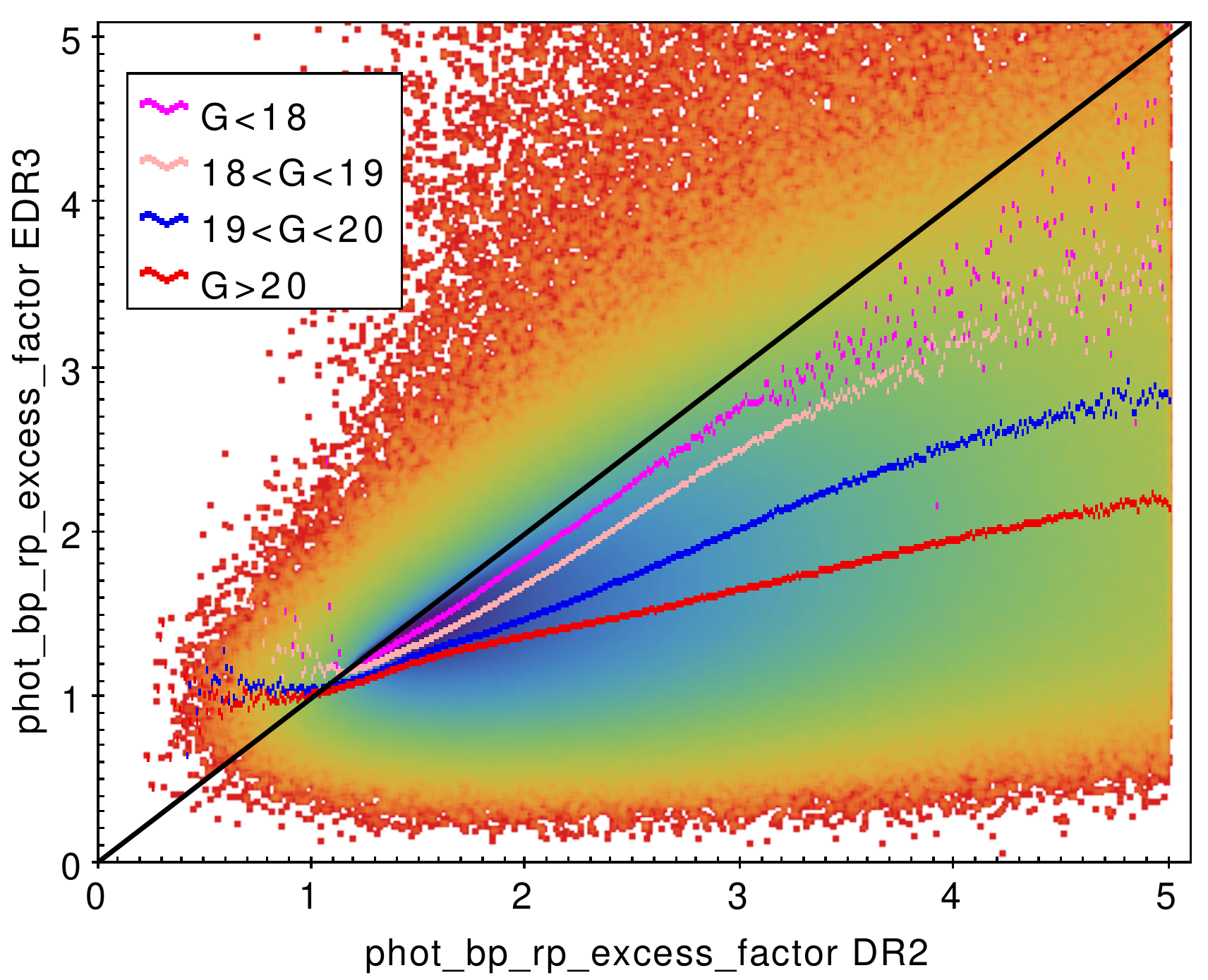}
\includegraphics[width=7.2cm]{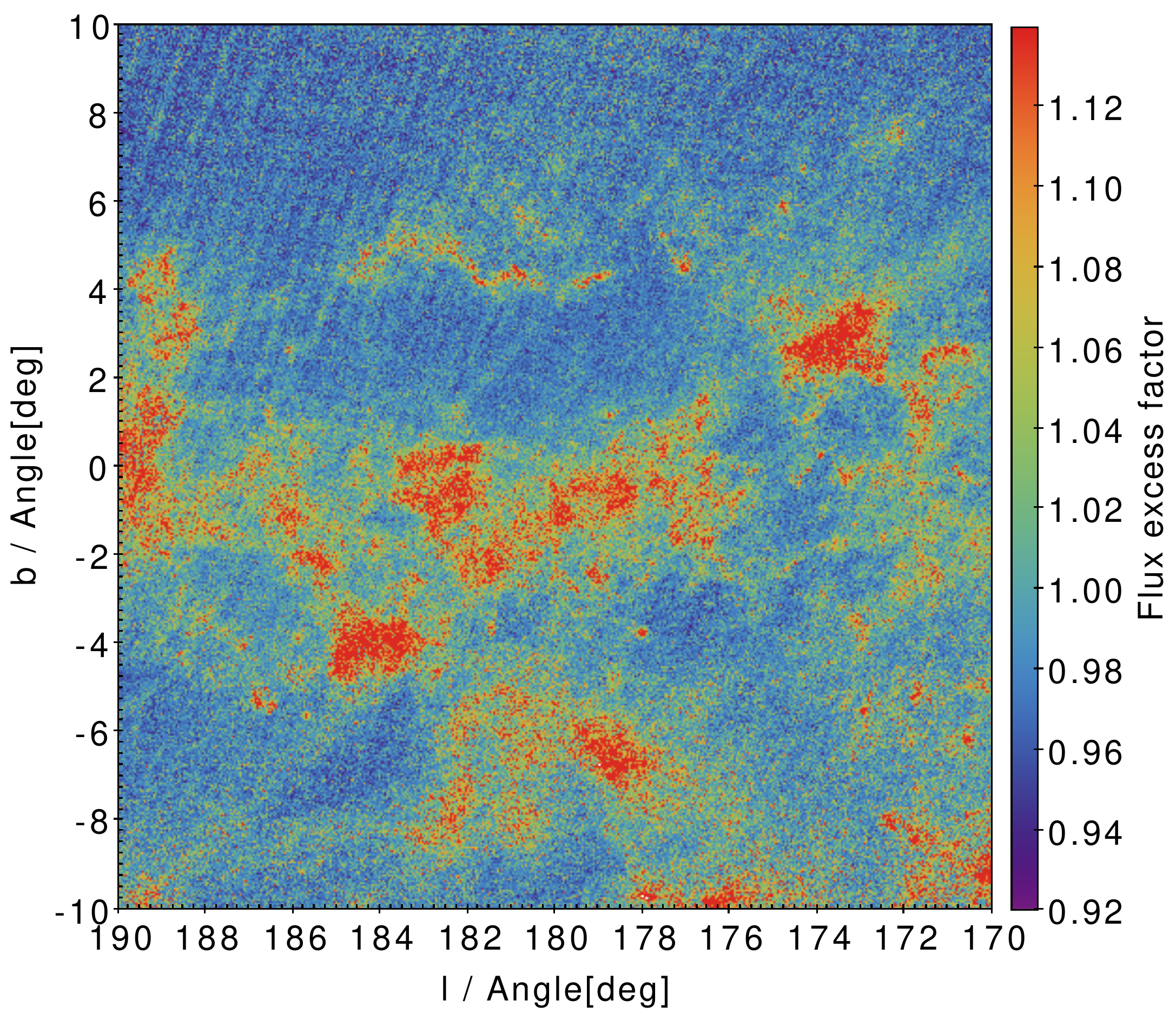}
\caption{
\label{fig:excess_flux_dr2_dr3}
{\em Top:} Excess flux for 36 million sources towards the direction $l\in(10^\circ,20^\circ)$ and the $b\in(-5^\circ,5^\circ)$
direction. The black line is the identity and the other ones are the median for different 
ranges of {\gmag}. 
{\em Bottom:} Excess flux for 14 million sources towards the anticentre direction.
The colour code accounts
for the ratio between the $\log_{\rm 10}$\,\dt{phot\_bp\_rp\_excess\_factor} and the relation
$0.05+0.039(G_{\rm BP}-G_{\rm RP}),$ which is the approximate locus of well behaved stars.
}
\end{figure}

\egdr{3}\ includes the flux excess factor,
\dt{phot\_bp\_rp\_excess\_factor}\footnote{\new{\cite{EDR3-DPACP-117} define 
a corrected factor $C^*$ that
takes the dependence of the \dt{phot\_bp\_rp\_excess\_factor} with
\bpminrp\ colour into account.}}
 as an indicator of the coherence among
{\gmag}, {\gbp}, and {\grp} fluxes.  It is sensitive to contamination by
close-by sources in dense fields, binarity, background subtraction problems, as
well as extended objects.
The improvement of
the full pipe-line calibrations in \egdr{3}\ yields a decrease in the excess
factor with respect to {\gdr{2}} as it can be seen in
Fig.~\ref{fig:excess_flux_dr2_dr3}. The fainter the magnitude is, the larger the
improvement, which is more noticeable for point-like sources with $G>19$~mag.
Sources with \dt{phot\_bp\_rp\_excess\_factor} $>5$ were published in {\gdr{2}}
without {\gbp} and {\grp} fluxes. In \egdr{3,}\ this filter has not been applied
and it is up to the users to decide on the use of the photometry in case of
a large excess of flux.  Small amplitude artefacts due to the scanning pattern
still remain (bottom panel in Fig.~\ref{fig:excess_flux_dr2_dr3}).
Criteria for filtering the incoherent fluxes are discussed in sect.~9.4 of \citep{EDR3-DPACP-117}.

\subsection{Photometry in crowded areas}

\begin{figure*}
 \begin{center}
\includegraphics[width=16cm]{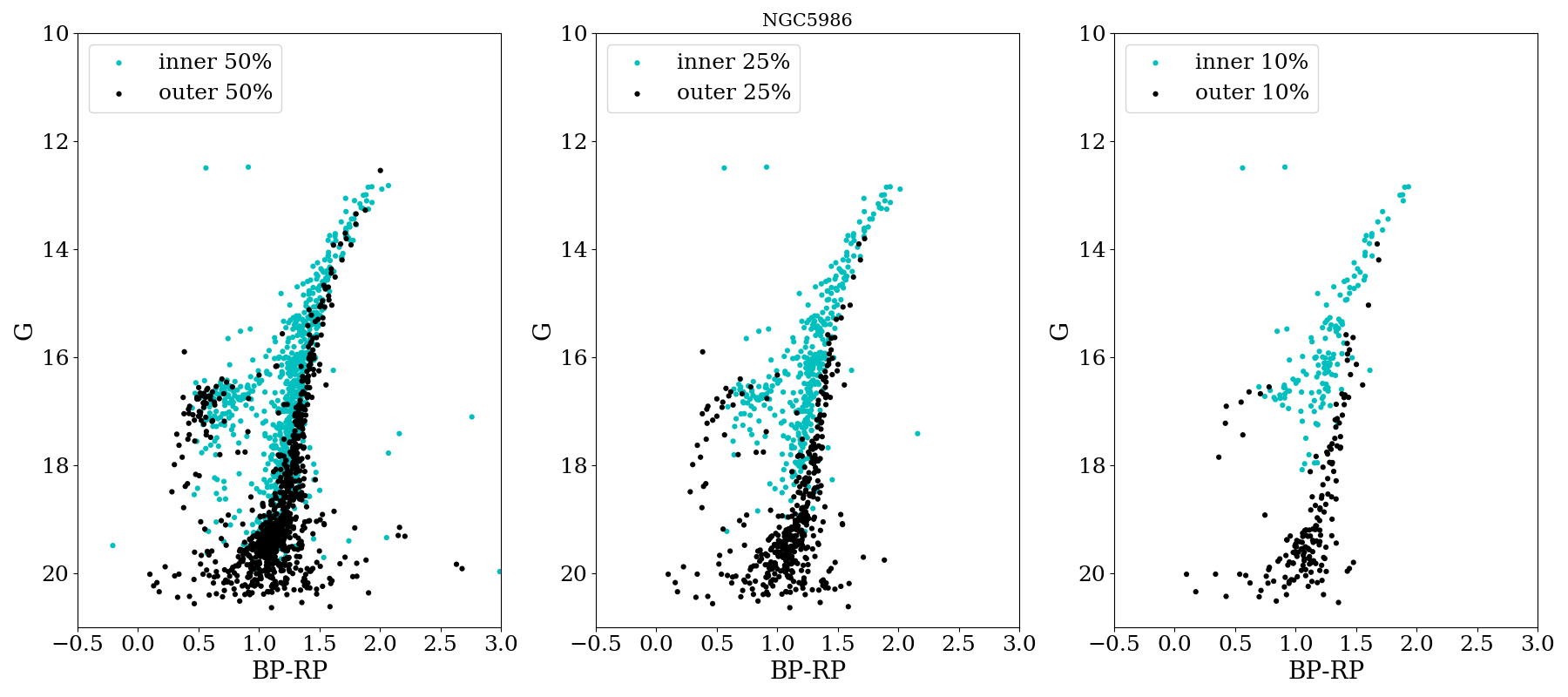}
\end{center}
\caption{Colour magnitude diagram in the inner (blue) and outer (black) regions of NGC~5986 in three different areas: inner 50\% and outer 50\% ({\em left}); inner 25\% and outer 25\% ({\em centre}); and inner 10\% and outer 10\% ({\em right}).  }\label{fig:WP947_NGC5986cmd}
\end{figure*}

Photometry in highly crowded areas is of lower quality than in non  crowded regions.  This is the case of all the globular clusters, where the photometry in the inner regions is shifted in colour and magnitude as an effect of crowding, with a large dispersion. This effect was visible in \gdr{2} and it is still present in \egdr{3}, in spite of the improvements in the number of observations and in the photometry.  See for instance Fig.~\ref{fig:WP947_NGC5986cmd}, showing the quality of the photometry in the inner and outer regions of  NGC~5986.

\subsubsection{Comparison with HST photometry in crowded areas}
\begin{figure}
 \begin{center}
\includegraphics[width=9.50cm]{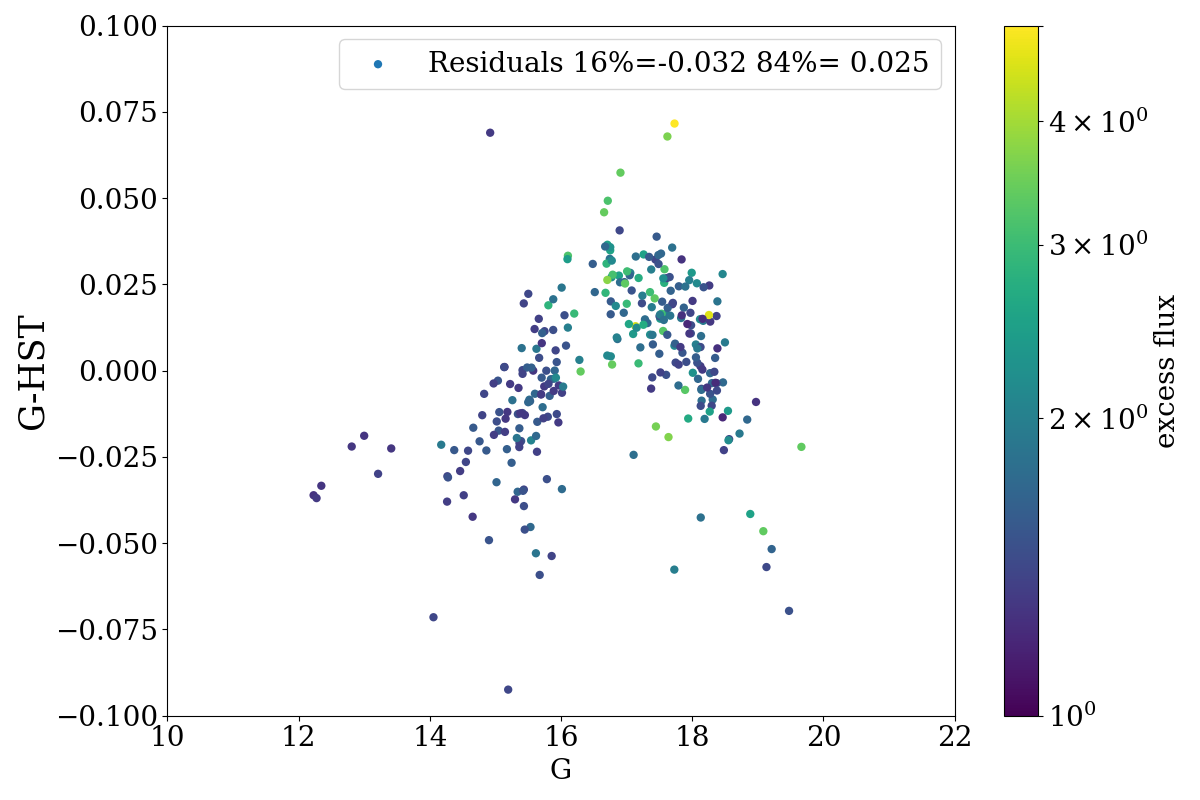}
\end{center}
\caption{Residuals of the difference between \gmag\ and  F606W HST magnitudes  as a function of \gmag\  in M4 in the colour range (\bpminrp) = 1.4-1.5\,mag (see text for detail).}\label{fig:WP947_M4f8}
\end{figure}

We compare  \gmag, \gbp, and \grp~  photometry with the high quality HST ACS/F606W magnitudes in M4 using the data set from  \cite{2013AN....334.1062B}. The precision of the HST photometry is at the level of a few milli-mag.
We subtracted the median difference between \gaia \ and HST photometry, and we calculated the variation of  the residuals  for stars having $1.4<\bpminrp<1.5$. We selected a fix   range to avoid colour effects.    The residuals around the median value  in \gmag\  are  of the order of 0.025-0.03 mag. These values are in agreement with the comparison with external catalogues presented in Sect.~\ref{phot_acc_ext}. However, a  clear trend with the magnitude and with the flux excess is visible (see Fig. \ref{fig:WP947_M4f8}). We recall here  that the flux excess can be considered as a proxy of the level of contamination due to neighbouring stars. The detected trend is an effect of the crowding, showing that faint stars located in regions with high contamination present larger residuals. No trend is present in \gbp~ and \grp,  albeit with a large dispersion due to the high level of contamination. 

\section{Global validation of \egdr{3}}\label{sec:global}
%
Here, we perform a global comparison of the statistical properties of
\egdr{3} to \gdr{2}.  This complements the detailed analyses presented
in the previous sections on the astrometry and the photometry. For the
statistical comparison, we use the {\sl Kullback--Liebler Divergence}
(KLD) to establish the degree of correlations and clustering between
observables or, more generally, entries in the catalogue. The KLD is defined as 
\begin{equation}
\label{eqn:kld_eq}
    KLD = - \int d^{n} x \, p({\bf x}) \, \log[p({\bf x})/q({\bf x})]
,\end{equation} 
where ${\bf x}$ is the $n$-dimensional vector of the observables
considered, $p({\bf x})$ is the distribution of these observables in
the dataset, and $q({\bf x})= \Pi_{i}p_{i}(x_{i})$, that is to say the product
of marginalised 1D distributions of each of the observables. Large
values of the KLD indicate highly structured data, while low values of the
KLD (below $\sim 0.5$) correspond to little information content. 

We compared the performance of \egdr{3} to \gdr{2} in two ways:
\textbf{1)} we considered all 2D subspaces ($n=2$), that is\  all possible
combinations of pairs of observables in the catalogues (e.g.
$(G, \varpi)$, $(\sigma_\varpi, \mu_\delta)$, etc); and \textbf{2)} we
computed the KLD for 3D subspaces ($n=3$) for small regions on the sky, with
two of the three observables being $\alpha$ and $\delta$. In both cases, we
excluded outliers by considering 99\% of the data for each
observable (e.g.\ for the $G$-magnitude, the range used for \egdr{3}
is 11.69 - 21.62).

\begin{figure}
\includegraphics[width=1.\columnwidth]{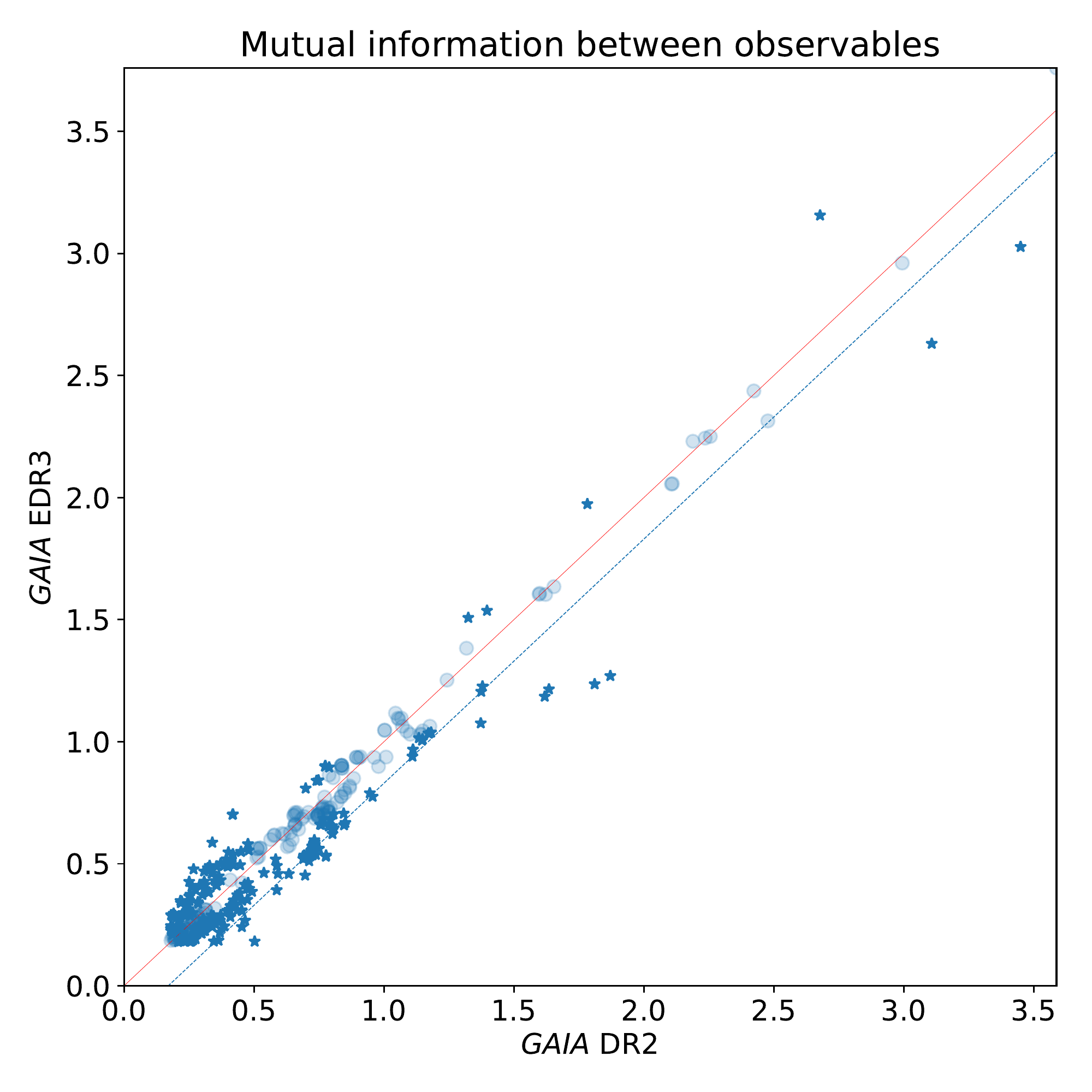} 
\caption{Comparison of 2D KLD between \egdr{3} and \gdr{2}. The 1:1 line is shown in red, while subspaces for which KLD was deviated by at least 10\% with respect to \texttt{DR2} are shown as `*', respectively. The blue dashed line is a guide to the low KLD sequence at an offset of about 0.17 from the 1:1 line. \label{fig:kld_int3_dr2_all}}
\end{figure}

Figure~\ref{fig:kld_int3_dr2_all} shows the distribution of KLD values
for the 2D comparison between \egdr{3} and \gdr{2}. While a few 2D
subspaces lie very close to the 1:1 line, indicating a similar behaviour
in \egdr{3} and \gdr{2}, there is a large set of subspaces for which
the KLD has decreased in \egdr{3}. Furthermore, these
sets follow a parallel track to the 1:1 line, with an offset of about
0.17. These are mostly subspaces involving astrometric uncertainties. This decrease can be fully attributed to the
smaller uncertainties since the KLD is the relative
entropy $\log[p(x)/q(x)]$ weighted by the distribution of
$p(x)$ values, which have become smaller in the case of the
uncertainties. The subspaces with the largest decrease in the KLD for
\egdr{3} are those including photometric uncertainties, as may be
expected. On the other hand, a few subspaces depict an increase in the
KLD, and these are mostly subspaces combining the number of observations
or visibility periods. In this case, the range of the parameter has
increased significantly from \gdr{2} to \egdr{3}.

We also computed the KLD in 3D in four circular patches on the sky,
each with a radius of 5~deg, and centred on ($l$,$b$) = $(-90\degr, -45\degr)$,
$(-90\degr, 45\degr)$, $(90\degr, -45\degr),$ and $(90\degr, 45\degr),$ respectively. As for \gdr{2},
patches that are symmetric with respect to the Galactic plane exhibit
a similar behaviour.  However, there is also a strong dependence on
their location with respect to the scanning law. When compared with
\gdr{2}, we find \egdr{3} to be systematically less clustered in
subspaces containing astrometric or photometric
information, for example (see e.g. Fig.~\ref{fig:espatial_pattern}). We interpret
this as an improvement on the systematics introduced by the scanning
law. Nonetheless, we still find some amount of residual clustering in
the astrometric parameters which appear to be sensitive to the
orientation of the visits, and not just the number of observations (as
seen, for example, in Fig.~\ref{fig:patternPlx}).

The global analyses performed here thus confirm that the quality of
\egdr{3} has improved significantly compared to that of \gdr{2}. The
largest improvement is found for the photometry, since some
`systematics' associated to the imprint of the scanning law pattern
are still present in the astrometric parameters and their uncertainties.

\section{Conclusions and recommendations}\label{sec:conclusions}
%
 
\egdr{3}\ provides updated astrometry and photometry for 1\,811\,709\,771 sources.
Of these, 19.0\% have a 2p astrometric solution, that is\ to say only a position,
32.3\% have a 5p solution, that is\ also including the parallax and proper motion, and
48.7\% have a 6p solution where also a colour parameter, the pseudo-colour, is determined.
For the photometry, the catalogue gives \gmag\ magnitudes for 99.7\% of the sources, \gbp\ for
85.1\%,  \grp\ for 85.8\%, and a \bpminrp\ colour for 85.0\%. 

In this paper, we have presented a series of tests aimed at illustrating the quality of the
catalogue, with an emphasis on known issues as it is natural for a validation paper. The
idea has been that these examples can serve as a guide for actual use cases. In many
tests, we have used rather strict selection criteria in order to better answer certain
questions, but selection criteria should always be chosen to fit the case at hand. \new{For convenience, we summarise the principal recommendations from this
validation exercise in Table~\ref{tab:recom}. Additional advice and recommendations can be found in the astrometric and photometric processing papers \citep{EDR3-DPACP-128, EDR3-DPACP-117}.}

\begin{table}
\begin{center}
\caption{Principal recommendations for using \egdr{3}.\label{tab:recom}}
{\small
\begin{tabular}{p{25mm}p{39mm}p{13mm}}
\hline\hline
{Item}  &  {Recommendation} & {Reference}  \\ \hline
Find a DR2 source & Use dedicated neighbourhood table & Sect.~\ref{sec:comp_source_id}\\
Isolated source? & For astrometry: check \dt{ruwe} and image parameter statistics. For \gbp\ and \grp: check statistics on blended and contaminated transits & Fig.~\ref{fig:spuriousGCpm}, Table~\ref{tab:acro}\\
Parallax zero point & Consider applying the correction by \cite{EDR3-DPACP-132} & Sect.~\ref{sec:ast}\\ 
Spurious astrometry & Check -- statistically -- for the presence of spurious parallaxes and proper motions & Sect.~\ref{sec:ast_spur}\\
Missing \gmag & Alternative \gmag\ photometry is available & Sect.~\ref{phot_spur}\\
\gmag\ magnitude & A small correction should be added for 6p solutions & Sect.~\ref{phot_6p}\\
\gbp\ magnitude & Be aware of a strong bias for \gbp\ $> 20.5$\,mag & Sect.~\ref{phot_thres}\\
\grp\ magnitude & Be aware of a strong bias for \grp\ $> 20$\,mag & Sect.~\ref{phot_thres}\\
Colour & Use $G-G_{\rm RP}$ instead of $G_{\rm BP}-G_{\rm RP}$ when faint red sources are included & Sect.~\ref{phot_thres}\\
\gmag, \gbp, \grp\ coherence& Check the $C^*$ value defined by \cite{EDR3-DPACP-117} & Sect.~\ref{phot_qual}\\
\hline
\end{tabular}
}
\end{center}
\end{table}

The catalogue is the third in a series, and it is natural to check where it stands
compared to the previous release,\ \gdr{2}.
In particular, we note the following.\ 
\begin{itemize}
\item It contains 7\% more sources.
\item It presents parallaxes and proper motions for 10\% more sources.
\item It has a high completeness until $G \sim 19$\,mag, cf.\ Sect.~\ref{sec:comp_large}.
\item The completeness has improved in dense areas, cf.\ Sect.~\ref{sec:compleGlobular}.
\item The angular resolution has improved but is still
dropping fast below 0\farcs7, cf.\ Sect.~\ref{sec:comp_small}.
\item Source identifiers from \gdr{2}\ have, in general, been maintained for 97\% of the sources, cf.\
Sect.~\ref{sec:comp_source_id}, but
it is not advisable to rely on the identifier alone.
If the counterpart of a \gdr{2}\ source is sought, we recommend using 
the table \dt{gaiaedr3.dr2\_neighbourhood} instead, which is available in the \gaia\ archive.
\item A little more than half a percent of \gdr{2}\ sources are not found in \egdr{3} within 50\,mas, cf.\ Sect.~\ref{sec:comp_source_id}. They are mostly
faint and some are likely to be spurious sources.
\item The catalogue contains several fields that help to identify sources
with a close neighbour. In addition to the parameters detailing the internal 
consistencies of the astrometric and photometric solutions already in
\gdr{2}, the catalogue now also provides statistics on the images themselves, such as \dt{ipd\_frac\_multi\_peak}, \dt{ipd\_harmonic\_gof\_amplitude}, etc., cf.\ Table~\ref{tab:acro}.
\item For a guide to the full list of parameters, we recommend the datamodel description in
the online documentation.\footnote{\url{https://gea.esac.esa.int/archive/documentation/GEDR3/Gaia\_archive/chap\_datamodel/}}
\end{itemize}


For the astrometry, the precision has significantly benefited from the 
additional year of observations. In addition, we notice the following. \begin{itemize}
\item There are fewer sources than in \gdr{2}\ with incomplete astrometry (2p),
for example\ 1.5\% as compared to 2.4\% for $G<19$\,mag, which is mainly due to 
the increase in number of observations, cf.\ Sect.~\ref{sec:2p}. 
\item A significant number of 2p solutions, more than half for $G<19$\,mag, are caused by the insufficiency
of a 5p or 6p source model, cf.\ Sect.~\ref{sec:2p}.
\item In specific sky areas, there can be mean differences in position with 
respect to \gdr{2}\ at the 1\,mas level.
\item The systematic errors in parallax -- as shown by QSOs -- are significantly
smaller than in \gdr{2}, cf.\ Sect.~\ref{sec:ast_large}. They can be further diminished by applying the 
corrections detailed
in \cite{EDR3-DPACP-132}, and we recommend following the guidelines from
that paper. 
\item High proper motion sources now have a much improved reliability with
no negative parallaxes for sources with motions above 300\masyr, cf.\
Sect.~\ref{sec:hpm}.
\item We have 1.6\% spurious solutions among sources with 
$\varpi/\sigma_\varpi > 5$, cf.\ Sect.~\ref{sec:ast_spur}, but this fraction is much less for brighter sources 
and for 5p solutions. 
We recommend testing any specific sample, designed to contain only positive parallaxes, by also selecting the corresponding
sample with negative parallaxes.
\item Parallax uncertainties are underestimated, but less than for \gdr{2},
cf.\ Sect.~\ref{sec:ast_precision}.
\item Parallaxes have up to a +0.02\,mas level offset in zero point for sources brighter than
$G=13$\,mag as shown in binaries and in clusters. It is 
mostly removed when applying the \cite{EDR3-DPACP-132} parallax correction.
\item Parallaxes for 6p solutions show a clear correlation with the
pseudo-colour in the LMC, which is largely removed with the
\cite{EDR3-DPACP-132} correction.
\item Proper motions in right ascension have a small offset of 
0.01\,mas\,yr$^{-1}$ for sources brighter than $G=13$\,mag, cf.\ Figs.~\ref{fig:DiffBinaryPlx} and \ref{fig:wp947pmra}. This is seen
in proper motion pairs as well as in clusters.
\item The quality indicators \dt{ruwe} and \dt{astrometric\_gof\_al} are 
strongly underestimated for bright sources in crowded areas, cf.\ Sect.~\ref{sec:vbs}.
\end{itemize}



For the photometry, nearly all issues found with \gdr{2}\ \citep{2018A&A...616A..17A} are either solved
or have improved significantly. Still, the
blue and red \gbp\ and \grp\ photometry has a series of issues of its
own. This photometry is based on prism spectra and has therefore -- by design -- a limited
angular resolution. Each observation contains a substantial flux from the sky
background, limiting the performance for faint sources.
To help judge the
reliability of the photometry, the excess factor,
\dt{phot\_bp\_rp\_excess\_factor}, gives a simple measure of the consistency
between the three fluxes. For \gbp\ and \grp, which collect the flux in a
relatively wide window, the number of transits with other sources within
(\dt{phot\_bp\_n\_blended\_transits}) or close to
(\dt{phot\_bp\_n\_contaminated\_transits}) the window, can be useful but
it is important to only count, of course, well known sources.

For the photometry we notice the following.
\begin{itemize}
\item The trend in \gmag\ as a function of \gmag, which was pronounced in
\gdr{2}, has been significantly reduced in \egdr{3}.
\item The indications of a discontinuity in the \gmag\ magnitude  around $G=10.87$ and 13\,mag are
much weaker in \egdr{3} than in \gdr{2}, cf.\ Sect.~\ref{phot_acc_int}.
\item Sources, where a reliable colour was not known at the beginning of the
\egdr{3}\ processing were processed using a default colour.
They constitute a significant fraction of the catalogue and
\cite{EDR3-DPACP-117} recommend a correction to the \gmag\ photometry for
such sources, in particular those with astrometric 6p solutions, see\ Sect.~\ref{phot_6p} for example, where it is demonstrated to work.
The correction is typically a hundredth of a magnitude and can of course 
only be applied if a reliable colour is now known, which is the case for 86\% of the 6p solutions.
\item The \gbp\ and \grp\ photometry is much less affected by systematic errors
in the background subtraction than they were in \gdr{2}.
\item The completeness of \gbp\ and \grp\ is reduced in dense fields.
\item Photometry in \gbp\ that is fainter than about 20.5\,mag and in \grp\ that is fainter
than about 20\,mag is heavily biased towards brighter values as illustrated in
the simulations in Figs.~\ref{fig:lowFlux1} and \ref{fig:lowFlux2}. The cause is
discussed in \cite{EDR3-DPACP-117} and well understood, cf.\ 
\secref{phot_thres}. We therefore recommend
the use of the colour $G-G_{\rm RP}$ instead of $G_{\rm BP}-G_{\rm RP}$ 
for samples including faint, red sources (\gbp\ fainter than about 20.5\,mag).
\end{itemize}

Also a global, statistical analysis, cf.\ Sect.~\ref{sec:global}, confirms that
the systematics of \egdr{3} have improved significantly compared to that of
\gdr{2}. The more notable improvement, seen in this way, is found for the
photometry, since some `systematics' associated to the imprint of the
scanning law pattern are still present in the astrometric parameters and their
uncertainties.

\begin{acknowledgements}

This work has made use of data from the European Space Agency (ESA) mission Gaia (\url{https://www.cosmos.esa.int/gaia}), processed by the Gaia Data Processing and Analysis Consortium (DPAC, \url{https://www.cosmos.esa.int/web/gaia/dpac/consortium}). Funding for the DPAC has been provided by national institutions, in particular the institutions participating in the Gaia Multilateral Agreement.

This work was supported by the Spanish Ministry of Science, Innovation and
University (MICIU/FEDER, UE) through grants RTI2018-095076-B-C21,
ESP2016-80079-C2-1-R, and the Institute of Cosmos Sciences University of
Barcelona (ICCUB, Unidad de Excelencia 'Mar\'{\i}a de Maeztu') through grants
MDM-2014-0369 and CEX2019-000918-M.

TM, DB, AG and EL acknowledge financial support from the Agenzia Spaziale Italiana (ASI) provided through contracts 2014-025-R.0, 2014-025-R.1.2015 and 2018-24-HH.0 to the Italian Istituto Nazionale di Astrofisica. 

PK acknowledges support from the European Research Council (ERC) under the European Union’s Horizon 2020 research and innovation programme under grant agreement No 695099 (project CepBin), and the French Agence Nationale de la Recherche (ANR), under grant ANR-15-CE31-0012-01 (UnlockCepheids).

This research has made an extensive use of Aladin and the SIMBAD, VizieR
databases operated at the Centre de Donn\'ees Astronomiques (Strasbourg) in
France and of the software TOPCAT \citep{2005ASPC..347...29T}.

We finally wish to thank Alessandro Spagna for his comments to an earlier version
of this paper.
\end{acknowledgements}

\bibliographystyle{aa} 
\bibliography{biblio} 

\appendix

\section{Completeness in globular clusters}\label{AppendixA}

\begin{table*}
\scriptsize
\begin{center}
        \caption{ \label{tab:wp947completeness26gcs} Completeness tables for 26 GCs. Core: $r<0.5$ arcmin. Out: $0.5<r<2.2$ arcmin. }
        \addtolength{\tabcolsep}{-2pt}
        \begin{tabular}{l l c c c c c c c c c c}
        \hline
        \hline
   Name & Region &                                    \multicolumn{10}{c}{$G$}                                                 \\
        &        & 11 -- 13 & 12 -- 14 & 13 -- 15 & 14 -- 16 & 15 -- 17 & 16 -- 18 & 17 -- 19 & 18 -- 20 & 19 -- 21 & 20 -- 22 \\
        \hline
LYN07 & core     & -- & -- & -- & -- & -- & -- & 50 & 37 & 14 & 4 \\ 
LYN07 & out      & -- & -- & -- & -- & 57 & 52 & 39 & 34 & 19 & 6 \\  
\hline
NGC0104 & core   & 40 & 7 & 2 & 0 & 0 & 0 & 0 & 0 & 0 & 0 \\ 
NGC0104 & out    & 100 & 75 & 68 & 47 & 22 & 3 & 1 & 0 & 0 & 0 \\ 
\hline
NGC0288 & core   & -- & -- & -- & -- & -- & -- & 72 & 51 & 21 & 3 \\ 
NGC0288 & out    & -- & -- & -- & 99 & 92 & 89 & 81 & 67 & 37 & 10 \\ 
\hline
NGC1261 & core   & -- & -- & -- & 82 & 77 & 57 & 20 & 2 & 0 & 0 \\ 
NGC1261 & out    & -- & -- & -- & 100 & 100 & 94 & 73 & 42 & 18 & 4 \\ 
\hline
NGC1851 & core   & -- & -- & 50 & 42 & 31 & 13 & 2 & 0 & 0 & 0 \\ 
NGC1851 & out    & -- & -- & -- & 99 & 94 & 80 & 50 & 25 & 10 & 2 \\ 
\hline
NGC2298 & core   & -- & -- & -- & -- & 89 & 85 & 52 & 23 & 8 & 2 \\ 
NGC2298 & out    & -- & -- & -- & -- & 98 & 96 & 92 & 83 & 58 & 19 \\ 
\hline
NGC4147 & core   & -- & -- & -- & -- & -- & 73 & 50 & 24 & 9 & 2 \\ 
NGC4147 & out    & -- & -- & -- & -- & -- & 100 & 87 & 85 & 44 & 12 \\ 
\hline
NGC5053 & core   & -- & -- & -- & -- & -- & -- & -- & 96 & 61 & 20 \\ 
NGC5053 & out    & -- & -- & -- & -- & -- & 100 & 96 & 97 & 65 & 19 \\ 
\hline
NGC5139 & core   & -- & -- & 25 & 19 & 6 & 0 & 0 & 0 & 0 & 0 \\ 
NGC5139 & out    & 97 & 64 & 51 & 38 & 13 & 1 & 0 & 0 & 0 & 0 \\ 
\hline
NGC5272 & core   & -- & -- & 74 & 62 & 45 & 12 & 1 & 0 & 0 & 0 \\ 
NGC5272 & out    & -- & 100 & 100 & 93 & 84 & 56 & 25 & 9 & 2 & 0 \\ 
\hline
NGC5286 & core   & -- & -- & 68 & 55 & 39 & 19 & 3 & 0 & 0 & 0 \\ 
NGC5286 & out    & -- & -- & 91 & 94 & 89 & 74 & 50 & 21 & 6 & 1 \\ 
\hline
NGC5466 & core   & -- & -- & -- & -- & -- & -- & -- & 79 & 46 & 14 \\ 
NGC5466 & out    & -- & -- & -- & -- & 100 & 100 & 98 & 97 & 67 & 21 \\ 
\hline
NGC5927 & core   & -- & -- & -- & 79 & 77 & 61 & 16 & 1 & 0 & 0 \\ 
NGC5927 & out    & -- & -- & 91 & 83 & 87 & 80 & 58 & 19 & 5 & 0 \\ 
\hline
NGC5986 & core   & -- & -- & -- & 66 & 64 & 41 & 10 & 0 & 0 & 0 \\ 
NGC5986 & out    & -- & -- & 93 & 90 & 92 & 79 & 54 & 22 & 7 & 1 \\ 
\hline
NGC6121 & core   & -- & -- & -- & 76 & 64 & 47 & 23 & 3 & 0 & 0 \\ 
NGC6121 & out    & -- & 96 & 93 & 86 & 85 & 74 & 58 & 36 & 12 & 1 \\ 
\hline
NGC6205 & core   & -- & -- & 89 & 65 & 35 & 7 & 0 & 0 & 0 & 0 \\ 
NGC6205 & out    & 98 & 100 & 100 & 95 & 81 & 47 & 16 & 4 & 0 & 0 \\ 
\hline
NGC6366 & core   & -- & -- & -- & -- & -- & -- & 78 & 69 & 52 & 20 \\ 
NGC6366 & out    & -- & -- & -- & 89 & 85 & 82 & 85 & 82 & 62 & 24 \\ 
\hline
NGC6397 & core   & -- & -- & 68 & 55 & 41 & 21 & 6 & 0 & 0 & 0 \\ 
NGC6397 & out    & -- & 100 & 98 & 90 & 84 & 73 & 53 & 30 & 9 & 1 \\ 
\hline
NGC6656 & core   & -- & -- & 82 & 63 & 24 & 3 & 0 & 0 & 0 & 0 \\ 
NGC6656 & out    & 94 & 93 & 85 & 75 & 55 & 17 & 3 & 0 & 0 & 0 \\ 
\hline
NGC6752 & core   & -- & 85 & 61 & 39 & 14 & 3 & 0 & 0 & 0 & 0 \\ 
NGC6752 & out    & -- & 100 & 100 & 93 & 78 & 48 & 22 & 6 & 0 & 0 \\ 
\hline
NGC6779 & core   & -- & -- & -- & -- & 83 & 72 & 37 & 7 & 1 & 0 \\ 
NGC6779 & out    & -- & -- & 92 & 90 & 88 & 86 & 72 & 51 & 24 & 5 \\ 
\hline
NGC6809 & core   & -- & -- & -- & -- & -- & 59 & 29 & 6 & 0 & 0 \\ 
NGC6809 & out    & -- & -- & 100 & 100 & 93 & 81 & 53 & 19 & 2 & 0 \\ 
\hline
NGC6838 & core   & -- & -- & -- & -- & -- & 79 & 68 & 44 & 15 & 1 \\ 
NGC6838 & out    & -- & -- & 93 & 91 & 78 & 87 & 84 & 70 & 41 & 9 \\ 
\hline
NGC7099 & core   & -- & -- & -- & 56 & 41 & 17 & 5 & 1 & 0 & 0 \\ 
NGC7099 & out    & -- & -- & -- & 100 & 91 & 77 & 56 & 34 & 13 & 2 \\ 
\hline
PAL01 & core     & -- & -- & -- & -- & -- & -- & -- & -- & 86 & 47 \\ 
PAL01 & out      & -- & -- & -- & -- & -- & -- & -- & 84 & 80 & 28 \\ 
\hline
PAL02 & core     & -- & -- & -- & -- & -- & -- & 93 & 80 & 37 & 12 \\ 
PAL02 & out      & -- & -- & -- & -- & -- & -- & 92 & 91 & 62 & 22 \\ 
\hline
\end{tabular}
\end{center}
\end{table*}

%
\section{\gaia-specific terms}\label{sec:acronyms}
%
Below, in \tabref{tab:acro}, we give short definitions of \gaia -related terms
appearing in this paper.  Several are fields in the \gaia\ catalogue and
detailed explanations are available in the \egdr{3}\ datamodel description.

\begin{table*}
\begin{center}
\caption{Short definitions for \gaia -related terms.\label{tab:acro}}
{\small
\begin{tabular}{ll}
\hline\hline
\textbf{Term}  &  \textbf{Description}  \\ \hline
AC & {\gaia} ACross scan (direction) \\
\dt{astrometric\_gof\_al} & GoF for the astrometric solution\\
GoF & goodness of fit, reduced $\chi^2$, e.g.\ for image fitting or for the astrometric solution\\
IPD & Image Parameters Determination, estimating location and flux for each CCD window  \\
\dt{ipd\_frac\_multi\_peak} & percentage of CCD transits where additional images were seen\\
\dt{ipf\_frac\_odd\_win} & percentage of incompletely sampled CCD transits\\
\dt{ipd\_harmonic\_gof\_amplitude} & scan angle dependent variation of GOF for the image fitting\\
\dt{ipd\_harmonic\_gof\_phase} & phase of the IPD GoF versus position angle of scan\\
\dt{parallax\_over\_error} & $\varpi/\sigma_\varpi$ \\
\dt{PhotPipe} & photometric pipeline, software system for the photometric calibration\\
\dt{phot\_bp\_rp\_excess\_factor} & sum of \gbp\ and \grp\ fluxes divided by the \gmag\ flux\\
\dt{pseudocolour} & wave number determined as the 6th parameter for 6p astrometric solutions\\
\dt{ruwe} & renormalised unit weight error\\
SSC & spectral shape coefficient: the flux in a certain section of the \gbp\ or \grp\ spectrum\\
visibility period & set of transits for a source with no time gap exceeding four days\\
\hline
\end{tabular} 
}
\end{center}
\end{table*}


\end{document}